\documentclass{aa}
\usepackage{hyperref}
\hypersetup{
    colorlinks=true,
    linkcolor=purple,
    citecolor=blue,   
    urlcolor=cyan,
    }
\usepackage[utf8]{inputenc}
\usepackage{amsmath,amssymb}
\usepackage[varg]{txfonts}
\usepackage{xcolor}
\usepackage{natbib}
\bibpunct{(}{)}{;}{a}{}{,}
\usepackage{float}
\usepackage{xspace}
\usepackage{placeins}
\setcitestyle{notesep={; }}
\usepackage{orcidlink}
\graphicspath{{./plots/}}

\title{BlazEr1: The eROSITA blazar catalog}
\subtitle{Blazars and blazar candidates in the first eROSITA survey}
\author{S.~H\"ammerich\inst{\ref{inst:remeis}}
\and A.~Gokus\inst{\ref{inst:wustl}}
\and F.~McBride\inst{\ref{inst:bowdoin}}
\and P.~Weber\inst{\ref{inst:remeis}}
\and L.~Marcotulli\inst{\ref{inst:desy}, \ref{inst:clemson}}
\and A.~Zainab\inst{\ref{inst:remeis}}
\and W.~Collmar\inst{\ref{inst:mpe}}
\and M.~Salvato\inst{\ref{inst:mpe}}
\and J.~Wolf\inst{\ref{inst:mpe},\ref{inst:mpia}}
\and T.~Sbarrato\inst{\ref{inst:INAF_merate}}
\and S.~Belladitta\inst{\ref{inst:mpia},\ref{inst:INAF_bologna}}
\and J.~Buchner\inst{\ref{inst:mpe}}
\and S.~Saeedi\inst{\ref{inst:remeis}}
\and L.~Dauner\inst{\ref{inst:remeis}}
\and M.~Lorenz\inst{\ref{inst:remeis}}
\and O.~K\"onig\inst{\ref{inst:CfA}}
\and C.~Kirsch\inst{\ref{inst:remeis}}
\and K.~Berger\inst{\ref{inst:remeis}}
\and S.~Bahic\inst{\ref{inst:aip}}
\and D.~Tubín-Arenas\inst{\ref{inst:aip}}
\and M.~Krumpe\inst{\ref{inst:aip}}
\and D.~Homan\inst{\ref{inst:cam},\ref{inst:aip}}
\and A.~Markowitz\inst{\ref{inst:ncac}}
\and P.~Benke\inst{\ref{inst:JMU},\ref{inst:MPIfR},\ref{inst:GFZ}}
\and F.~R\"osch\inst{\ref{inst:JMU},\ref{inst:MPIfR}}
\and P.~Rajasekar~Kavitha\inst{\ref{inst:remeis}}
\and H.~Tambe\inst{\ref{inst:remeis}}
\and M.~Kadler\inst{\ref{inst:JMU}}
\and E.~Ros\inst{\ref{inst:MPIfR}}
\and R.~Ojha\inst{\ref{inst:nasaHQ}}
\and J.~Wilms\inst{\ref{inst:remeis}}
}
\date{Received 27 October 2025 / Accepted 4 March 2026}
\authorrunning{H\"ammerich et al.}

\institute{\label{inst:remeis}Dr.\ Karl Remeis-Sternwarte and Erlangen Centre for Astroparticle Physics, Friedrich-Alexander Universit\"at Erlangen-N\"urnberg, Sternwartstr.~7, 96049 Bamberg, Germany\\
    \email{steven.haemmerich@fau.de}
\and\label{inst:wustl}Department of Physics \& McDonnell Center for the Space Sciences, Washington University in St.\ Louis, One Brookings Drive, St.\ Louis, MO 63130, USA
\and\label{inst:bowdoin}Department of Physics and Astronomy, Bowdoin College, Brunswick, ME 04011,
USA
\and\label{inst:desy}Deutsches Elektronen-Synchrotron DESY, Platanenallee~6, 15738 Zeuthen, Germany
\and\label{inst:clemson}Department of Physics and Astronomy, Clemson University, Clemson, SC 29631, USA
\and\label{inst:mpe}Max-Planck-Institut für extraterrestrische Physik, Gießenbachstraße 1, 85748 Garching, Germany
\and\label{inst:mpia}Max-Planck-Institut f{\"u}r Astronomie, K{\"o}nigstuhl 17, 69117 Heidelberg, Germany
\and\label{inst:INAF_merate}INAF - Osservatorio Astronomico di Brera, via E. Bianchi 46, 23807 Merate (LC), Italy
\and\label{inst:INAF_bologna}INAF -- Osservatorio di Astrofisica e Scienza dello Spazio, via Gobetti 93/3, 40129, Bologna, Italy
\and\label{inst:CfA}Center for Astrophysics | Harvard \& Smithsonian, 60 Garden Street, Cambridge, MA 02138, USA
\and\label{inst:aip}Leibniz-Institut für Astrophysik Potsdam, An der Sternwarte~16, 14482 Potsdam, Germany
\and\label{inst:cam}Institute of Astronomy, University of Cambridge, Madingley Road, Cambridge, CB3 0HA, UK
\and\label{inst:ncac}Nicolaus Copernicus Astronomical Center, Polish Academy of Sciences, ul.\  Bartycka 18, 00-716 Warsaw, Poland
\and\label{inst:JMU}Julius-Maximilians-Universität Würzburg, Fakultät für Physik und Astronomie, Institut für Theoretische Physik und Astrophysik,
Lehrstuhl für Astronomie, Emil-Fischer-Str.~31, 97074 Würzburg, Germany
\and\label{inst:MPIfR}Max-Planck-Institute for Radio Astronomy, Auf dem H\"ugel 69, 53121 Bonn, Germany
\and\label{inst:GFZ}GFZ Helmholtz Centre for Geosciences, Telegrafenberg, 14476, Potsdam, Germany
\and\label{inst:nasaHQ}NASA HQ, 300 E St.\ SW, DC 20546-0002, Washington DC, USA
}

\makeatletter
\renewcommand*\aa@pageof{, page \thepage{} of \pageref*{LastPage}}
\makeatother
\begin{document}

\abstract 
{}
{\textit{eROSITA,} on board the Spectrum Roentgen Gamma (SRG) spacecraft, performed its first X-ray all-sky survey (eRASS1) between December 2019 and
June 2020. It detected about 930000
sources, providing us with an unprecedented opportunity
for a detailed blazar census. We present the properties of blazars and blazar candidates in
eRASS1 and the compilation of the \textit{eROSITA} blazar catalog.} {We compiled a
list of blazar and blazar candidates from the literature and matched it
with the eRASS1 catalog, constructing the Blazars in
eRASS1 (BlazEr1) catalog. For sources with more than 50 counts,  we obtained their
X-ray spectral properties. We compiled multiwavelength data from
the radio to the $\gamma$-ray regimes for all sources, including multiwavelength spectral indices and redshifts. The full catalog is available online.} {We
present the BlazEr1 catalog, containing 5865 sources, of which 2106
are associated with confirmed blazars. For 2966 sources, \textit{eROSITA} provides
the first X-ray data. The
contamination from non-blazar sources
of the entire sample is less than 11\%. Most candidates exhibit
properties typical for blazars. We present the properties of the entire
X-ray detected blazar population, including the distributions of X-ray
luminosities and photon indices, multiwavelength properties, and the
blazar $\log N$-$\log S$ distribution. Our catalog provides
follow up targets, such as potential MeV and TeV blazars.}
{The BlazEr1 catalog provides a compilation of X-ray
detected blazars and blazar candidates. The catalog serves as a starting
point for exploiting further \textit{eROSITA} surveys using the same methodology, enabling us to study the X-ray variability and a large number of spectral energy distributions of blazars in the
future.}

\maketitle
\section{Introduction}
Active galactic nuclei (AGNs) with relativistic jets with a
line-of-sight orientation toward Earth are referred to as blazars
\citep{blandford:1978,blandford:1979,antonucci:1993,urry:1995,schlickeiser:1996}.
Relativistic beaming makes blazars the most luminous
persistent sources in the Universe and the dominant source type in X-ray
and $\gamma$-ray wavelengths at high Galactic latitudes
\citep{mattox:1993}, with high degrees of variability across the
entire electromagnetic spectrum on timescales from minutes to years
\citep[e.g.,][]{urry:1996a,tanihata:2001,ciaramella2004,agarwal:2015,rajput:2020}.

Blazars emit a double-peaked multiwavelength
spectral energy distribution \citep[SED; see,
  e.g.,][their Fig.~1]{middei:2022}, which is often modeled by two
log-parabola components \citep[][and references
  therein]{massaro:2006,hinton:2009,madejski:2016,krauss:2016}. The
low-energy peak originates from synchrotron
emission of electrons (and possibly positrons) in the relativistic
jets \citep[e.g.,][]{marscher:1985}. The peak of this emission
component is typically located between the radio and the optical bands
\citep{fossati:1998}. In contrast, the high-energy component, with the highest level of emission in the
$\gamma$ rays, 
can be explained using leptonic, hadronic, or
lepto-hadronic models \citep{bottcher:2013}. In the leptonic scenario, the jet
is assumed to consist of electrons and positrons.  Photons inverse Compton scatter
off of the relativistic electrons and positrons in the jet to higher energies. The
up-scattered photons could originate from the same population of
synchrotron photons \citep[synchrotron self-Compton radiation;
SSC;][]{ginzburg:1965,rees:1967,jones:1974,maraschi:1992,dermer:1993,bloom:1996}
or other photon fields, such as thermal emission from the accretion disk, the broad line region or the
torus \citep[external Compton; EC;][and references therein]{sikora:1994,ghisellini:1996,finke:2016}.
In hadronic models, the high-energy peak would be produced exclusively by relativistic protons
\citep[e.g.,][]{mannheim:1992,mannheim:1993,mucke:2001,mucke:2003},
resulting in pions from proton-photon interactions. 
Neutral pions and
their subsequent decay cascades then produce the observable X-ray and
$\gamma$-ray photons \citep{liodakis:2020}.
Leptonic and hadronic SED
models are able to describe multiwavelength data adequately well due
to degeneracies, incomplete multiwavelength coverage,
nonsimultaneous data, and additional systematic uncertainties \citep[e.g.,][]{bottcher:2013}.

The spectral position of the peaks is a useful tool to classify blazars based on the energy
of their synchrotron peak
\citep{fossati:1998,abdo:2010a}. Sources with low-energy peaks at
$\nu_{\mathrm{peak}}\leq10^{14}\,\mathrm{Hz}$, above
$\nu_{\mathrm{peak}}\geq10^{15}\,\mathrm{Hz}$, and in the intermediate
range are called low (LSPs), high (HSPs), and intermediate peaked
blazars (ISPs), respectively \citep{padovani:1996}. 
The X-ray
band tends to fall in the energy range near the transition between the
synchrotron and the high-energy peak. In a $\nu F_\nu$ representation,
this results in a falling X-ray spectrum ($\Gamma > 2.0$) for a
higher-peaked source, where the X-rays probe the synchrotron emission.
A low-peaked source exhibits a rising ($\Gamma < 2.0$)
X-ray spectrum, that is, X-rays are part of the high-energy peak \citep[][and references therein]{blandford:2019}.
This basic classification with regard to the photon index 
from a single snap shot observation can be systematically affected
by source variability, as during outburst and flares, sources can exhibit a peak-shift
behavior, including possible extreme HSP behavior \citep[e.g.,][]{pian:1998,
giommi:2000, ahnen:2018, sahu:2021, gokus:2024}.

Blazars have historically been classified based on optical spectra:
sources exhibiting emissions lines with widths of $>5$\AA\ are called
flat spectrum radio quasars \citep[\textit{FSRQ}s,][]{stickel:1991},
while sources with weaker or even no emission lines are classified as
BL Lacs (\textit{BLL}s). In recent years, however,
  obvious weaknesses of the emission line classification scheme have
  been pointed out by multiple authors. Some sources clearly exhibit
  features of both classes, which led to the introduction of
  intermediate classes in some classification schemes
  \citep{turriziani:2007}, while other sources showed strong
  contributions from the host galaxy
  \citep[\textit{BZG;}][]{massaro:2015}. By studying a sample of 354
  individual blazars with available multi-epoch spectroscopy,
  \citet{ruan:2014} found six cases of transition, attributed
  to the continuum of the jet sometimes outshining the emission lines,
  making these transitional objects likely \textit{FSRQ}s. As shown by
  \citet{delia:2015}, especially for sources located at redshifts of
  $z>0.7$, some emission lines that are redshifted into the infrared
  might be missed due to the lack of infrared coverage. From a sample of
  five sources classified as \textit{BLL}s, as predicted by
  \citet{giommi:2012}, \citeauthor{delia:2015} found two sources which
  could be \textit{FSRQ}s, based on the line width. In addition,
  multiple changing-look blazars (i.e., sources transitioning
  between the \textit{FSRQ} and \textit{BLL} class) have been
  reported. These changes are believed to be related to changes in
  accretion rate \citep{kang:2024}.

In the
context of the peak energy classification, \textit{FSRQ}s are
predominately LSPs, while \textit{BLL}s are distributed among all these
different categories \citep{ghisellini:1998}. Overall,  \textit{FSRQ}s are more bolometrically luminous
\citep{ghisellini:2013}.  In fact, the position of the synchrotron peak
is thought to be connected to the overall luminosity via the so-called
``blazar sequence.'' In other words, it might be that due to more efficient cooling the more
luminous sources peak at shorter frequencies
\citep{fossati:1998,ghisellini:2013}.  It is not clear, however, if this
sequence is due to selection effects \citep[see
e.g.,][]{giommi:2012,keenan:2021}.

To understand the population and the X-ray properties of
blazars as a whole, it is necessary to systematically study a large
sample. Blazars have been targets of many X-ray observations, mainly
focused on bright or variable sources, due to
the extensive multiwavelength campaigns required for SED modeling. Different X-ray observatories
have been used to build X-ray catalogs. For these catalogs of known blazars,
identified through optical, radio, and $\gamma$-ray surveys \citep[e.g., catalogs compiled by, e.g.,][]{massaro:2015,ackermann:2015}, were matched against the
observational data. Samples obtained using only one observatory
were built using \textit{Einstein} \citep[][55 sources]{worrall:1990},
\textit{EXOSAT} \citep[][26 sources]{sambruna:1994,sambruna:1994a},
\textit{Beppo-SAX} \citep[][86 sources]{donato:2005}, \textit{ROSAT}
(\citealt{urry:1996}, 36 \textit{BLL}s, 
\citealt{perlman:1996}, 23 \textit{BLL}s, 
\citealt{turriziani:2007}, 510 confirmed and 173 new
  blazars), \textit{Swift}-XRT \citep[][OUSXB\footnote{Throughout the paper, we use DR3, based on 15\,years of \textit{Swift}-XRT data, with 2831 distinct blazars.
  }: 2308 sources]{giommi:2019},
\textit{XMM-Newton} \citep[][103 sources]{delacalleperez:2021}, and \textit{NuSTAR}
\citep[][126 sources]{middei:2022}. Other studies have used data from
multiple X-ray missions \citep[e.g.,][${>}500$
  sources]{comastri:1997,donato:2001,kadler:2005,fan:2012,kapanadze:2013,yuan:2014},
often aiming at the multiwavelength properties of blazars. All these
catalogs only cover previously observed areas of the sky; therefore,
these observations are often biased toward the preselected sources proposed as observation targets, which can introduce further biases in sky
coverage. The few X-ray catalogs providing nearly all sky coverage
suffer from limited flux sensitivity and the number of sources in these
samples is therefore small compared to the number of blazars known in other bands.

A first X-ray sample with all-sky coverage and a deeper X-ray flux
limit was obtained with the \textit{ROSAT} all-sky survey \citep[RASS;][]{truemper:1982,truemper:1993,voges:1999,voges:2000}. The newest
RASS catalog \citep[2RXS;][]{boller:2016} offers information for
roughly 135000 sources with a limiting sensitivity of $F_{\mathrm{X},\,0.1-2.4\,\mathrm{keV}}{\sim}10^{-13}\,\mathrm{erg}\,\mathrm{cm}^{-2}\,\mathrm{s}^{-1}$ in the 0.1--2.4\,keV band.
Shortly after its publication, RASS was used to study the photon index
distributions of blazars \citep{urry:1996,perlman:1996} and to derive
blazar catalogs \citep[e.g.,][]{turriziani:2007}. For a long time, the
RASS was the most comprehensive X-ray all-sky survey. This changed
with the advent of the extended ROentgen Survey with an Imaging
Telescope Array (\textit{eROSITA}) on the Russian Spectrum-Roentgen-Gamma
(\textit{SRG}) satellite \citep{merloni:2012,sunyaev:2021}. Launched
in July 2019 from Baikonur, the mission began all-sky-survey
operations in December 2019 \citep{predehl:2021,merloni:2024}.
Consisting of seven nearly identical Wolter type 1 X-ray telescopes,
called telescope modules (TMs), with a total field of view of 1\degr\,
and frame store CCDs in the focal plane, \textit{eROSITA} is sensitive in the 0.2--10.0\,keV
band. \textit{eROSITA} performed an all-sky slew survey as \textit{SRG}, which
orbits $L_2$, constantly rotated around the spacecraft-Earth axis with a period of 4\,h. Therefore, distinct positions on the sky were,
on average, observed about six times in consecutive spacecraft rotations, while the source remained in the field of view for about 40\,s during every visit.
Close to the ecliptic poles,
which coincide with the survey poles, the number of consecutive observations is a lot higher.
Due to its $L_2$-orbit and its rotation around the Sun the whole sky could be covered within half a year.
Therefore, \textit{eROSITA} allows us to observe the entire sky in a systematic,
unbiased (unaffected by triggering on sources of interest) way and to
investigate source variability on timescales of hours and months. In
total, \textit{eROSITA} observed the full sky four times
as operations had
to be halted for political reasons during the fifth all-sky scan in
late February 2022. During the first all-sky scan by \textit{eROSITA} (eRASS1),
between December 2019 and June 2020, nearly 930000 individual sources
were detected on the Western Galactic hemisphere, which is accessible
to the German \textit{eROSITA} consortium \citep{merloni:2024}. This makes the
eRASS1 catalog the largest X-ray source catalog to date. In the
0.5--2.0\,keV band, 50\% completeness across the entire sky is
achieved at a flux of
$\lesssim5\times10^{-15}\,\mathrm{erg}\,\mathrm{cm}^{-2}\,\mathrm{s}^{-1}$.
The vast majority of eRASS1 sources are AGNs ($\sim$80\%), 
enabling a census of accreting supermassive black holes of
unprecedented completeness. \textit{eROSITA} data therefore provide a great
opportunity to investigate the X-ray properties of the blazar
population. The first \textit{eROSITA} all-sky survey has already been used to
identify potential TeV blazars for follow-up
\citep{marchesi:2025,metzger:2025}, to investigate neutrino events
\citep{km3netcollaboration:2025}, to identify high-redshift blazars
\citep{wolf:2024}, and to study the intergalactic medium with blazars
\citep{gatuzz:2024}.

We present the first \textit{eROSITA} eRASS1 blazar catalog, including the X-ray
identification and the X-ray and multiwavelength properties of blazars
and blazar candidates found during eRASS1. In Sect.~\ref{sec:sample},
we discuss the preparation of a sample of previously known blazars and
blazar candidate sources, which is matched against the eRASS1 all-sky data.
In Sect.~\ref{sec:erodata}, we describe our identification of blazars
and the analyses of the \textit{eROSITA} data. Additional multiwavelength data are
discussed in Sect.~\ref{sec:mwldata}. The properties of the blazars
and blazar candidates observed by \textit{eROSITA} are discussed in
Sect.~\ref{sec:properties}. We summarize results and provide future
prospects in Sect.~\ref{sec:outlook}. Throughout the paper, we
 assume a flat $\Lambda $ cold dark matter ($\Lambda$CDM) cosmology with $\Omega_{\Lambda}=0.7$,
$\Omega_\mathrm{m}=0.3$, and $H_0=70\,\mathrm{km}\,\mathrm{s}^{-1}\,\mathrm{Mpc}^{-1}$
\citep{beringer:2012}.

\section{The BLAZE catalog: Catalog of blazars from the literature}
\label{sec:sample}
\begin{table*}\small
    \centering
    \caption{Blazar and blazar candidate catalogs used to build the
      BLAZE catalog.}
    \label{tab:blazar_catalogs}
    \begin{tabular}{lllllll}
    \hline 
    \hline
      Catalog & $N_\mathrm{sources}$  & $N_\mathrm{sources,initial}$ & $N_\mathrm{sources,original}$ & Match radius & Included classes & Reference\\
       & & && $''$ & & \\
     \hline
        4FGL-DR4  & 3934 & 3934 & 3810 & -- & \textit{BLL}, \textit{BCU}, \textit{FSRQ} & 1 \\
        BZCAT & 3561 & 1830 & 1724 & 4 & \textit{BLL}, \textit{BCU}, \textit{FSRQ}, \textit{BZG}, \textit{BLLC} & 2 \\
        3HSP & 2013 & 777 & 773 & 3 & \textit{BCU} & 3\\
        HighZ & 48 & 33 & 33 & 4 & \textit{BCU} & 4, 5\\
\hline
    Milliquas  & 2814 & 305 & 303 & 4 & \textit{BLLC} & 6 \\
    KDEBLLACS  & 5525 & 5035 & 4992 & 5 & \textit{BLLC} & 7 \\
    WIBRaLS2  & 9541 & 7340 & 6165 & 2 & \textit{BLLC}, \textit{BCUC}, \textit{FSRQC} & 7\\
    ABC  & 1580 & 975 & 961 & 2 & \textit{BLLC}, \textit{BCUC} & 8\\
    BROS  & 88211 & 83269 & 81607 & 5 & \textit{BCUC} & 9 \\
    \hline
    \end{tabular}
\tablefoot{The order in which the catalogs are listed reflects the
      order in which we merged the catalogs. As the 4FGL-DR4 catalog
      is used as the first added catalog, there is no catalog against
      which it would be checked for duplicates; therefore, no match
      radius is listed. The radius given for the other catalogs is
      then used to filter for duplicates against all the previously
      added catalogs (therefore, e.g., only 1724 sources from the BZCAT are added, since the other sources are already part
      of the 4FGL-DR4 catalog within $4''$). The horizontal line
      divides the catalogs of confirmed blazars from those containing
      only blazar candidates.
$N_\mathrm{sources}$: Total number of sources contained in
  each catalog. $N_\mathrm{sources,inital}$: number of initially new sources
  added to BLAZE catalog, taking into account duplicates from the previous
  catalogs within the match radius listed in column ``Match radius''. $N_\mathrm{sources,original}$: number of sources from each catalog after applying quality cuts.}
  \tablebib{(1)~\citet[4FGL-DR4;][]{abdollahi:2022}; (2)~\citet{massaro:2015}; (3)~\citet{chang:2019}; (4)~\citet{sbarrato2026}; (5)~\citet{marcotulli:2025}; (6)~\citet{flesch:2023}; (7)~\citet{dabrusco:2019}; (8)~\citet{paggi:2020}; (9)~\citet{itoh:2020}}
\end{table*}

As no recent standard catalog of blazars exists that includes all
claimed or confirmed candidates in the literature, here we
describe how we created a ``master'' catalog of blazars and blazar
candidates by cross-matching existing catalogs from the literature.
This catalog has been released at the time of this paper's publication.

\subsection{Construction of the catalog}
\label{subsec:blaze-construction}
Older blazar compilations, such as the 5th Roma-BZCAT catalog
\citep[][we refer to this catalog simply as BZCAT throughout the paper]{massaro:2015} miss a large number of newer
sources. These catalogs are also significantly biased in terms of flux or
region on the sky observed \citep[e.g.,][]{bellenghi:2023a}. To search for X-ray counterparts of known blazars, we constructed a
catalog of blazar and blazar candidates from catalogs found in the
literature. As many sources are  included in multiple catalogs, we
filtered for duplicates by positional matching, taking into account the
accuracy of the individual catalogs. The catalogs used to build the
``master'' list, the number of sources provided by the catalogs and
the number of sources added, and the radii used to identify duplicates
with respect to other catalogs and the spectral classes provided, are
listed in Table~\ref{tab:blazar_catalogs}.

Following, for instance, \citet{giommi:2019} and \citet{bellenghi:2023a}, we
combined all the blazars and blazar candidates from the latest data release
of the fourth \textit{Fermi}-LAT source catalog
\citep[4FGL-DR4;][]{abdollahi:2020, abdollahi:2022}, the BZCAT multi-frequency catalog \citep{massaro:2015}, and the 3HSP
catalog \citep{chang:2019}. For
the source positions of the \textit{Fermi}-LAT blazars, we used the coordinates of the associated counterparts
provided in the 4FGL catalog, since the $\gamma$-ray positions were not well constrained enough. Duplicates were identified by position matching
or using associations provided by the input catalogs. Since the
catalogs have different spatial accuracy, each catalog was assigned an
individually selected maximum radius within which the sources were considered duplicates (see Table~\ref{tab:blazar_catalogs}). In addition, we added 48 high-redshift
blazars reported in the literature
\citep{yuan:2000,yuan:2003,sowards-emmerd:2003,romani:2004,worsley:2004,shemmer:2006,healey:2008,sbarrato:2012,sbarrato:2013,sbarrato:2015,sbarrato:2022,ghisellini:2014,ghisellini:2015,ghisellini:2015a,massaro:2015,coppejans:2016,belladitta:2019,belladitta:2020,caccianiga:2019,ighina:2019,khorunzhev:2021,an:2023,marcotulli:2025}.
This sample of high-redshift sources was  compiled by \citet{sbarrato2026}\footnote{Catalog available at \url{https://blaz4r.brera.inaf.it/}} and extended by us with the source discussed by \citet{marcotulli:2025}. This sample will be referred
to as the HighZ sample.

In addition to sources with a confirmed blazar designation, we added
objects with properties that are similar to those of blazars, with
varying criteria depending on the input catalog. We started with the
Milliquas catalog \citep[Version 8,][]{flesch:2023}. This catalog contains mainly
AGNs but also lists BL Lac-like objects identified via various
detection methods. 
Based on WISE data, the KDEBLLACS and WIBRaLS2 catalogs provide candidate
blazars of various spectral types \citep{dabrusco:2019},
while the ABC catalog \citep{paggi:2020} uses ALMA calibration data as
well as other multiwavelength information to characterize blazar
candidates. The largest catalog used to build our candidate sample is
the BROS catalog \citep{itoh:2020}. This catalog lists
objects which exhibit a flat radio spectrum and a counterpart in
Pan-STARRS1. Compared to other catalogs, BROS sources are not
homogeneously spread across the entire sky, but cover areas with
Galactic latitude $|b|\ge 10\degr$ and declination $\delta> -40\degr$.
A small window centered around Galactic coordinates $b\,\sim\,40\degr$
and $l\,\sim\,220\degr$ was excluded due to a lack of radio
coverage.
To create
the master list, we started with the first catalog shown in Table~\ref{tab:blazar_catalogs}, then cross-matched with
the next catalog and added any previously not included sources. We continued this process down the list of catalogs given in Table~\ref{tab:blazar_catalogs}.

Our initial list after positional cross-matching contained 103498
individual blazars and blazar candidates spread over the entire sky, of which 43148 (${\sim} 41\%$) are located on the western Galactic hemisphere due to the
inhomogeneity of the BROS catalog.  Based on the input catalogs, we
classified the blazars and blazar candidates into the following classes:
(1) a BL Lac object is listed as \textit{BLL}; (2) a galaxy-dominated BL
Lac object is denoted as \textit{BZG}; (3) a flat spectrum radio quasars
is abbreviated as \textit{FSRQ}; and (4) a confirmed blazar of an unknown type
is called \textit{BCU}. We appended the letter "C" to the abbreviation to
indicate that a source is a blazar candidate (\textit{BLLC},
\textit{BCUC}, \textit{FSRQC}). If no spectral classification is listed,
sources were denoted as \textit{BCUC}, which included all entries originating from the BROS catalog.
The classifications from the input catalogs for \textit{BLL}s and \textit{FSRQ}s, and of course the corresponding candidates, have to be taken with caution, since roughly 2--5\% of these sources are probably misclassified \citep{ruan:2014,delia:2015,kang:2024}.
Therefore, we estimate that at most 180 confirmed \textit{BLL}s and \textit{FSRQ}s and 560 corresponding candidates are wrongly grouped in the initial list.
An extensive and detailed spectroscopic study would be needed to further investigate the correct classification of sources on the list, which is beyond the scope of this study. 

\subsection{Assessment of quality}
\label{subsec:quality}
No source in the initial master list is guaranteed to be a
blazar, especially since there are many candidates. Thus, it is
important to assess the contamination and remove as many non-blazars
as possible. 

Due to the extremely complex selection function, assessing the
contamination is not straightforward. We therefore investigated a few
indicators and assigned upper limits to the level of contamination and
checked the purity of the input catalogs. In the 4FGL, about 98\% of AGNs are
confidently classified as blazars and, hence, we would expect a very low level of
contamination. A similar level of purity is expected from BZCAT;
however, this catalog also contains radio galaxies such as Cen~A.
According to \citet{xie:2024} about 5\% of BZCAT sources are
non-blazars. The 3HSP contamination is expected to be $<2\%$
\citep{chang:2019} and similar levels are expected from the HighZ
sample. Unfortunately, no level of contamination is listed for the
sample of the Milliquas and the ABC. \citet{demenezes:2019} assessed the
contamination of the WIBRaLS2 and KDEBLLACS catalogs using SDSS.
They found that
31\% and 30\%, respectively for these catalogs, are blazars.
The main contamination is caused by QSOs (${\sim}69\%)$,
which could also be blazars, representing a loosely constrained upper
limit. These catalogs were also tested by \citet{xie:2024}. They find
14\% and 12\% to be non-blazars, respectively.
The difference between the estimated levels of contamination for the WISE catalogs might be related to the entire sample not having available data in both approaches and due to other thresholds set to distinguish a blazar from a non-blazar.
For the BROS catalog,
\citet{itoh:2020} estimated a contamination of about 10\%. Due to flux
and spatial limits in the BROS data, only 60\% of the BZCAT sources are also present in this
catalog.  Therefore, we expected roughly 5\% of the blazars and about 14\%
of the candidates listed in the master list to be non-blazars, when utilizing the most conservative estimates presented above.

To identify non-blazar contamination by nearby galaxies in our
list (and since blazars have higher redshifts), we matched against the HECATE catalog of nearby galaxies
\citep{kovlakas:2021},  one of the most complete catalogs of galaxies in the
local Universe ($D\lesssim200\,\mathrm{Mpc}$) at the time of writing.  We obtained 1227 matches within $12''$ of the galaxy
center, of which 1215 are located within the $D_{25}$-ellipse.  Out of these, 57
positional matches are associated with confirmed blazars,
including well known sources with redshifts consistent with the matched
HECATE galaxy; hence, it is clear that the position alone is not enough to identify
non-blazars.  HECATE also offers a Hubble galaxy classification
\citep{devaucouleurs:1976}. Out of the positional matches 689
correspond to a spiral galaxy.  Since blazars are typically not hosted in
spiral galaxies \citep{urry:2000,odowd:2002} and a location within the
$D_{25}$-ellipse with a maximum separation of $12''$ indicates that in X-rays
the source is undistinguishable from the center of the Galaxy, we excluded these 689 objects, almost 99\% of which 
originate
from the BROS catalog.

\citet{xie:2024} used moderate resolution radio images from the VLA Sky Survey \citep[VLASS;][]{lacy:2020} to classify the sources from BZCAT,
WIBRaLS2, and KDEBLLACS based on their morphology.  We matched the initial list
with their results using the match radii for the individual catalogs
(see Table~\ref{tab:blazar_catalogs}). If a visual assessment of the
morphology exists we used this classification instead of the automated
one.  A two-sided radio morphology, which is inconsistent with the source
being classified as a blazar, was found for 1139 objects. These outliers were then removed.

Due to their different optical and X-ray spectra, as
  well as due to their typical radio-quietness, we removed known
  narrow-line Seyfert~1 (NLSy1) galaxies, even though a fraction of
  the NLSy1 galaxy population has been detected in the radio band
  \citep[e.g.,][]{komossa:2006,singh:2018}, with a small number also
  having been detected at $\gamma$ rays
\citep[e.g.,][]{abdo:2009,paliya:2018}. Several authors have shown that
$\gamma$-loud NLSy1s exhibit blazar-like characteristics including
bright flaring episodes
\citep[e.g.,][]{dammando:2015,paliya:2016,gokus:2021}, but ended up arriving at
the conclusion that these objects resemble less powerful (i.e., younger sources). We excluded $\gamma$-ray emitting NLSy1 galaxies from
the initial list to consider only "full-scale" blazars. Two NLSy1s
listed in the 4FGL are located on the western Galactic hemisphere and
detected by \textit{eROSITA} (1eRASS\,J094857.1$+$002226 and
1eRASS\,J200754.9$-$443446). To exclude NLSy1s which do not show
$\gamma$-ray emission, we matched with the catalog of NLSy1 galaxies by
\citet{rakshit:2017}. We found 55 matches within $5''$ of initial list
sources, the maximum match radius used during the construction of the
master list, of which 37 were associated with candidates that were
also removed.

To identify other types of radio galaxies, we matched against the
high-fidelity sample from \citet{gordon:2023}, made up of double radio sources, which is a
morphology not expected to be observed for blazars.  We again used a
maximum separation of $5''$, identifying 1243 matches.  More than 76\% of
them were associated with a BROS source and overall 96.4\% of the
matches were blazar candidates. We also removed these.

Finally, we removed individual objects which are known to be non-blazars.
BROS\,J0729.1$+$2054 is the counterpart of the Galactic planetary nebula NGC~2392.  The nature of the source 4FGL\,J0647.7$-$4418 is debated in the
literature  either as being a blazar \citep{marti:2020} or a B-type
subdwarf and white dwarf binary
\citep[HD\,49798;][]{mereghetti:2009,rigoselli:2023}.  The young radio galaxy PMN\,J1603$-$4904
\citep{muller:2015,krauss:2018} is falsely classified in 4FGL as the blazar
4FGL\,J1603.8$-$4903.  Finally, the BZCAT catalog erroneously includes the radio galaxy Cen A as
a blazar (5BZU\,J1325$-$4301).

\subsection{Release and comparison with other catalogs}
\label{subsec:blaze-compare}
\begin{figure*}
\sidecaption
\includegraphics[width=12cm]{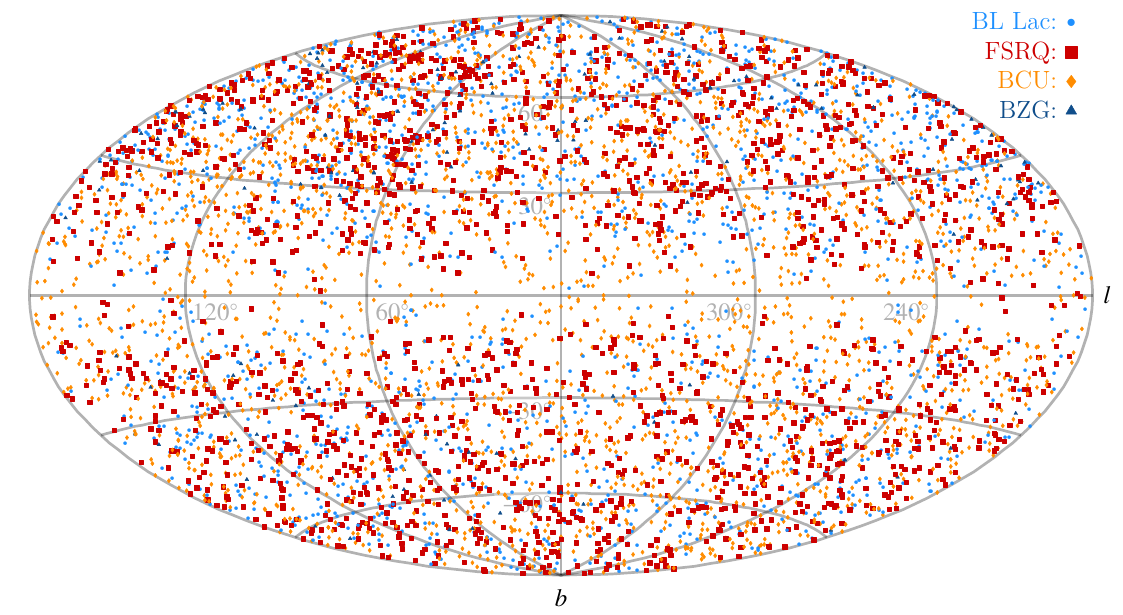}
\caption{All-sky distribution of the BLAZE catalog gold sample (confirmed blazars) in Galactic coordinates (\textit{BLL}: 1697, \textit{FSRQ}: 1937, \textit{BCU}: 2503, \textit{BZG}: 164).
        The different blazar classes are shown color coded and with different
        symbols.
        }
\label{fig:all_sky_secure_map}
\end{figure*}

After removing obvious non-blazars, 100368 out of the initial 103498
objects remained.  We call our master list of blazars and blazar
candidates the BLAZars from litErature catalog, or BLAZE catalog.  The BLAZE catalog
can be split into a ``gold sample,'' which includes 6301 confirmed
blazars with or without a known type (3031 are located on the footprint of the
eRASS1 survey), and a ``silver sample'' of 94067 candidates (38905 in the
footprint of the eRASS1 survey), containing the blazar candidates.
Figure~\ref{fig:all_sky_secure_map} shows the spatial distribution of
the gold sample.  Due to the general utility of a compilation of blazars
for the field, the BLAZE catalog was published with this paper and made
available online on Vizier. The description of the BLAZE catalog is given in
Appendix~\ref{sec:blazeinfo}, as the catalog is enriched with redshifts
for the confirmed blazars and \textit{eROSITA} exposure times and upper limits for flux and luminosity. The list of the likely non-blazar objects, referred to as the "unverified" BLAZE catalog, has been released as a separate file.
Because the BLAZE catalog was compiled from multiple input catalogs
with different wavelength selections and flux limits, and  because
obvious non-blazars sources were removed introducing new selection
cuts, the population of blazars and blazar candidates listed in BLAZE
is neither complete nor statistically well defined.
Extensive simulations would be required to assess the completeness and the flux limits in different bands for the BLAZE catalog, which is beyond the scope of this study.

In comparison to previous studies, we found 6307 individual blazars from
the 4FGL, BZCAT, and 3HSP catalogs; this is a slightly different number
than in earlier catalogs (5340\,sources, \citealt{giommi:2019} and
6425\,sources, \citealt{bellenghi:2023a}).
To avoid source confusion,
we applied a stricter angular limit for cross-matching, leading to a difference between BLAZE and \citet{giommi:2019} and \citet{bellenghi:2023a}.
The deviation of 15\% between BLAZE and \citet{giommi:2019} is also related to this estimate being based on the 3LAC catalog which only contains 1591 sources in total compared to the 3934 included in the 4FGL catalog.
However, the deviation between the BLAZE and \citet{bellenghi:2023a} is only 2\% mainly driven by the exclusion of contaminants. 
\citet{marchesi:2025} found 1772 matches between the BZCAT and the 4FGL,
of which 1725 are within $2''$, while our analysis returns 1625  within
$4''$ (1731 without quality cuts).  This difference is ascribed to our
match radius, as \citet{marchesi:2025} also accept wider separations
for counterparts between the catalogs and our filtering.  A total of 615
out of 651\,objects in the isotropic catalog of \citet{kudenko:2024} are
contained in the BLAZE catalog, which includes 409 of the 433 blazars and
blazar-like sources from the isotropic catalog. The objects not
contained simply do not have a counterpart in \citet[][to within $6'$]{kudenko:2024}.  Out of the matches, 204 are classified as quasars, AGNs or based on an emission band
by \citet{kudenko:2024} and 19 of these are associated with blazar
candidates in the BLAZE catalog.
A radio flux-limited sample of HSP sources was constructed by
\citet{giommi:2020} based on \citet{puccetti:2011}.
Of the 23 sources in this sample, 15 are included in the BLAZE catalog, including all sources associated with the 3HSP catalog.
Again, this stresses that the BLAZE catalog, although it is the largest catalog to date, is not complete.

\section{Matching \textit{eROSITA} to the BLAZE catalog}
\label{sec:erodata}
In this section, we describe the construction of our \textit{eROSITA} blazar and
blazar candidates catalog based on the BLAZE catalog. Out of the 100368 blazar and blazar candidate sources in the BLAZE catalog, 41936 are located on the western Galactic hemisphere and were matched with \textit{eROSITA}. We also assess the
level of contamination of the catalog and the analysis of the X-ray data.

\subsection{The \textit{eROSITA} observed blazars and blazar candidates}
\label{subsec:eroob}
\begin{table*}\footnotesize
    \centering
    \caption{Breakdown of the blazar source classes during the construction of the BlazEr1 catalog.}
        \label{tab:blazar_classes}
    \begin{tabular}{llllllllll}
    \hline 
    \hline
      Class & Acronym & $N_\mathrm{BLAZE}$  & $N_\mathrm{eRASS1}$ & $N_\mathrm{match,15}$ & $N_\mathrm{match,8}$ & $N_\mathrm{sample}$ &  $N_\mathrm{new}$ & $N_\mathrm{spec}$ & $N_\mathrm{spec, new}$\\
     \hline
        BL Lac object  & \textit{BLL} & 1697 & 802 & 682 & 641 & 597  & 90 & 353 & 40  \\
        BL Lac candidate & \textit{BLLC} & 7750 & 3188 & 1269 & 1099 & 954 & 574  & 185 & 85 \\
        Galaxy-dominated BL Lac object & \textit{BZG} & 164 & 59 & 33 & 32 & 28 & 10 & 8 & 1 \\
        FSRQ & \textit{FSRQ} & 1937 & 929 & 844 & 796 & 769 & 252 & 220 & 43 \\
        FSRQ candidate & \textit{FSRQC} & 3444 & 1799 & 1135 & 1011 & 913  & 569 & 109 & 63 \\
        Blazar of unknown type & \textit{BCU} & 2503 & 1241 & 849 & 766 & 712 & 311 & 253 & 86 \\
        Blazar candidate of unknown type & \textit{BCUC} & 82873 & 33918 & 3305 & 2507 & 1892 & 1160 & 145 & 73 \\
        \hline
        Total &  & 100368 & 41936 & 8117 & 6852 & 5865  & 2966 & 1273 & 391 \\
        \hline
    \end{tabular}
\tablefoot{ $N_\mathrm{tot}$ and $N_\mathrm{eRASS1}$: number of
  sources contained in the BLAZE catalog across the entire sky and in the
  western Galactic hemisphere, respectively. $N_\mathrm{match,15}$ and
  $N_\mathrm{match,8}$: Number of \textit{eROSITA} sources within $15''$ and $8''$
  of a BLAZE catalog source. $N_\mathrm{sample}$: final number of \textit{eROSITA}
  counterparts after accounting for the detection likelihood and
  quality flags. $N_\mathrm{new}$: Number of sources of each class
  with no previous exposure time with \textit{XMM-Newton}, \textit{Chandra}, \textit{ASCA}, \textit{NuSTAR},
  \textit{Suzaku}, \textit{Swift}-XRT, and \textit{ROSAT} pointed observations.
  $N_{\mathrm{spec}}$: Number of sources with $N>50$ counts, for which
  X-ray spectral analysis is possible. $N_\mathrm{spec,new}$: Number
  of sources from the spectroscopic subsample without prior X-ray
  data available.}
\end{table*}

\subsubsection{X-Ray counterparts}
\label{subsubsec:xraycounterparts}
\begin{figure}
\includegraphics[width=\linewidth]{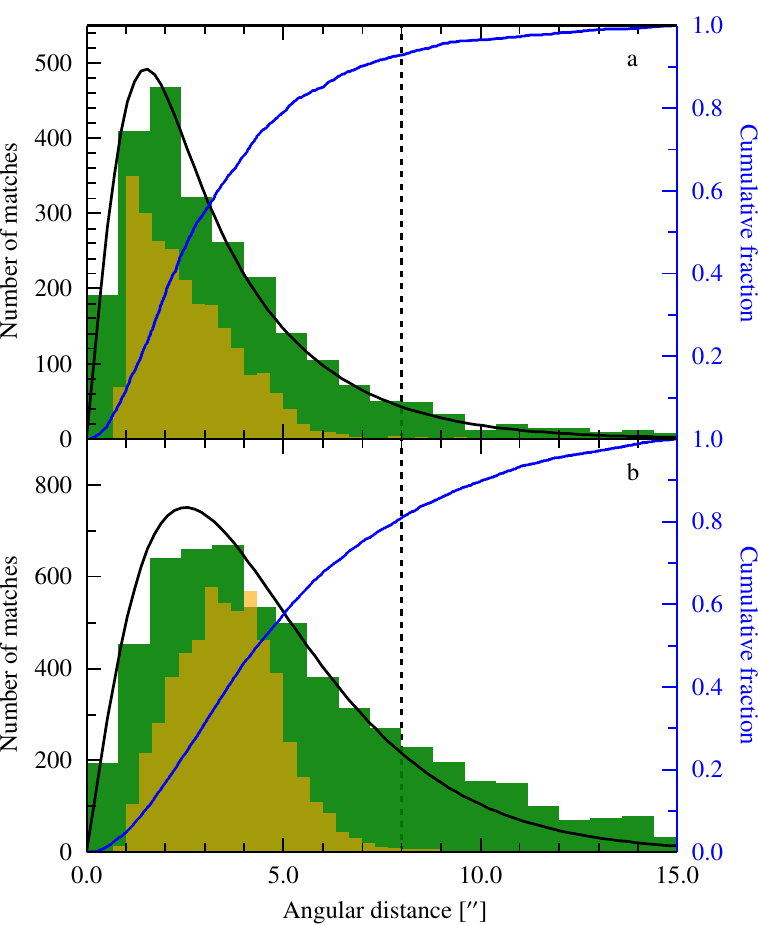}
    \caption{Histograms of angular separation between BLAZE catalog and \textit{eROSITA}
      positions (green) with the distribution of the matched \textit{eROSITA}
      source positional uncertainty overlaid (yellow). The vertical
      line at $8''$ indicates the distance threshold for the final
      sample. The cumulative distribution of the angular separation is
      shown in blue. \textbf{a} Sample of \textit{eROSITA} observed blazars
      with 93\% of the matches found within $8''$, and 55\% and 79\%
      within $3''$ and $5''$, respectively. \textbf{b} \textit{eROSITA}
      observed blazar candidates. Only 81\% of matches are within
      $8''$, and 31\%/58\% within $3''$/$5''$. We show Rayleigh
      distributions (black) for illustrative purposes, calculated
      following \citet[][Eq.~3, assuming $F=0.8$]{merloni:2024}.}
\label{fig:blazar_erass1_angsep_corrected}
\end{figure}

To identify the X-ray counterparts of the BLAZE catalog, we matched
 the BLAZE catalog and the eRASS1 catalog positions
\citep{merloni:2024}. We used the BLAZE catalog and the attitude-corrected
positions from \textit{eROSITA},  only considering point sources
\citep[$\mathrm{EXT}=0.0$, see][]{merloni:2024} and identified an
initial number of 8117 matches with angular separation ${\le} 15''$. This
separation limit is based on the accuracy of the astrometric
correction of \textit{eROSITA} \citep{merloni:2024} and the point spread function
of the \textit{eROSITA} telescopes. The histogram of the angular distance in
Fig.~\ref{fig:blazar_erass1_angsep_corrected} shows that most matches
are within the positional accuracy of the matched \textit{eROSITA} sources (shown
in yellow, the normalized version of the histogram is displayed in
Fig.~\ref{fig:mara_sep_combi_blaze}). About 84\% of associations have
an angular separation of $8\arcsec$ or less, and 38\% and 64\% of the
sample are located within $3''$ and $5''$, respectively. The
distributions of separations for blazars and blazar candidates are
different; the candidates exhibit a broader peak, possibly due to
contamination. Based on the distribution of angular separation and to avoid unnecessary source confusion and false identifications
we conservatively cut our final sample at a separation of $8\arcsec$ between the BLAZE and eRASS1 source position.
This cut reduced the sample to 6852 blazars and blazar candidates.

\begin{figure}
\includegraphics[width=\linewidth]{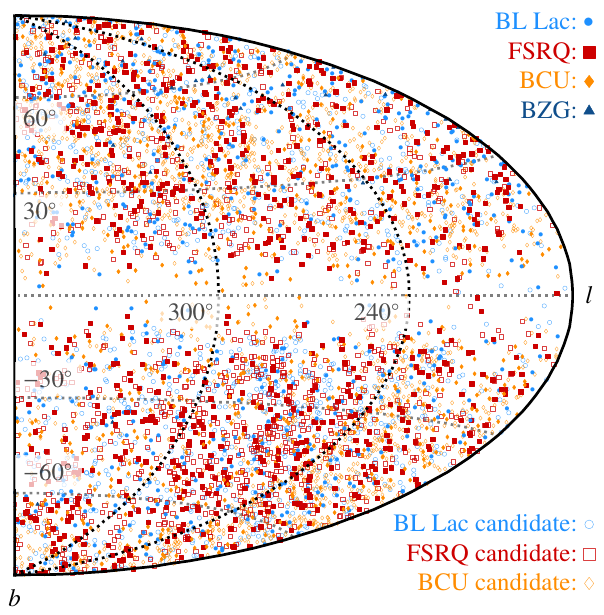}
\caption{Positions of the BlazEr1 sources in Galactic coordinates. The
  different colors and symbols indicate the blazar types in the
  catalog. Candidates, especially \textit{BCUC}s, dominate the sample.
  The overdensity along $l\sim 240\degr$ coincides with the footprint
  of BROS. Confirmed blazars are displayed by filled symbols,
  candidates are shown in the same color without filling. All the
  following figures will adapt the same color and symbol scheme.}
\label{fig:blazar_erass1_map_paper}
\end{figure}

Since the \textit{eROSITA} exposures were still quite low (${\sim}240\,\mathrm{s}$), many sources have a low
detection likelihood in \textit{eROSITA}. To avoid including possibly spurious
detections, we removed all matches with a detection likelihood,
$\mathrm{DET\_LIKE\_0}<10$\,\footnote{$\mathrm{DET\_LIKE\_0}$ equals
the negative logarithm of the probability of detected counts being
caused by fluctuations of the background}, which reduces the fraction
of spurious sources to $\sim$1\% \citep{seppi:2022,merloni:2024}. We
also removed all entries with uncertain positions, that is, those without
values for $\mathrm{RA\_LOWERR}$, $\mathrm{RA\_UPERR}$,
$\mathrm{DEC\_LOWERR}$, and $\mathrm{DEC\_UPERR}$ in the eRASS1
catalog, and excluded all objects where \textit{eROSITA} quality flags indicate
issues in the source detection ($\mathrm{FLAG\_SP\_SNR}$,
$\mathrm{FLAG\_SP\_BPS}$, $\mathrm{FLAG\_SP\_SCL}$,
$\mathrm{FLAG\_SP\_LGA}$, $\mathrm{FLAG\_SP\_GC\_CONS}$,
$\mathrm{FLAG\_NO\_RADEC\_ERR}$, $\mathrm{FLAG\_NO\_CTS\_ERR}$, and
$\mathrm{FLAG\_NO\_EXT\_ERR}$). 
Additionally, we removed all blazars and blazar candidates with X-ray luminosities too low to actually be
a blazar ($L_{\mathrm{X},
    0.2-2.3\,\mathrm{keV}} < 10^{41}\,\mathrm{erg}\,\mathrm{s}^{-1}$), lowering our contamination rate.
This reduced the sample size to 5865
sources of which 2106 are associated with confirmed blazars, and the
remainder with blazar candidates. The normalized separation
distributions shown in Fig.~\ref{fig:mara_sep_combi_blaze} clearly
indicate that after the applied cuts the agreement with the
theoretical Rayleigh distribution is significantly improved.
Table~\ref{tab:blazar_classes} lists the number of sources after each
cut and for each blazar type, and
Fig.~\ref{fig:blazar_erass1_map_paper} displays the sky distribution
of the final sample.

The matches outlined above are the basis for the Blazars in
eRASS1 (BlazEr1) catalog. This catalog combines
X-ray and multiwavelength information that we compiled for all catalog
sources. A detailed description of the content of the catalogs is
given in Appendix~\ref{sec:cataloginfo}. We refer to the sample
of sources detected by \textit{eROSITA} and associated with ``gold sample''
BLAZE sources as "\textit{eROSITA} observed blazars," while we refer to the \textit{eROSITA}
matches with an entry from the BLAZE ``silver sample'' as \textit{"eROSITA}
observed blazar candidates."
The 2252 objects removed from the catalog are included in a separate catalog, as 
this sample is expected to be less pure, but
it might still be of value for follow-up studies. 
Due to the survey mode of \textit{eROSITA}, we are
able to obtain a view unbiased by variability or source
luminosity, and only limited by \textit{eROSITA} sensitivity variations and exposure in eRASS1.

\subsubsection{Validation of positional matching}
\label{subsubsec:eros-xraymatching2}
Since our association of blazars with \textit{eROSITA} sources is based on
positions, it was necessary to check whether our positional matching is
a valid approach. We used two independent methods to derive an upper
limit on the possible contamination of matching by pure chance.

We first estimated the possibility of randomly associating an input position with an eRASS1 source by uniformly
distributing $5\times 10^5$ (which is five times larger but has the same order of magnitude in numbers as the BLAZE catalog) sources at random position on the entire
sky and matching these positions to the eRASS1 catalog. Within
$8\arcsec$, we found 346 matches whereas within $15\arcsec$ 1205
matches were found, indicating a probability for random association of
0.07\% and 0.24\% for these separations, respectively. Alternatively,
conserving the density properties of the eRASS1 survey we rotated the
positions in the eRASS1 catalog around the ecliptic poles by a few
degrees in ecliptic longitude and then matched them with the BLAZE catalog. For
rotations by $1\degr$ and $10\degr$, 89 and 104
sources matched within $15\arcsec$, respectively. When reducing the separation to
$8\arcsec$, only 34 and 23 matches remained, implying a random match
probability of about 0.08\% for a separation of $<8\arcsec$,
regardless of the applied rotation.

Results from both methods suggest a probability below 0.1\% of
accidental source confusion. Hence, we expect at most 42 \textit{eROSITA} detections to be
randomly assigned to BLAZE catalog sources. Given the size of the
BlazEr1 catalog, the rate of source confusion is therefore at
a negligible level of 0.7\%. 
Our cross-matching angular
distance is well justified, as the loss of
matches when reducing the acceptable angular
distance from $15\arcsec$ to $8\arcsec$ indicates that the probability
of including random matches drops significantly.
We emphasize that this low level of accidental source confusion is due
to us using pre-existing positions of objects with known
multiwavelength properties, including a very high probability that
these sources are X-ray bright. This approach is different from
approaches that attempt to find multiwavelength counterparts for new
X-ray detected sources in multiwavelength catalogs, where little is
known about the properties. These approaches have a much higher
probability of misidentifying the multiwavelength counterparts.

\subsection{Contamination}
\label{subsec:contamination}
The sample size makes a study of the source nature on an individual basis
too time consuming.
Therefore, we try to asses upper limits of the contamination.  We derived
the level of contamination caused by non-blazars in the sample by comparing the BlazEr1 catalog with the
catalog of Legacy Survey data release 10 \citep[LS10;][]{dey:2019}
counterparts of eRASS1 point sources \citep[][hereafter S25]{salvato:2025}. This catalog is based on an identification of optical
counterparts with the Bayesian algorithm NWAY \citep{salvato:2018},
using a combination of astrometric information such as separation,
positional error, and source number density \citep[similar to
Xmatch;][]{pineau:2017}.  The completeness and purity of the LS10
catalog of counterparts to eRASS1 is $\sim$93\% (S25).

We matched by using the detection ID of eRASS1. If more than one possible
counterpart was listed for a given ID, we used the one with the closest
position. In total, we can match 84\% of the BlazEr1 sources (4924
objects). We only considered counterparts to be a match between both
catalogs if the association of S25 and the BlazEr1 catalog have
an angular separation of at most $2''$, and if
$p_\mathrm{any}\ge0.1$\footnote{$p_\mathrm{any}$ is the probability
given by NWAY that any of the matches found is the correct
counterpart.}. For 508 matches (10.3\% of the sources with a match based
on detection ID) we associated a different counterpart as S25 or
$p_\mathrm{any}$ is very low. About 89\% of the disagreed upon objects
are associated with blazar candidates and almost 80\% originate
from the BROS catalog. This leaves us with 4416 counterparts in common,
which corresponds to a 90\% agreement. Therefore, when comparing with
the results of S25 and assuming that their assigned counterparts to the \textit{eROSITA} sources
are all correct, a disagreement between the counterpart by S25 and our catalog and hence a contamination of at maximum 3.3\% for the \textit{eROSITA}
observed blazars and 14.0\% for the \textit{eROSITA} observed blazar candidates is
expected.  Sometimes the X-ray source might be the sum of multiple
emitters. This is the case for 32 sources, of which in 22 cases the
counterparts by S25 and our source are identical. 
We also compared the matches obtained by matching the \emph{unverified} BLAZE catalog with eRASS1, using the same methodology as for the BlazEr1 sample, with the counterpart catalog.
Given the same criteria as listed in Sect.~\ref{subsec:eroob}, the \emph{unverified} BLAZE catalog has 491 matches with the eRASS1, of which 397 have a counterpart in S25.
For 33 of these matches, all of them candidates, S25 assigns a different counterpart. Therefore, a slightly lower (but similar) level ($\sim8\%$) of contamination has been reached for this low-quality sample as for the BlazEr1 catalog.
This also indicates that cleaning of the BLAZE catalog is useful.

For all \textit{eROSITA} sources listed in the BlazEr1 and the counterpart 
catalog, we compared the angular separation between the \textit{eROSITA} position
and the counterpart by S25 with the separation between the BLAZE catalog
source and \textit{eROSITA}, normalized by the positional uncertainty of \textit{eROSITA}.
Some BlazEr1 counterparts tend to be too far away as they lie
significantly above the theoretical Rayleigh distribution, whereas few counterparts
assigned by S25 seem to be closer or more distant 
than expected (see Fig.~\ref{fig:mara_sep_combi_LS10}). 
This indicates that some
BLAZE catalog objects are erroneously associated to the \textit{eROSITA} source or 
that some of the associations of S25 might not be the correct
counterpart. The estimated number of contaminating sources
based on the assumption that S25 provides the correct
counterpart and disagreement indicates contamination therefore represents an
upper limit. We also note that ours and Salvato et al.'s counterpart
associations methodologies differ significantly. While we focus on
selecting blazar specific objects using multiwavelength information
and pre-existing catalogs, the approach from Salvato et al. has been finetuned
toward general X-ray emitters, where non-jetted AGNs dominate the
population of X-ray emitting extragalactic objects. However, an
in-depth comparison between the approaches is beyond the scope of this
paper.

The contamination is mainly driven by the BLAZE catalog selection function and
not by random matches (see
Sect.~\ref{subsubsec:eros-xraymatching2}).  The most pessimistic
estimates from the purities of the input catalogs (see
Sect.~\ref{subsec:quality}) and the comparisons with the counterpart
catalog indicate that the contamination for confirmed blazars is at most
5\%, and $\sim$14\% for the candidate blazars.
The limit for the confirmed blazars is based on the purity estimate of the BZCAT by \citet{xie:2024}, which has the highest impurity of the confirmed blazar catalogs.
Although sources are excluded from the BLAZE catalog based on \citet{xie:2024}, they do not provide a classification for all catalog entries.
We therefore expect the subsample not covered by \citet{xie:2024} in the BZCAT to have a similar level of contamination as the overall catalog, and since we do not have information available for all BZCAT objects we adapt the 5\% limit as an upper limit for the entire sample of confirmed blazars.
For the candidates the limit is based on the contamination estimate based on the comparison with S25.
Therefore, at most 106 of
the thought to be confirmed blazars and 527 of the candidates are non-blazars. In total, less than 11\% of the
BlazEr1 catalog is contaminated, corresponding to a purity of almost
90\%. 

\subsection{Survey sensitivity}
\label{subsec:sens}
Since the sensitivity of the \textit{eROSITA} survey is not uniform across the
sky, it is crucial to determine the lowest flux sources that could
have been detected as a function of position and exposure time. We
used the official \textit{eROSITA} mission simulator, \texttt{SIXTE}
\citep{dauser:2019} to determine this sensitivity limit. Using the
as-flown \textit{eROSITA} attitude of the first all-sky survey and a diffuse
foreground from \texttt{ROSAT}, we simulated \textit{eROSITA} observations of
$10^5$~sources randomly and uniformly distributed across the western
Galactic hemisphere. Based on the expected blazar spectrum averaged
across the entire population, each source was assigned an absorbed
power law spectrum ($\Gamma_\mathrm{X} = 2.0$, $N_\mathrm{H} =
1\times10^{21}\,\mathrm{cm}^{-2}$). We varied the 0.2--2.3\,keV flux
between $1\times10^{-16}$ and
$1\times10^{-11}\,\mathrm{erg}\,\mathrm{cm}^{-2}\,\mathrm{s}^{-1}$,
with 20000 random flux values per flux decade. The simulated data were
then sorted into event lists for each \textit{eROSITA} sky tile and processed
using the official source detection pipeline contained in the \textit{eROSITA} data
analysis software \texttt{eSASS} \citep[version
  211214,][]{predehl:2021,brunner:2022,merloni:2024}. The processing
yielded a mock catalog containing information on the detection
likelihood, exposure times, and fluxes. To ensure the validity of the
simulations, we checked that all fluxes of the detected source are
consistent with
the input
values within uncertainties; and since this is
the case we used the input fluxes.

\begin{figure}
\includegraphics[width=\linewidth]{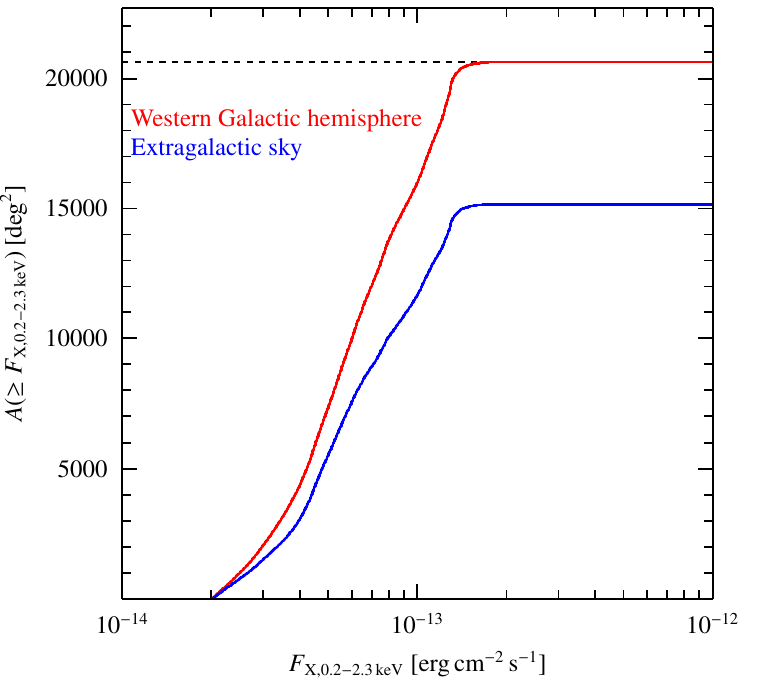}
\caption{Sky area in which 95\% of all sources above a minimum flux will
        be detected with DET\_LIKE\_0\,$\geq10$. The dashed line
        indicates the total area of the western Galactic hemisphere.
        The red line displays the sensitivity over the entire western
        Galactic hemisphere, while the blue line shows the extragalactic
        sky excluding an area of 15\degr\ below and above the Galactic
        plane and regions with radii of $5.5\degr$ around the Large
        Magellanic cloud (LMC) and $3.0\degr$ around the Small
        Magellanic cloud (SMC).}
\label{fig:erosita_sensitivity_minflux_lin}
\end{figure}

To estimate the completeness and sensitivity, we defined the survey as
complete if, for a given exposure time and minimum detection
likelihood of 10, 95\% of sources with a given flux were detected. For
a given minimum detection likelihood, we found that we can approximate the
minimum detectable flux by
\begin{equation}\label{eq:pwrlw_sim}
    F_{\mathrm{X},\,0.2\mbox{--}2.3\,\mathrm{keV, min}} = I_\mathrm{sens} \times \left(\frac{t_\mathrm{expo} - t_\mathrm{0}}{1\,\mathrm{s}}\right)^{-\Gamma_\mathrm{sens}} + F_\mathrm{0} ,
\end{equation}
where for the given minimum $\mathrm{DET\_LIKE\_0} = 10$,
$I_\mathrm{sens} =
2.14\times10^{-9}\,\mathrm{erg}\,\mathrm{cm}^{-2}\,\mathrm{s}^{-1}$,
$t_\mathrm{0} = -13.90\,\mathrm{s}$, $\Gamma_\mathrm{sens} = 2.20$,
and $F_\mathrm{0} =
2.02\times10^{-14}\,\mathrm{erg}\,\mathrm{cm}^{-2}\,\mathrm{s}^{-1}$.
Combining Eq.~\eqref{eq:pwrlw_sim} with the \textit{eROSITA} exposure map, we were able to
determine the area of the sky at which the survey is sensitive to a
specific flux. We divided the sky into equally-sized pixels with an
area of $0.002\,\mathrm{deg}^2$ and determined the minimum flux within
that pixel. Figure~\ref{fig:erosita_sensitivity_minflux_lin} shows
that the survey is complete
on the whole western sky down to a flux of
$F_{\mathrm{X},\,0.2\mbox{--}2.3\,\mathrm{keV}} \sim
2\times10^{-13}\,\mathrm{erg}\,\mathrm{cm}^{-2}\,\mathrm{s}^{-1}$. For
all sources with a lower flux, corrections using the derived
sensitivity will have to be applied, for instance when deriving $\log
N$-$\log S$-distributions.

\subsection{X-ray spectral analysis} 
\label{subsec:ero-analysis} 
In order to gather information on the X-ray properties of our sample,
we
complemented the available eRASS1 catalog data
\citep{merloni:2024} with spectral information, using \texttt{eSASS}
version 211214, HEASOFT version 6.30, and \textit{eROSITA} event processing
version~020, which offers improved calibration compared to the release
version 010 used by \citet{merloni:2024}. While the processing version
affects the quality of spectral products, it does not affect the
source detection itself within the conservative selection criteria
applied above. Tests with different processing versions revealed that
their influence on the overall spectral fitting results does not change
the overall results. We extracted source spectra from circular
extraction regions centered on each catalog source, scaling the
extraction region radius, $R$, on the maximum likelihood count rate
for the 0.2--2.3\,keV band, given by \citet{merloni:2024},
\begin{equation}
    R = N \times (\mathrm{ML\_RATE\_1}\,[\mathrm{cts}/\mathrm{s}])^\gamma 
    \label{eq:radius_rate}
\end{equation}
where $N=85\farcs{}443$ and $\gamma=0.284$.
Higher count rate sources required larger regions to include
all source photons, however we required at least $23\arcsec$ and at
maximum a region with a radius of $200\arcsec$. Background regions are
annuli centered on the source position, using Eq.~\ref{eq:radius_rate}
to scale the inner and outer radius. Specifically, for the inner
radius we set $N=181\farcs{}963$, $\gamma=0.242$,
$R_\mathrm{inner,min}=54\arcsec$, and
$R_\mathrm{inner,max}=350\arcsec$. For the outer radius,
$N=1063\farcs211$, $\gamma=0.282$, $R_\mathrm{outer,min}=280\arcsec$,
and $R_\mathrm{outer,max}=2200\arcsec$. Within the background region
we excluded all neighboring eRASS1 sources, scaling the
exclusion radius of each source by their count rate. Spectra were then
extracted from event lists with the \texttt{eSASS} task
\texttt{srctool}. Since \textit{eROSITA} TM 5 and 7 are affected by light leaks
\citep{predehl:2021,merloni:2024}, these two modules required special
treatment in our analysis. Unless mentioned otherwise, for these TMs,
only data taken $>$1\,keV were considered. For the spectral analysis, we
used the Interactive Spectral Interpretation System
\citep[ISIS, version 1.6.2;][]{houck:2000} and quote all uncertainties at 90\%
confidence for one independent parameter, unless stated otherwise.

For all sources, we collected basic source information such as the total
amount of source counts and on-axis exposure times. The count information
is given as the number of photons detected by all TMs for the
0.2--2.3\,keV band, or as the combination of counts measured between
1.0--10.0\,keV for TMs 5 and~7 and 0.2--10.0\,keV for all other TMs.
We utilized the Bayesian approach of \citet{park:2006} to determine
hardness ratios,
\begin{equation}
    \mathrm{HR} = \frac{N_\mathrm{Ha}-N_\mathrm{So}}{N_\mathrm{Ha}+N_\mathrm{So}},
\end{equation}where $N_\mathrm{So}$ and $N_\mathrm{Ha}$ are the counts in the softer and harder band, respectively, and their uncertainties.
We ignored TMs 5 and~7, and used the energy bands
0.2--0.7\,keV, 0.7--1.2\,keV, and 1.2--5.0\,keV, hereafter bands 1, 2,
and 3. For an absorbed power law with $N_\mathrm{H}=1\times10^{21}\,\mathrm{cm}^{-2}$ and photon index
$\Gamma_\mathrm{X}=2$, these bands contain a similar number of
photons. The different hardness ratios are designated as follows: HR12
is the hardness calculated using bands~1 and~2, HR23 uses bands~2 and~3, and HR13 bands~1 and~3.
The uncertainties given for the hardness ratios correspond to the smallest credible interval at 90\% confidence around the most likely value.

The source fluxes reported in the eRASS1 catalogs were determined as
part of the \textit{eROSITA} source detection pipeline. They are based on maximum
likelihood methods, applying predetermined energy conversion factors
assuming an absorbed power law with $\Gamma_\mathrm{X}=2.0$ and an
absorption column density of
$N_\mathrm{H}=3\times10^{20}\,\mathrm{cm}^{-2}$. We determined source
intrinsic fluxes for the 0.2--2.3\,keV band based on spectral modeling
with fixed parameters. 
We used simple absorbed power laws
\texttt{tbabs*powerlaw} (\texttt{tbabs(1)*powerlaw(1)"}), with the abundances of \citet{wilms:2000} and
the cross-sections of \citet{verner:1996} and fix the equivalent
hydrogen column, $N_\mathrm{H}$, to the 21\,cm value for the source
position reported by the \citet{hi4picollaboration:2016}. We then
determined the 0.2--2.3\,keV fluxes from fits of the spectral continuum
to the full \textit{eROSITA} energy range from 0.2\,keV (1\,keV for TM5 and~7) to
10.0\,keV for four fixed photon indices, $\Gamma_\mathrm{X}$,
covering the range expected from blazar-like spectra
$\Gamma_\mathrm{X} = \{1.5 , 1.7 , 2.0 , 2.3\}$.

\begin{figure}
\includegraphics[width=\linewidth]{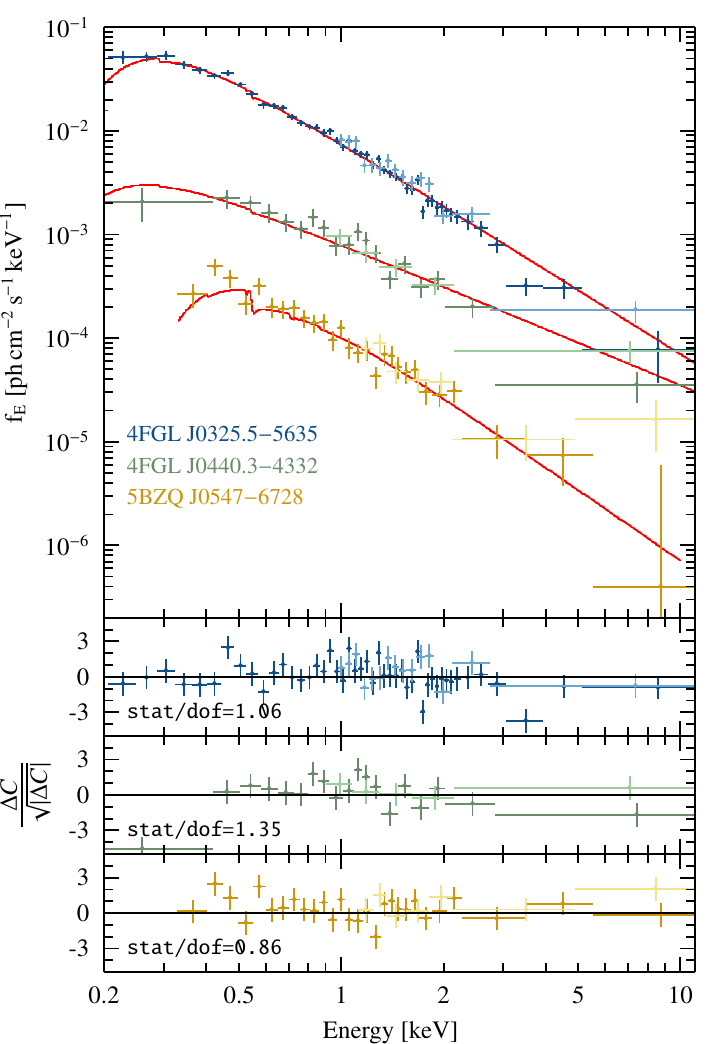}
\caption{Spectral fits of an absorbed power law model for three
  confirmed blazars with different fluxes. For 5BZQ\,J0547$-$6728, no
  prior X-ray observations are available. The three lower panels
  display the fit residuals for each source. Data from TM5 and~7 are
  displayed in a lighter color.}
\label{fig:comp_specs}
\end{figure}

For the 1273 sources with more than 50\,counts detected in the
0.2--2.3\,keV band, we also performed a more detailed spectral analysis,
again using an absorbed power law. For these, we dynamically
rebined each spectrum following \citet{kaastra:2016}, while requiring
that each bin contained at least one count, left the photon index
free, and fixed $N_\mathrm{H}$ to the 21\,cm value from the
\citet{hi4picollaboration:2016}. Spectral minimization is based on
Cash statistics \citep{cash:1979,kaastra:2017}.
Figure~\ref{fig:comp_specs} shows the spectra of three blazars with
our best fit model. 
The fluxes
derived from the fixed photon index power law fits, those determined
during the spectral fits, and the ones reported by
\citet{merloni:2024} are all similar within $\sim$18\%.
For high fluxes \citep[
  ${\ge}F_{\mathrm{X},\,0.2-2.3\,\mathrm{keV}} \sim
  10^{-11}\,\mathrm{erg}\,\mathrm{cm}^{-2}\,\mathrm{s}^{-1}$;][]{merloni:2024}
\textit{eROSITA} spectra are influenced by pileup \citep[see,
  e.g.,][]{konig:2022};
  we proceeded to check for pile-up in these sources.
Using \texttt{SIXTE} to simulate blazar-like spectra within different
flux ranges (also above the flux limit reported by
\citealp{merloni:2024}) we found that our spectra are not significantly
affected by pileup, with a maximum pileup fraction of 1.79\% for
the brightest source.

\section{Additional multiwavelength data} 
\label{sec:mwldata} 
To put the properties of the BlazEr1 catalog into the
multiwavelength context of these sources, we supplemented the eROSITA
information with data from other X-ray and multiwavelength catalogs.

\subsection{Observations of \textit{eROSITA} blazars with other X-ray missions} 
\label{subsec:other-xray} 
To collect soft X-ray flux and spectral information, we cross-matched
the BlazEr1 with the \textit{ROSAT} \citep[0.1--2.4\,keV,][]{boller:2016} and
the \textit{Swift}-XRT (0.3--10.0\,keV) source catalogs by \citet[][OUSXB
DR3]{giommi:2019} and \citet{evans:2020}. Although both catalogs use
\textit{Swift}-XRT, they were compiled independently and using different approaches.
The OUSXB catalog specifically targets blazars and treats each single
observation separately; whereas \citet{evans:2020} merge all exposures
of a given target. BlazEr1 counterparts are again identified by
positional cross-matching, using a maximum angular separation between the BlazEr1 position and \textit{ROSAT} of $40''$ and of $8''$ between BlazEr1 and \textit{Swift}-XRT. We find 1496 \textit{ROSAT} and 1249
\textit{Swift}-XRT \citep[for][]{evans:2020} counterparts. As the OUSXB catalog is
built on an observation basis with one entry per observation, we first matched it
with the BLAZE catalog. For each of the 1039 matches we then derived the
mean, median, minimum, and maximum fluxes and spectral indices from all
source visits.

In order to get information at higher energies, we cross-matched the BlazEr1 with the
\textit{NuSTAR} (3.0--10.0\,keV, 10.0--30.0\,keV) blazar catalog \citep{middei:2022}, using a maximum angular separation of $8\arcsec$
\citep[based on the positional accuracy of \textit{NuSTAR};][]{harrison:2013},
identifying 46 out of 126\,sources. Since this catalog contains one entry per observation, we included the fluxes and spectral
information of the observation closest in time to eRASS1 to the BlazEr1
catalog.  We also matched our detections with the \textit{Swift}-BAT catalog
\citep[14.0--195.0\,keV,][]{lien:2025}, assuming a maximum angular
separation of $60''$, and where available using the positions of
already known counterparts, obtaining 96 matches, for which we added the
flux and spectral information to the BlazEr1.

To identify objects with previous observations, we cross-matched
BlazEr1 with the observation catalogs of 
\textit{XMM-Newton}, \textit{Chandra}, \textit{ASCA}, \textit{NuSTAR}, \textit{Suzaku}, \textit{Swift}-XRT,
and \textit{ROSAT} available at HEASARC
\footnote{\href{https://heasarc.gsfc.nasa.gov/cgi-bin/W3Browse/w3browse.pl}{https://heasarc.gsfc.nasa.gov/cgi-bin/W3Browse/w3browse.pl} using the following distances between the source and all pointings: $0\fdg25$ for \textit{XMM-Newton}, $0\fdg25$ for \textit{Chandra}, $0\fdg417$ for \textit{ASCA}, $0\fdg1$ for \textit{NuSTAR}, $0\fdg15$ for \textit{Suzaku}, $0\fdg2$ for \textit{Swift}-XRT,
and $1.0\degr$ for \textit{ROSAT}.}.
BlazEr1 contains the total exposure time for each source and
mission until the end of the eRASS1 survey.

\subsection{The $\gamma$-ray counterparts of \textit{eROSITA} blazars} 
The Large Area Telescope \citep[LAT;][]{atwood:2009} on board
the \textit{Fermi} Gamma-ray Space Telescope 
(0.1--100.0\,GeV, 1.0--100.0\,GeV) has been observing the entire sky in
the $\gamma$-ray band since 2008. We used the $\gamma$-ray counterparts from the
fourth data release of the \textit{Fermi}-LAT Fourth Source catalog
\citep[4FGL-DR4;][]{abdollahi:2020,abdollahi:2022} listed in the BLAZE catalog, in total obtaining 1293 matches.
We additionally collected flux and spectral information, source classifications, and SED peak
positions provided by the third data release of the
Fourth Catalog of Active Galactic Nuclei Detected by \textit{Fermi}-LAT
\citep[4LAC-DR3;][]{ajello:2020,ajello:2022}.
For the comparison with the 4LAC catalog we used a positional
match between the BlazEr1 catalog and the position of the associated counterparts listed in the catalog with a maximum separation of $1\arcsec$\footnote{This leads to
seven candidates having 4LAC parameter values as well, however, given
the size of the catalog these data points can be ignored and do not
influence our results in any way.}, since it is based on an older 4FGL version as the one used for this paper. 

\subsection{Radio counterparts to \textit{eROSITA} blazars}
We searched for radio counterparts using Very Long Baseline Interferometry (VLBI) programs, since this implies that all flux densities are representative for the beamed compact jet rather than extended lobe emission.
Counterparts from the Radio Fundamental
Catalog \citep[\texttt{RFC}, version
\texttt{rfc\_2023c},][]{petrov:2025}\footnote{\url{http://astrogeo.org/rfc/},
covering the S (2.2--2.4\,GHz), C (4.1--5.0\,GHz), X (7.3--8.8\,GHz), U
(15.2--15.5\,GHz), and K (22.0--24.2\,GHz) bands} were identified using a maximum
angular separation of $1''$, since radio positional
uncertainties are low. We find 2620 matches.
Additionally, we cross-matched the BlazEr1 catalog with the blazars covered by the \texttt{TANAMI}
\citep[8.4\,GHz, 59\,sources;][]{ojha:2010,muller:2018} and \texttt{MOJAVE}
\citep[15\,GHz, 407\,sources][]{lister:2021,homan:2021} surveys. We used the flux density
values that are already public, since an extensive investigation
involving the more current data is beyond the scope of this paper.  With
a separation of at most $1''$ we find 52~\texttt{TANAMI} and
97~\texttt{MOJAVE} blazars within the BlazEr1 catalog. We assumed flux density
uncertainties of 5\% for \texttt{RFC} \citep{petrov:2025} and
\texttt{MOJAVE} \citep{homan:2002,lister:2021}, and 20\% for
\texttt{TANAMI} \citep{ojha:2010}.

\subsection{Infrared and optical counterparts to \textit{eROSITA} sources}
\label{subsec:optdata} 
We gathered counterpart optical and mid-infrared data (bands: g (4686\AA), r
(6165\AA), i (7481\AA), z (8931\AA), W1 (3.4\,$\mu$m), W2
(4.6\,$\mu$m), W3 (12\,$\mu$m), and W4 (22\,$\mu$m)) from S25.
Additionally, the catalog provides the publicly available spectroscopic
redshifts from \citet{kluge:2024}, photometric redshifts for AGNs
computed with CIRCLEZ \citep{saxena:2024}, and other multiwavelength
information.  
\textit{Gaia} information is also contained in the catalog from S25; however, the astrometric data were not used here since the proper motions and parallaxes listed for the brightest and well studied blazars (such as 3C\,273) would indicate a Galactic origin, which also enters the Galactic and extragalactic classification in S25.
This error in \textit{Gaia} is associated with the variability of these objects \citep{khamitov:2022}.
We only used the data for the 4416 sources where the BlazEr1 and the S25 catalogs agree with each other on the eRASS1 counterpart (for details, see Sect.~\ref{subsec:contamination}).

In order to obtain reliable redshifts for population studies we augmented the BLAZE and
BlazEr1 catalogs with redshifts given in the HighZ sample \citep{marcotulli:2025,sbarrato2026}
and the 4LAC
\citep{ajello:2022}. If no redshift is given in these two
catalogs we extended the catalogs to using un--flagged
redshifts from the BZCAT \citep[redshifts not considered to be spurious by][]{massaro:2015}, confirmed and reliable 
redshifts from the 3HSP \citep{chang:2019}, and redshifts from CGRaBS
\citep{healey:2008}, VERONCAT
\citep{veron-cetty:2010}, and SIMBAD \citep{wenger:2000}, along with spectroscopic redshifts from S25, prioritizing redshifts in the order of catalogs listed
here. For the high-redshift source BROS\,J1322.1$-$1323 we used the redshift reported by
\citet{belladitta:2025}, where the eRASS1 counterpart given in the BlazEr1 catalog has been previously reported.

\subsection{Broadband spectral indices}
\label{sec:bb_spec_idx}
We can calculate an estimated spectral index for a power law between different SED bands
\citep[e.g.,][]{tananbaum:1979,stocke:1991,perlman:1996a,giommi:1999,turriziani:2007},
\begin{equation}
    S = N \times E^{-\alpha},
    \label{eq:f_mwl}
\end{equation}
such that the slope, $\alpha$, between two bands characterized by reference energies. $E_i$
and $E_j$, is
\begin{equation}
\alpha_{ij} = - \frac{\log S_i - \log S_j}{\log E_i - \log E_j},
\label{eq:alpha}
\end{equation}
where $S_i=S(E_i)$ and $S_j=S(E_j)$ are the flux densities, converted to Janskys, at a
specific energy ($E_i$, $E_j$).
For the X-rays, we used the 1\,keV flux
density found from power law fits with $\Gamma_\mathrm{X}=2.0$ and a
fixed value of $N_\mathrm{H}$, since this flux estimate is available for all catalog sources. We computed $\alpha_\mathrm{X\Gamma}$ based on
the 0.1--100\,GeV $\gamma$-ray flux from \citet[4FGL-DR4;][]{abdollahi:2022}, assuming
the geometric mean between the energy band boundaries as reference energy $(0.1\,\mathrm{GeV}\times100.0\,\mathrm{GeV})^{1/2}$. For values of
$\alpha_\mathrm{IRX}$ we utilized the dereddened W1-band flux,
whereas for $\alpha_\mathrm{OX}$ we used the dereddened LS10 r-band
flux, both based on the transmission and flux values taken from S25. We accounted for the uncertainty of dust models in the dereddening
\citep[according to][]{fitzpatrick:1999} by applying a 20\% systematic
uncertainty to our estimate for the transmission. This estimate is
based on the different transmission factors obtained when assuming
$\mathrm{R}_\mathrm{V}=2.7$ or $\mathrm{R}_\mathrm{V}=3.1$ in the
extinction law. We also provide a set of values for $\alpha_\mathrm{RX}$,
determined using the \texttt{RFC} X-band (8.6\,GHz), \texttt{TANAMI}
\citep[8.4\,GHz;][]{ojha:2010,muller:2018} and \texttt{MOJAVE}
\citep[15\,GHz;][]{lister:2021,homan:2021} flux densities.
Uncertainties for the $\alpha$-values were estimated using Gaussian
error propagation and 68\% confidence intervals.

\section{The BlazEr1 catalog results and discussion}
\label{sec:properties}
The steps outlined in Sects.~\ref{sec:erodata} and~\ref{sec:mwldata}
led to the construction of the BlazEr1 catalog. We also performed the
same steps for objects which do not match our selection criteria
(Sect.~\ref{subsubsec:xraycounterparts}). These discarded sources
are included in a supplementary catalog (see
Appendix~\ref{sec:cataloginfo} for the catalog description). This sample
(referred to as the \emph{unverified} sample) contains 2252 objects, which we provide as a courtesy, but do not discuss
further.

In this section, we discuss the main
BlazEr1 results, focusing first on \textit{eROSITA} X-ray results, and then
presenting the multiwavelength picture. We also address nondetected
sources here. 

\subsection{Global properties of the BlazEr1 catalog}
The BlazEr1 catalog contains 5865 individual blazars and blazar candidates observed and
detected with \textit{eROSITA}. Of these, 2106 are associated with a confirmed
blazar, while the remaining are blazar candidates. We detect
roughly the same number of sources for each blazar subtype, with 597
\textit{BLL}s, 769 \textit{FSRQ}s and 712 \textit{BCU}s, and also find
28 \textit{BZG}s.
Given the large number of blazars in the catalog, the roughly 30 \textit{BLL}s and 40 \textit{FSRQ}s which might be misclassified or belong to a changing look class \citep{ruan:2014,kang:2024}, do not impact the overall sample statements made.
The blazar candidates are dominated by
\textit{BCUC}s (1892 sources, $\sim$50\% of the candidates), while the
remainder is composed of 25\% \textit{BLLC}s and 24\% \textit{FSRQC}s
(954 and 913, respectively). The origin of the \textit{BCUC}
population is dominated by the BROS catalog. The sky map
(Fig.~\ref{fig:blazar_erass1_map_paper}) shows a roughly evenly
distributed extragalactic distribution of objects. There are 5296
sources at Galactic latitudes $|b|>15^\circ$, corresponding to an area
density of $0.346\,\mathrm{sources}/\mathrm{deg}^2$.

\begin{figure}
\includegraphics[width=\linewidth]{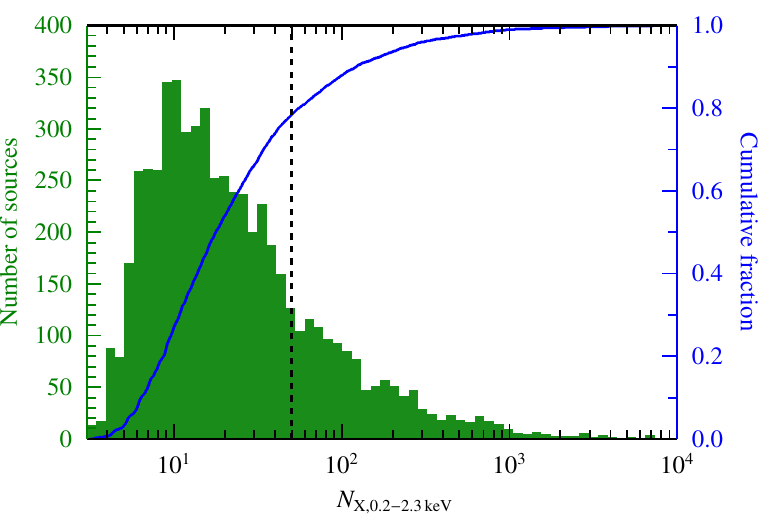}
    \caption{Distribution of source counts with its cumulative
      distribution shown by the solid line curve. The dashed vertical
      line marks the threshold of 50\,counts, above which we deem a
      spectral analysis meaningful. Only 20\% of all sources are above
      this limit, while about half have fewer than
      20\,counts, as shown by the cumulative distribution (blue).}
\label{fig:counts_hist_all}
\end{figure}

In total, the catalog is based on 487030 X-ray photons in the
0.2--2.3\,keV band. Most sources in the catalog have very few counts
(median value: 18\,counts; see Fig.~\ref{fig:counts_hist_all}). During eRASS1, \textit{eROSITA} detected more
than 50 counts per source for about 20\% of the BlazEr1 objects (i.e.,
1273 sources), which we select for a spectral analysis.
4FGL\,J0543.9$-$5531 has the most counts with 26819\,counts.
This source has a flux of $F_{\mathrm{X},\,0.2-2.3\,\mathrm{keV}} =
4.9\times10^{-11}\,\mathrm{erg}\,\mathrm{cm}^{-2}\,\mathrm{s}^{-1}$, which is increased by a factor of about two \citep[$F_{\mathrm{X},\,0.3-10.0\,\mathrm{keV}} =
4.5\times10^{-11}\,\mathrm{erg}\,\mathrm{cm}^{-2}\,\mathrm{s}^{-1}$;][]{giommi:2019} to six \citep[$F_{\mathrm{X},\,0.1-2.4\,\mathrm{keV}} =
9.0\times10^{-12}\,\mathrm{erg}\,\mathrm{cm}^{-2}\,\mathrm{s}^{-1}$][]{massaro:2015} compared to past observations. The long exposure time of 584\,s, given the position of the source, explains the high number of counts compared to equally bright sources.
About half
of the detections have fewer than 20\,counts, with some being registered with only three counts in the \textit{eROSITA} main band (0.2--2.3\,keV).

A vital characteristic of the sample is the sources' flux distribution. Unless stated
otherwise, for consistency and as not to mix different flux determination methods and since we do not expect robust
spectral fitting to be possible for the majority of sources (<50\,counts), we study the properties of the
blazar population in this section using the unabsorbed source intrinsic
flux in the 0.2--2.3\,keV band measured assuming a fixed photon index of
$\Gamma_\mathrm{X}=2.0$ and the 21\,cm $N_\mathrm{H}$ at the source
position for all sources, regardless if the counts are sufficient to determine the photon index or not. We use fluxes assuming $\Gamma_\mathrm{X}=2.0$ since this is
the expected photon index obtained when averaging across the entire
population. Fluxes determined with other values for a fixed photon index
($\Gamma_\mathrm{X} = \{1.5 , 1.7 , 2.3\}$) agree to within 9.1\%. The
fluxes from the eRASS1 catalog \citep{merloni:2024} are consistent
within the uncertainties, with an average deviation of 18\%, which might
be due to the different treatment of absorption and flux determination (on average the
$N_\mathrm{H}$ from the \citet{hi4picollaboration:2016} is higher than the
one assumed by \citealt{merloni:2024}). 
The measured fluxes span a range of
almost four decades; the brightest source is
3C\,273 (4FGL\,J1229.0$+$0202, $F_{\mathrm{X},\,0.2-2.3\,\mathrm{keV}} =
6.2\times10^{-11}\,\mathrm{erg}\,\mathrm{cm}^{-2}\,\mathrm{s}^{-1}$, with an exposure time of 111\,s and 7179\,counts),
the faintest is WIBRaLS2\,J045646.58$-$585411.7
($F_{\mathrm{X},\,0.2-2.3\,\mathrm{keV}} =
6.9\times10^{-15}\,\mathrm{erg}\,\mathrm{cm}^{-2}\,\mathrm{s}^{-1}$, with an exposure of 867\,s and 13\,counts).

\begin{figure}
\includegraphics[width=\linewidth]{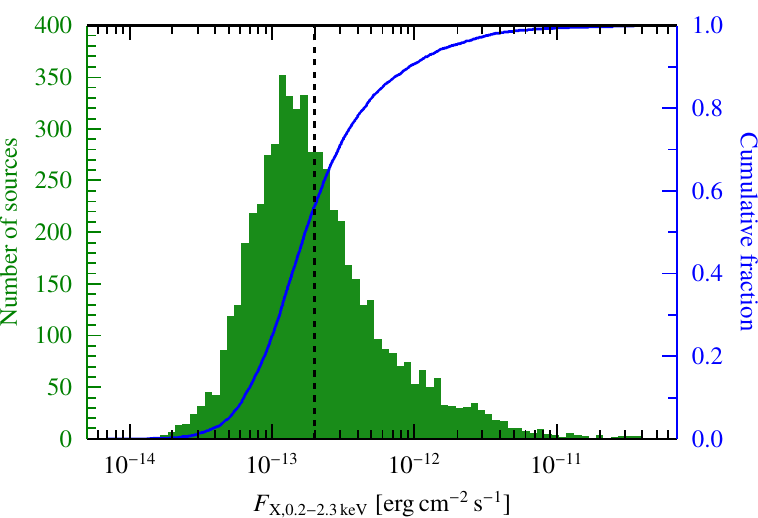}
    \caption{Flux distribution in the BlazEr1 catalog. The majority of
      sources have fluxes below the survey completeness limit on the
      western Galactic hemisphere, the vertical line marks this limit.
      The cumulative distribution is shown in blue.}
\label{fig:flux_hist_all}
\end{figure}

About half of the sources in the BlazEr1 catalog are brighter than the
flux level where completeness is reached across the western Galactic
hemisphere, $F_{\mathrm{X},\,0.2-2.3\,\mathrm{keV}} =
2\times10^{-13}\,\mathrm{erg}\,\mathrm{cm}^{-2}\,\mathrm{s}^{-1}$
(see Sect.~\ref{subsec:sens} and Fig.~\ref{fig:flux_hist_all}). Fainter
sources are located in parts of the sky where \textit{eROSITA} has deeper
exposure.

\begin{figure}
\includegraphics[width=\linewidth]{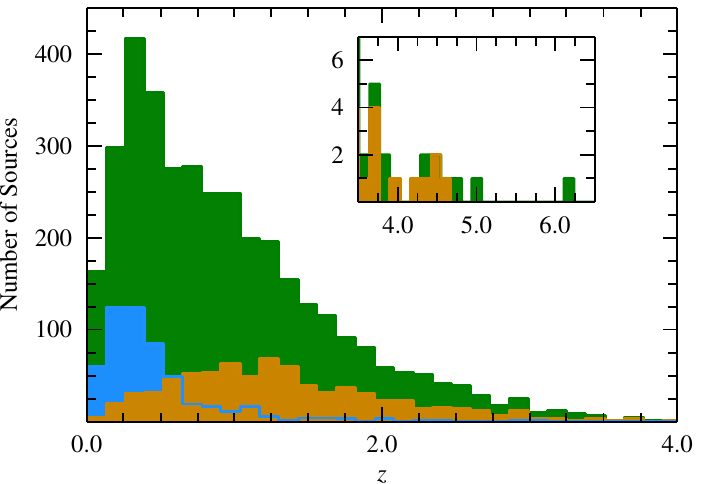}
\caption{Redshift distribution of BlazEr1 sources. The colors represent different subtypes.
  All BlazEr1 entries are shown in green (3627\,sources), \textit{BLL}s in blue,
  and \textit{FSRQ}s in orange. The inset shows the distribution for
  sources with $z>3.5$.}
\label{fig:z_hist}
\end{figure}

Reliable redshifts, $z$, are available for 3627 catalog objects
(Fig.~\ref{fig:z_hist}). Including the S25 photometric redshifts
increases the number of available values to 4795. The majority of
blazars and blazar candidates are found at $z<2$. The sample contains
155\,sources with $z\ge2.5$, including objects up to $z\sim6$, such as
the high-redshift blazar candidate BROS\,J1322.1$-$1323
\citep[$z=4.71$;][]{belladitta:2025,ighina:2025}.
Among these $z\geq2.5$ sources, we report the first X-ray detection of
the blazar 5BZQ\,J1007$+$1356, which was identified as a $\gamma$-ray
emitter by \citet{kreter:2020} during an a posteriori search for
transient $\gamma$-ray signals from high-$z$ blazars. Out of the 87
blazars on the western Galactic hemisphere studied by
\citet{kreter:2020}, \textit{eROSITA} detects 61 sources. Consistent
with previous results \citep{ajello:2012a,ajello:2014,ajello:2020},
the observed distribution of redshifts indicates that \textit{BLL}s
tend to be at low $z$, while \textit{FSRQ}s peak toward $z\sim1.0$.
\textit{BCU}s can be found at all redshifts. About
  13\% of the \textit{BLL}s with measured redshifts are found at
  $z>0.7$, which could be potentially misclassified if the infrared band is
  not covered \citep{delia:2015}. If two out of five high redshift
  \textit{BLL}s would be misclassified, the resulting effect would have a similar impact as the contamination by changing look blazars \citep{delia:2015}. Therefore,
  given the size of the sample the overall results will not be
  significantly altered by not considering this in detail.

Several remarkable \textit{eROSITA}-detected blazars located in the early Universe
are not part of BlazEr1.  \textit{eRASSU}\,J020916$-$562650 is one of the
most distant blazars known \citep[$z=5.6$;][]{wolf:2024,ighina:2024}, but was not
detected in eRASS1, only in subsequent surveys in eRASS2, eRASS3, and eRASS4 \citep[the name of the source deviates from the name used by][]{wolf:2024}. The $z=6.19$ quasar
CFHQS\,J142952+544717, which was detected in the first \textit{eROSITA} survey
\citep{medvedev:2020}, is located on the eastern Galactic hemisphere.
Recent observations with \textit{Chandra} and \textit{NuSTAR} revealed rapid
variability, identifying it as a likely blazar
\citep{marcotulli:2025}.  Similarly, the blazars TXS\,1508+572 ($z=4.3$)
and GB6\,1428+4217 ($z=4.7$), which have shown remarkably luminous
$\gamma$-ray flaring events in recent years \citep{Gokus:2024b, Gokus:2025},
are also located on the eastern Galactic hemisphere.

To assess the quality of the photometric redshifts by S25 we compared them to
the reliable literature (spectroscopic and photometric from, for instance, the BZCAT or 4LAC catalogs) redshifts where both
parameters are available. For almost 67\% of the sources the
difference in redshift is $>$10\%. We therefore limit our population
study involving distance-dependent properties
(such as e.g., luminosity) to objects with
reliable literature redshifts and omit the photometric redshifts of S25.

\subsection{Luminosity distributions of \textit{eROSITA} detected blazars}
\label{subsec:lums}
\begin{figure}
\includegraphics[width=\linewidth]{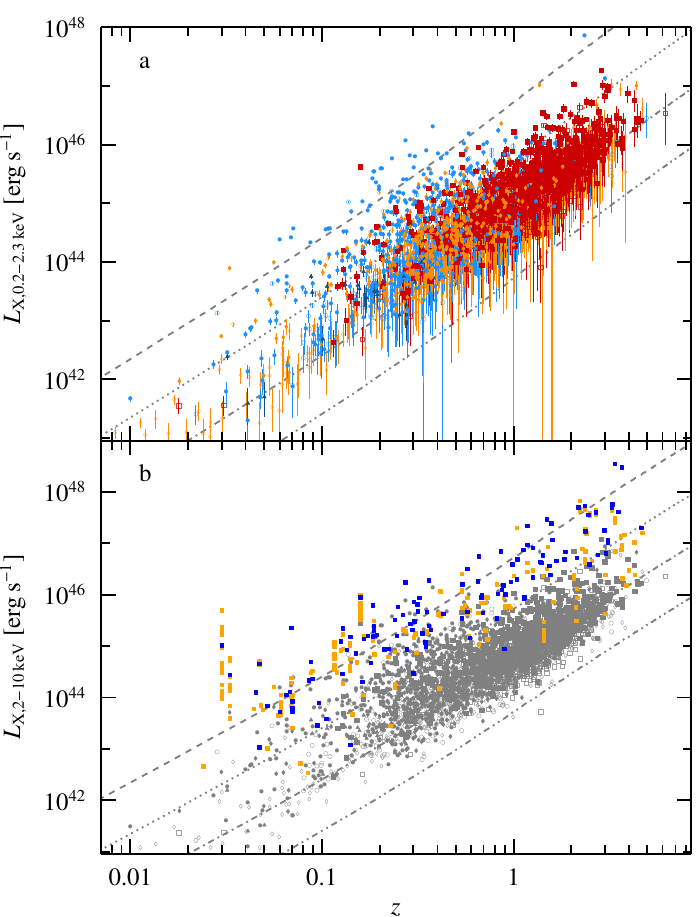}
\caption{X-ray luminosity as a function of redshift. The different
  dashed lines display luminosities of sources with
  $\Gamma_\mathrm{X}=2.0$ for increasing $z$ and a fixed X-ray flux:
  from top (dashed line) to bottom (dash-double dot line), we display
  $F_{\mathrm{X}, 0.2-2.3\,\mathrm{keV}}$ from
  $10^{-11}\,\mathrm{erg}\,\mathrm{cm}^{-2}\,\mathrm{s}^{-1}$ down to
  $10^{-14}\,\mathrm{erg}\,\mathrm{cm}^{-2}\,\mathrm{s}^{-1}$ in steps
  of 10.
  We show the main band
  \textit{eROSITA} luminosity in panel \textbf{a}, following the color scheme introduced in
  Fig.~\ref{fig:blazar_erass1_map_paper}; and the 2-10 keV X-ray luminosity in panel \textbf{b}.
  In gray the
  BlazEr1 sample is shown. Orange points are based on \textit{NuSTAR} data
  \citep{middei:2022}, blue points on \textit{Swift}-BAT \citep{marcotulli:2022}, using the redshifts from BZCAT and 3HSP for the \textit{NuSTAR} and the redshifts from \citet{marcotulli:2022} and adopting their best fit models.}
\label{fig:lum_z_log}
\end{figure}
We derived rest-frame luminosities for the
0.2--2.3\,keV band, $L_{\mathrm{X},
    0.2-2.3\,\mathrm{keV}}$\footnote{Since we use
$\Gamma_\mathrm{X}=2$, the $K$-correction to transform to the
rest-frame, $(1+z)^{\Gamma_\mathrm{X} - 2}=1$ \citep{langejahn:2020}.}. Not surprisingly, the
luminosity of the detected blazars increases with redshift, and the
catalog is also biased toward higher luminosities due to the flux
limit while sampling a larger volume. This bias becomes evident when
comparing with other missions (see Fig.~\ref{fig:lum_z_log}). The
\textit{NuSTAR} sample of \citet{middei:2022} tends to contain sources with
higher luminosities and fluxes, covering similar values as \textit{eROSITA}.
Additionally, with the \textit{NuSTAR} sample it can be seen that blazars are highly
variable. Due to the lower sensitivity of the \textit{Swift}-BAT sample
\citep{marcotulli:2022}, luminosities from that sample are biased
toward high luminosities, as only high-flux sources are part of this
sample. The comparison with these hard X-ray samples indicates that
\textit{eROSITA} samples intermediate luminosities between the shallow all-sky
hard X-ray samples and deeper surveys, and also extends to higher
redshifts (see above).

\begin{table}
\footnotesize
    \centering
   \caption{Characterization of the distribution of luminosities in
     BlazEr1.}
    \label{tab:Lum_means}
    \renewcommand*\arraystretch{1.2} 
    \begin{tabular}{llll}
      \hline\hline
     Population & 1. quartile  & 2. quartile  & 3. quartile  \\
    \hline
    All & 0.21 & 0.73 & 2.17 \\
    Confirmed & 0.18 & 0.74 & 2.57 \\
    Candidates & 0.23 & 0.72 & 1.89 \\
    \textit{FSRQ} & 0.83 & 2.17 & 5.00 \\
    \textit{BLL} & 0.09 & 0.33 & 1.19 \\
    \textit{BCU} & 0.11 & 0.28 & 0.86 \\
    \textit{FSRQC} & 0.67 & 1.28 & 2.48 \\
    \textit{BLLC} & 0.10 & 0.21 & 0.58 \\
    \textit{BCUC} & 0.27 & 0.86 & 2.19 \\
    \hline
    \end{tabular}
    \tablefoot{We list the quartiles for luminosities
     $L_{\mathrm{X}, 0.2-2.3\,\mathrm{keV}}$, in units of
     $10^{45}\,\mathrm{erg}\,\mathrm{s}^{-1}$ for the different
     classes of blazars.}
\end{table}

\begin{figure}
\includegraphics[width=\linewidth]{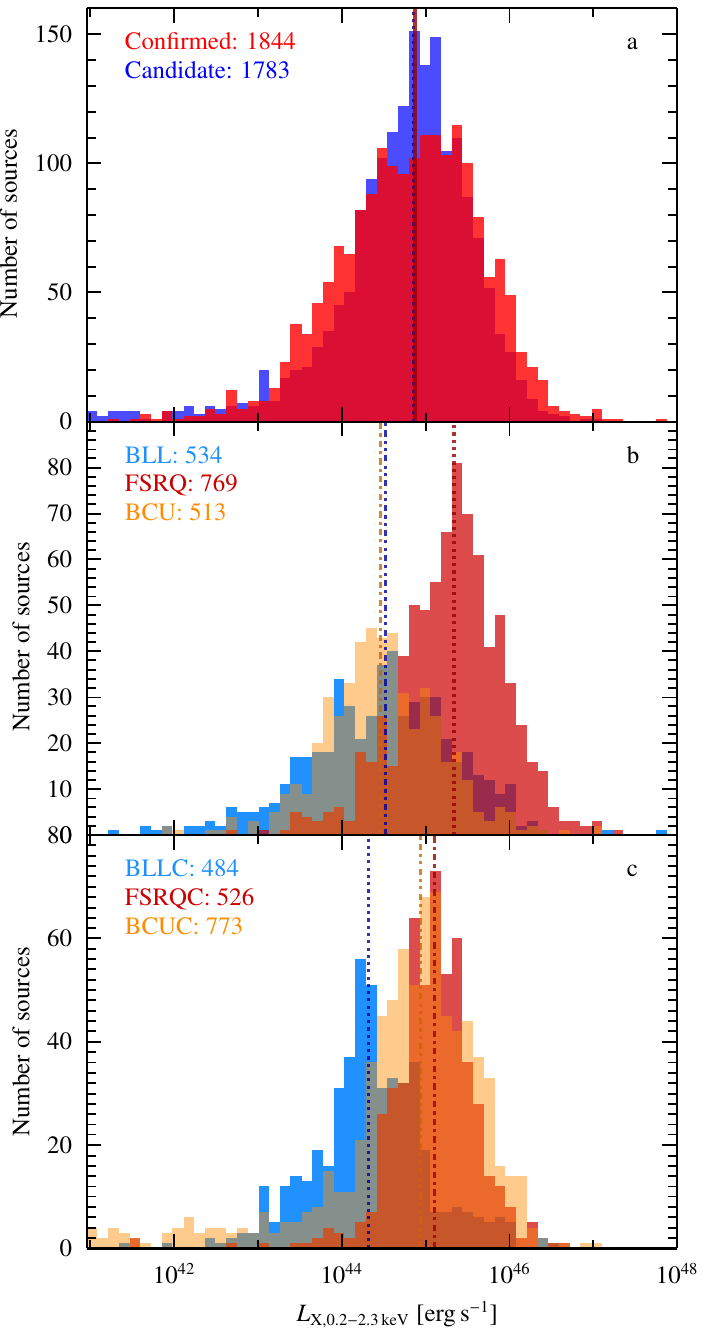}
\caption{Luminosity distribution for all sources with reliable
  redshifts. The panels show distributions for different subgroups of
  the \textit{eROSITA} blazars and \textit{eROSITA} blazar candidates, with vertical lines
  indicating the median luminosity for each subgroup (see
  Table~\ref{tab:Lum_means}).}
\label{fig:lum_hist}
\end{figure}
The median luminosity for all objects is $L_{\mathrm{X},
0.2-2.3\,\mathrm{keV}} \sim7.3\times10^{44}\,\mathrm{erg}\,\mathrm{s}^{-1}$.
Confirmed blazars show slightly higher luminosities, due to a tail
toward lower luminosities apparent in the luminosity distributions
(Fig.~\ref{fig:lum_hist}). On
average \textit{FSRQ}s are more luminous by almost an order of magnitude
compared to \textit{BLL}s and \textit{BCU}s, which have similar values.
The median values for all source subgroups are given in Table~\ref{tab:Lum_means}.
Since redshifts are more difficult to determine for \textit{BLL}s,
luminosities are available for more \textit{FSRQ}s than \textit{BLL}s. The
luminosities inferred with \textit{eROSITA} are consistent with previous studies
\citep[e.g.,][]{donato:2001,fan:2012}. Overall, luminosities determined
for blazar candidates agree with those of confirmed blazars and no stark
difference is found between $\gamma$-detected and nondetected sources.
Kolmogorov-Smirnov tests indicate that most distributions differ
significantly at the 95\% confidence level, all combinations result in a
$p$-value below 0.05, with the exception of the \textit{BCU} and
\textit{BLL} distributions, which likely have the same underlying
distribution (95\%, $p$-value of 0.09). This result is consistent with
previous results which hinted that most \textit{BCU}s are \textit{BLL}s
\citep{kang:2019,pena-herazo:2020,chiaro:2021}.

The low-energy peak frequencies of the LSPs (as identified in the
4LAC) increase with decreasing luminosity. For HSPs and ISPs no
significant trend is observed. This observation is consistent with the
origin of the X-rays within the SED coming from different processes.
The result would also be consistent with the blazar
  sequence, however, since the input catalog is neither complete nor
  statistically well defined, and since the blazar sequence might also
  be due to selection effects, no further conclusion can be drawn from
  this matching behavior.

\subsection{Hardness ratios} 
\begin{figure}
        \includegraphics[width=\linewidth]{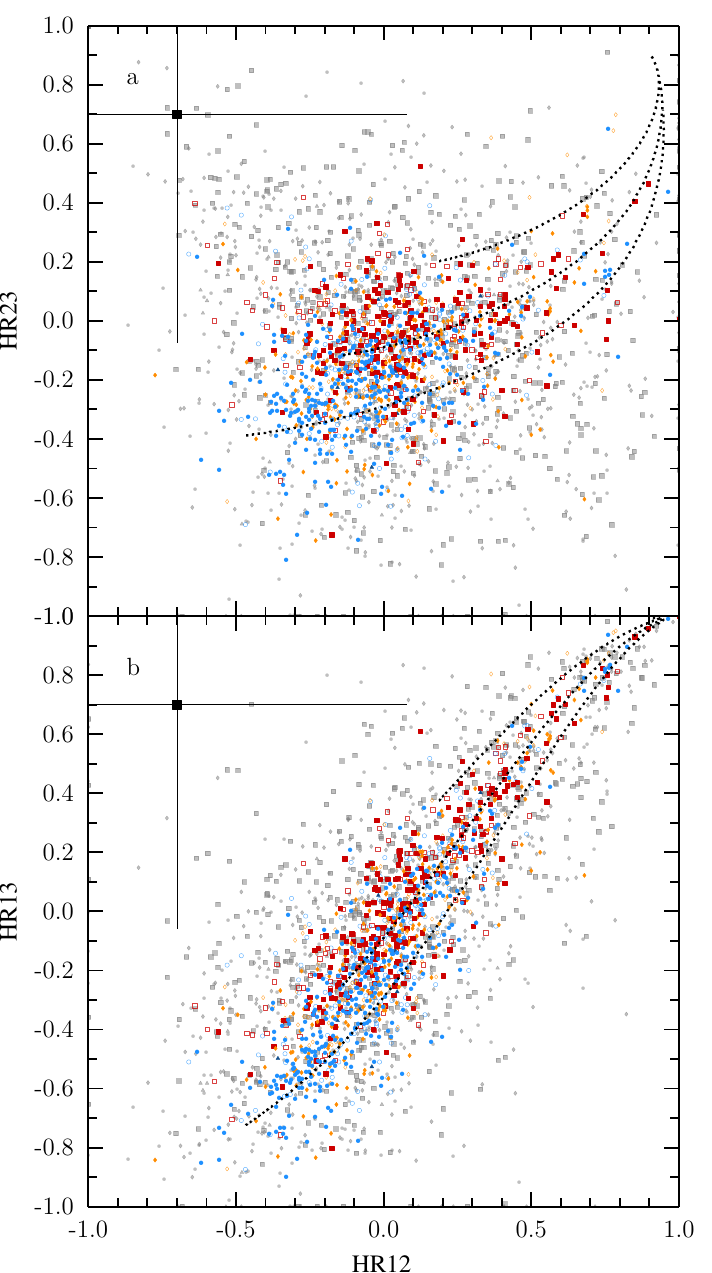}
        \caption{Scatter plot of hardness ratios between energy
          bands~1 (0.2--0.7\,keV), 2 (0.7--1.2\,keV), and 3
          (1.2--5.0\,keV). Sources with $\geq50$ counts are shown in
          the color scheme of Fig.~\ref{fig:blazar_erass1_map_paper},
          fainter sources in gray. These sources show a higher
          dispersion and larger uncertainties. Dotted lines show, from
          top to bottom, expected hardness ratios for photon indices
          $\Gamma_\mathrm{X}=1.0$, $\Gamma_\mathrm{X}=2.0$, and
          $\Gamma_\mathrm{X}=3.0$. The absorption increases along the
          tracks toward the top-right corner. The majority of
          source hardnesses are consistent with power law spectra in
          the expected range of spectral shapes due to the large
          uncertainties. The black data point in the top-left
          corners displays the median uncertainties of
            all, including the low count sources.}
\label{fig:HR_scatterplots_combined_mincounts50_all_in_bkg} \end{figure}
In Fig.~\ref{fig:HR_scatterplots_combined_mincounts50_all_in_bkg}, we
show the distribution of hardness ratios (see Sect.~\ref{subsec:ero-analysis} for the definition) based on the source counts of the BlazEr1 objects. The hardness ratio diagrams highlight
the difference in spectral properties among the blazar classes,
revealing softer spectra for \textit{BLL}s (blue) in comparison to
\textit{FSRQ}s (red). 
The obtained hardness ratios have values expected if sources have absorbed power law spectra. Sources in Fig.~\ref{fig:HR_scatterplots_combined_mincounts50_all_in_bkg} cluster toward the left of the diagrams and the tracks, therefore most objects are consistent with
lower $N_\mathrm{H}$, as expected for the blazar population.

\subsection{Photon indices}
\label{subsec:photonindices}
\begin{figure}
        \includegraphics[height=0.95\textheight]{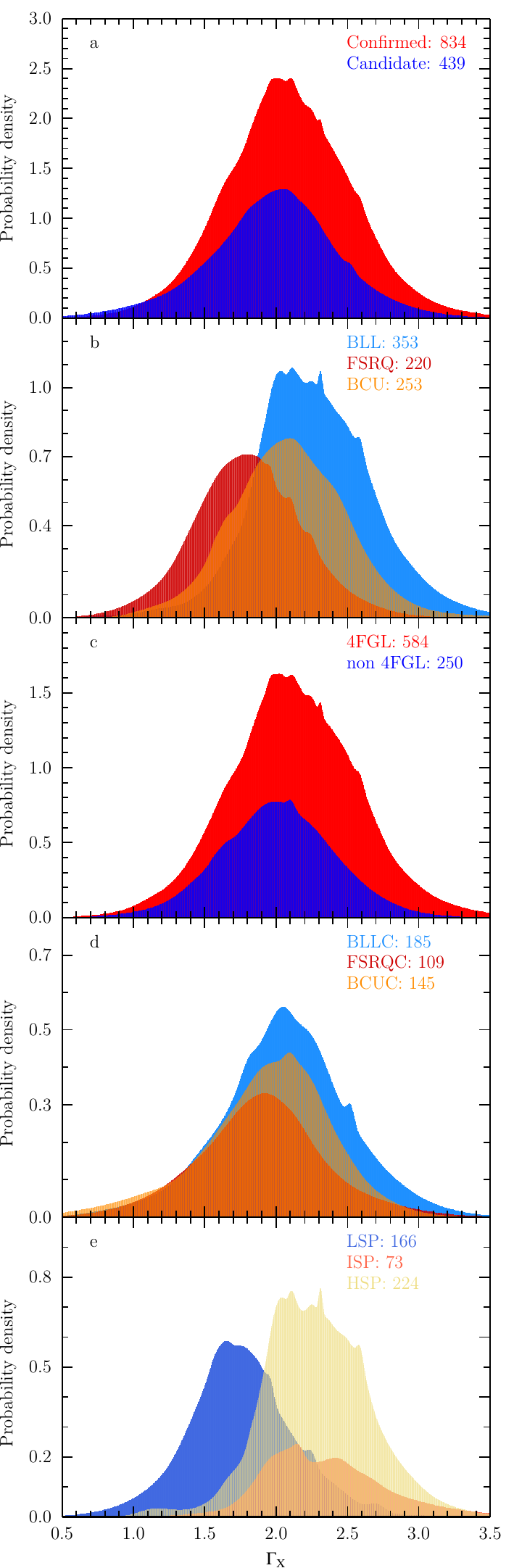}
        \caption{Probability distribution of photon indices for
        different types of \textit{eROSITA} observed blazars and blazar candidates.
        The distribution takes into account the uncertainties of
individual measurements of the photon index.}
\label{fig:gamma_master_gaus} 
\end{figure}
For the 1273 blazars with $\geq50$ counts, more detailed information
about the spectral shape is available from our spectral fits
(Sect.~\ref{subsec:ero-analysis}), since for these sources constraints on the photon index can be derived\footnote{For the spectroscopic sample we use the fluxes determined assuming $\Gamma_\mathrm{X}=2.0$ for consistency.}. Figure~\ref{fig:gamma_master_gaus}
shows probability distributions of the photon indices of all of
these sources as well as for the various subclasses, taking into account
the (sometimes large) uncertainties of the spectral fits. The
distributions are approximately Gaussian; Table~\ref{tab:Gammas_means}
lists the results from fitting Gaussian functions to the distributions.
In order to see how similar the distributions are, we performed KS-tests
on all the possible combinations, requiring a 95\% confidence level.  In
some cases subsamples from the BlazEr1 catalog are likely to have the same underlying distribution
and we list the $p$-values for those.  All not listed combinations have
different underlying distributions ($p<0.05$).

The distribution of X-ray photon indices of the whole spectroscopic sample ($\geq50$ counts) peaks at
around $\Gamma_\mathrm{X}=2.0$. This is observed for the candidates as
well, indicating that a sufficient fraction of these candidates have
X-ray spectra similar to blazars. A clear separation is seen between
\textit{FSRQ}s and \textit{BLL}s.  Consistent with the blazar sequence
\citep{fossati:1998}, \textit{FSRQ}s with a higher luminosity peak at
lower energies due to more efficient cooling \citep[][see
Fig.~\ref{fig:gamma_master_gaus}b]{ghisellini:1998}, resulting in harder
X-ray spectra with a higher chance of the emission not being part of the low
energy SED peak compared to \textit{BLL}s.  \textit{BCU}s are overall
similar to all confirmed blazars ($p=0.707$), confirming their blazar
nature.  In Fig.~\ref{fig:gamma_master_gaus}c, we compare confirmed blazars detected in $\gamma$ rays
to those not detected in $\gamma$ rays, noting that the distributions are quite similar.  We find no difference between
sources detected in the 4FGL and the overall (including those from the 4FGL catalog and those not detected in $\gamma$ rays) distribution of confirmed blazars ($p=0.429$)
or \textit{BCU}s ($p=0.138$). Confirmed blazars with no $\gamma$-ray
detection have similar photon indices as the overall candidate
distributions ($p=0.648$), those of the \textit{BLLC} ($=0.280$) and
\textit{BCUC} ($p=0.589$) classes and the \textit{BCU}s ($p=0.076$).
The distributions of the 4FGL and non-4FGL confirmed blazars do not originate from
the same underlying distribution ($p<0.001$). 
This is also observed for the \textit{BLL}s ($p=0.093$) and \textit{BCU}s ($p=0.972$) subclasses, however, we find no difference between the $\gamma$-ray detected and the nondetected \textit{FSRQ}s ($p=0.008$).
Although the X-ray photon indices of $\gamma$-ray detected and non-$\gamma$-ray detected sources have very similar values, consistent
with previous studies \citep[e.g.,][]{yuan:2014}, the comparison of the distribution indicates different distributions, with the exception of \textit{FSRQ}s. 
A possible reason is
Compton dominance and the general decoupling of X-rays and $\gamma$ rays
due to the different physical mechanisms.
Figure~\ref{fig:gamma_master_gaus}d compares the power law slopes of
different candidate populations. Although \textit{FSRQC}s exhibit
slightly harder spectra, the distributions of \textit{FSRQC}s,
\textit{BLLC}s (which is similar to the confirmed distribution $p=0.705$
and to \textit{BCU}s $p=0.701$), and \textit{BCUC}s are more similar than for the confirmed objects.  The
\textit{BCUC} distribution is similar to the other candidate subclasses
($p=0.066$ for \textit{BLLC}s, and 0.064 for \textit{FSRQC}s).
Potentially this is due to an uncertain source classification.  All
candidates are similar to the \textit{BLLC}s ($p=0.198$), and
\textit{BCUC}s ($p=0.893$).

Figure~\ref{fig:gamma_master_gaus}e shows the photon index by SED type. LSP blazars tend to have
indices below $\Gamma_\mathrm{X} \lesssim 2.0$, indicating a rising
X-ray spectrum, HSPs have photon indices $\Gamma_\mathrm{X} \gtrsim
2.0$, therefore the X-rays are part of the synchrotron peak.  ISPs peak
around $\Gamma_\mathrm{X} \sim 2.0$. We emphasize again that all
distributions have a large overlap with each other, indicating again
that $\Gamma$ alone is not a good measure of source type.  As expected,
\textit{FSRQ}s and LSPs have the same underlying photon index
distribution ($p=0.272$), whereas the HSPs and \textit{BLL}s share
similar values ($p=0.569$), which is also the case for the ISPs
($p=0.680$), which also exhibit a similar distribution as the HSPs
($p=0.841$).
\begin{table*} \centering \caption{Fitting results for the distributions of photon indices.}
        \label{tab:Gammas_means} 
     \renewcommand{\arraystretch}{1.1}
    \begin{tabular}{lllllll} 
    \hline
    \hline
    Population &
                $\langle\Gamma_{\mathrm{X}}\rangle$ & $\sigma_{\Gamma_{\mathrm{X}}}$ & $\langle\Gamma_{\mathrm{X,Comastri}}\rangle$ & $\langle\Gamma_{\mathrm{X,Donato}}\rangle$ & $\langle\Gamma_{\mathrm{X,Kadler}}\rangle$ & $\langle\Gamma_{\mathrm{X,Giommi}}\rangle$ \\
       Reference &  &  & 1 & 2 & 3 & 4\\
        \hline All
                & 2.05 & 0.42 & - & - & - & -\\ Confirmed & 2.08 & 0.43
                & - & - & - & -\\ Candidates & 1.99 & 0.42 & - & - & - &
                -\\ \textit{BLL} & 2.26 & 0.38 & $2.43\pm0.21$ &
                $2.28\pm0.04$$^{*}$, $2.39\pm0.08$$^{**}$ &
                $\sim1.4-2.5$ & $\sim2.0$\\ \textit{FSRQ} & 1.80 & 0.37
                & $1.67\pm0.13$ & $1.76\pm0.06$ & $1.61\pm0.20$$^{***}$
                & $\sim1.6$\\ \textit{BCU} & 2.08 & 0.39 & - & - & - &
                -\\ \textit{BLLC} & 2.07 & 0.42 & - & - & - & -\\
                \textit{FSRQC} & 1.89 & 0.42 & - & - & - & -\\
                \textit{BCUC} & 1.96 & 0.41 & - & - & - & -\\ 4FGL &
        2.12 & 0.44 & - & - & - & -\\ Non-4FGL & 2.00 & 0.40 & - & - & -
        & -\\ HSP & 2.28 & 0.35 & - & - & - & -\\ ISP & 2.32 & 0.39 & -
        & - & - & -\\ LSP & 1.75 & 0.36 & - & - & - & -\\ 
    \hline
    \end{tabular}
        \tablefoot{We fit the distributions in Fig.~\ref{fig:gamma_master_gaus} assuming a gaussian distribution and list the mean values and the standard
        deviations and values from the literature. $^{*}$ HBLs, $^{**}$ LBLs, $^{***}$ their value
for quasars} 
\tablebib{(1)~\citet{comastri:1997}; (2)~\citet{donato:2001}; (3)~\citet{kadler:2005}; (4)~\citet{giommi:2019}}
\end{table*}

It is interesting to compare our average X-ray photon indices for the
different blazar types with those derived previously.
Table~\ref{tab:Gammas_means} lists a few literature values.  Our results
are consistent with the previous studies, despite these samples being
significantly smaller.  Comparing our values to the larger sample by
\citet{giommi:2019} we notice that their distributions peak at different
positions \citep[][their Fig.~10]{giommi:2019}; \textit{eROSITA} spectra
are slightly softer. 
We see an overall trend that \textit{FSRQ}s and
\textit{BLL}s show different spectral slopes, which
is also observed in the literature \citep[e.g.,][]{giommi:2019}.

\begin{figure*} \includegraphics[width=\linewidth]{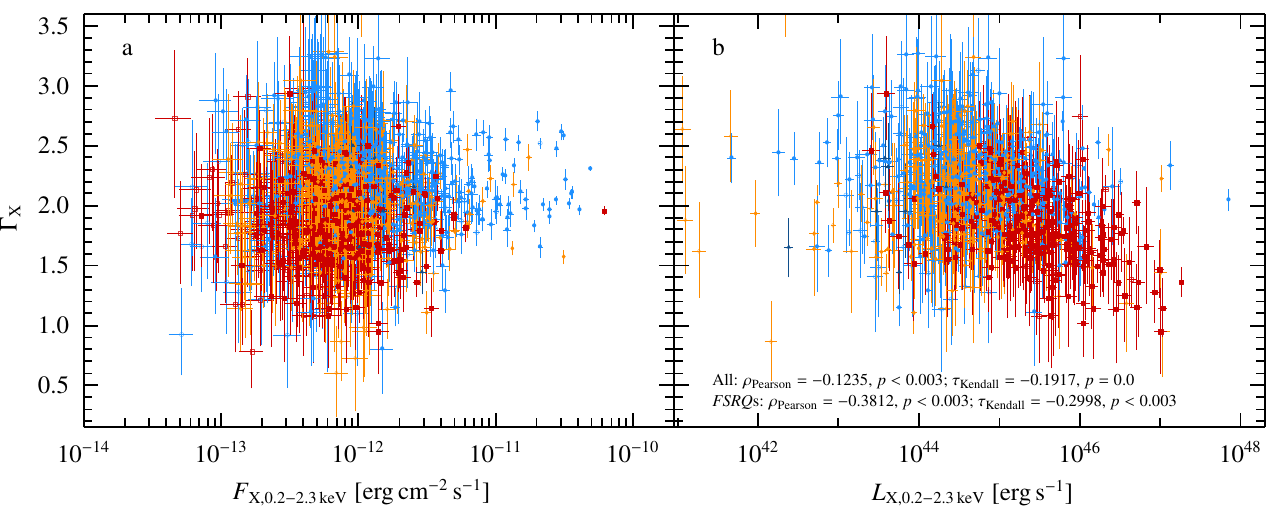}
        \caption{Photon index as a function of \textbf{a} flux and \textbf{b} luminosity, following the
        color scheme introduced in
        Fig.~\ref{fig:blazar_erass1_map_paper} (\textit{BLL(C)}s: blue, \textit{FSRQ(C)s}: red, \textit{BCU(C)s}: orange).}
\label{fig:blazar_gamma_flux} \end{figure*}
In our
study, we also observe the split between \textit{FSRQ}s and \textit{BLL}s in the distribution of $\Gamma$ with flux
(Fig.~\ref{fig:blazar_gamma_flux}a). \textit{BLL}s tend to have
higher indices as well as higher fluxes, while \textit{FSRQ}s tend to
show a harder index but extending toward lower fluxes. 
In the entire sample, we detect no
correlation between flux and photon indices for the entire
sample or any sub sample with the exception of 
\textit{BLL}s showing a very weak but statistically significant anticorrelation\footnote{For the correlation $p$-values we adapt the following: $>0.10$: correlation does not exist, 0.10--0.05: exists but weak, 0.05--0.003: exists with moderate strength, $<0.003$: exists with high significance, if numbers are orders of magnitude smaller for the high significance we just give the threshold value. If based on the $p$-value a correlation does not exist or the coefficient is within the range $-0.1$ and $+0.1$, we do not list the correlation coefficient and $p$-value. Values are listed in Table~\ref{tab:correlations}.} (see Table~\ref{tab:correlations} for correlations coefficients).
Figure~\ref{fig:blazar_gamma_flux}b shows $\Gamma$ as a function of
luminosity. These parameters seem to be anti-correlated, as sources with harder spectra have
higher luminosities, a trend already described by \citet{kadler:2005}.
The anticorrelation is strongest for the \textit{FSRQ}s; all
other sub samples have less statistically significant anticorrelations and some are not statistically significant at all. This
observation is consistent with the blazar sequence, where \textit{FSRQ}s
tend to have higher luminosities and harder indices, as the X-rays are
more likely to originate from inverse Compton emission.  \textit{BLL}s
tend to display lower luminosities and softer indices, which has already
been described in the literature \citep[e.g.,][]{fan:2012}.
Despite the
higher X-ray luminosity of \textit{FSRQ}s, the highest observed fluxes
(${>}5\times 10^{-12}\,\mathrm{erg}\,\mathrm{cm}^{-2}\,\mathrm{s}^{-1}$)
are almost
exclusively from \textit{BLL}s due to their prevalence at lower redshifts.
Additionally, the hard spectral index of
\textit{FSRQ}s makes it easier to detect low-flux \textit{FSRQ}s, while
\textit{BLL}s with very soft spectra and low fluxes are likely
underrepresented due to the detection limits. This indicates a bias in
our sample toward high-flux \textit{BLL}s. For \textit{BLL}s no
correlation exists between the photon index and the redshift of the
source, while for FSRQs we find increasingly
hard spectra and high luminosities with increasing redshift. A
similar observation is made for the \textit{FSRQC}s.
These results resemble those of \citet{giommi:2019}, and are due to a selection effect.

\subsection{The flux-dependent source population of \textit{eROSITA} blazars}
\label{subsec:lognlogs}
\begin{figure*}
\sidecaption
\includegraphics[width=12cm]{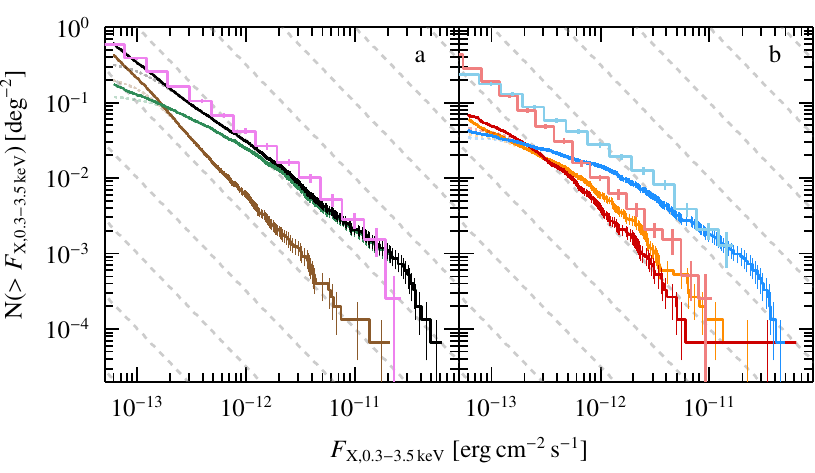}
\caption{$\log N$-$\log S$ distributions for sources within the
  extragalactic sky. Panel \textbf{a}: comparison of the total sample of \textit{eROSITA}
  observed blazars and blazar candidates (black) with the distribution
  by \citet[][purple]{giommi:2015}. Green and brown show the $\log
  N$-$\log S$ for both classes separately (green: \textit{eROSITA} observed
  blazars, brown: observed candidates). Panel \textbf{b}: comparison of the
  predictions of \citet{giommi:2015} for \textit{BLL}s (light blue)
  and \textit{FSRQ}s (light red) with \textit{eROSITA} measurements for
  \textit{BLL}s, \textit{FSRQ}s, and \textit{BCU}s, using the color
  code of Fig.~\ref{fig:blazar_erass1_map_paper}).}
\label{fig:logn_logs_fin}
\end{figure*}
We generated $\log N$-$\log S$ distributions using fluxes and the
sensitivity, giving us the sky area, $A$, covered at a given limiting
flux, $F_{\mathrm{X}}$ (see Sect.~\ref{subsec:sens}), 
\begin{equation}
\label{eq:logNlogS}
    N(>F_X) = \sum_{i} \frac{1}{A_i/1 \mathrm{deg}^2}
\end{equation}
where areas sensitive to a certain flux level were obtained from the absorbed
sensitivity limits of our simulation by correcting for the
absorption used ($N_\mathrm{H} =
1\times10^{21}\,\mathrm{cm}^{-1}$). To avoid contamination by Galactic point sources and those in neighboring galaxies, we excluded regions at Galactic
latitudes $\vert b \vert\leq15\,\degr$ and circular regions of radius $5\fdg5$ and
$3\fdg0$ around the Large and Small Magellanic Clouds, respectively
(Fig.~\ref{fig:erosita_sensitivity_minflux_lin}, blue line). Also, due to the large uncertainties, 
we also excluded all sources fainter than
$F_{\mathrm{X},\,0.2-2.3\,\mathrm{keV}} \sim
6\times10^{-14}\,\mathrm{erg}\,\mathrm{cm}^{-2}\,\mathrm{s}^{-1}$,
such that 4785 objects were finally included for the $\log N$-$\log S$ calculation,
including 1804 blazars (519 \textit{BLL}s, 682 \textit{FSRQ}s, and 575
\textit{BCU}s) and 2981 blazar candidates.

The $\log N$-$\log S$ distributions are shown in
Fig.~\ref{fig:logn_logs_fin}. 
The faded color and dashed lines show the distributions without using the correction obtained from the simulation.
These uncorrected
distributions stress how important it is to take the \textit{eROSITA} survey
sensitivity into account.
The $\log N$-$\log S$ distribution shown here is not binned but cumulative, therefore giving uncertainties is not straightforward. The uncertainties displayed are the square root of the variance of the source number count calculated using Eq.~4 of \citet{cappelluti:2009}.
There is a slight deviation, or ``dip'', in
the distribution at fluxes around
$F_{\mathrm{X},\,0.2-2.3\,\mathrm{keV}} \simeq
1\times10^{-11}\,\mathrm{erg}\,\mathrm{cm}^{-2}\,\mathrm{s}^{-1}$.
Tests with different selection criteria, such as restricting the sources to the common footprint of the
BLAZE catalog input catalogs, did not remove this feature.  Based on the
simulation presented in Sect.~\ref{subsec:sens}, however, we were able to exclude
\textit{eROSITA}'s sensitivity as the cause of this feature, as it is not present in the simulated data, which uses the as-flown \textit{eROSITA} attitude. Computing $\log N$-$\log S$
distributions for various ecliptic latitude bins for the eRASS1 catalog
resulted in noisier distributions in this particular flux range, which is
characterized by a small number of objects. The deviation from a smooth
function became more pronounced, and we therefore concluded that this
feature is most likely due to small number statistics, possibly in
combination with source inhomogeneities in the sky, as also documented
by \citet[][Sect.~5.5]{merloni:2024}.

In order to compare our results to the only available predictions for the X-ray band for
the $\log N$-$\log S$ of blazars \citep{giommi:2015}, we converted all
fluxes to the 0.3--3.5\,keV band, assuming an unabsorbed power law with $\Gamma_\mathrm{x}=2.0$, since the fluxes used throughout the paper and those from the simulation have already been corrected for absorption.  Our $\log N$-$\log S$ distributions
agree quite well with theoretical expectations and agree in terms of
the slope. \citet{giommi:2015} did not assume the blazar sequence but
the blazar-simplified view, which provides a reasonable explanation
for the observed $\log N$-$\log S$ distribution. However, no
definitive statement about whether the blazar sequence or the
blazar-simplified view is true can be made, as the overall results are
consistent with both scenarios. A detailed investigation is beyond the
scope of this paper. Assuming that the blazar candidates are indeed
predominantly blazars, the incompleteness of the BlazEr1 catalog at high fluxes is low with respect to the prediction by \citet{giommi:2019}. At the highest fluxes the
prediction is lower than the observed $\log N$-$\log S$ distribution.
This discrepancy is caused by the 20 brightest blazars, which are
predominantly \textit{BLL}s. After cross-checking the fluxes with the
unpublished eRASS2 data, we can rule out flaring as the cause of this
feature. Fluxes this high are not covered by \citet{giommi:2015}. 
Since the number of sources in this flux
regime is negligible the discrepancy at high fluxes is due to small
number statistics and thus not relevant to the overall result.

We find that confirmed blazars dominate the source numbers in the high-flux regime, with a similar trend as in the combined $\log N$-$\log S$
distribution, with incompleteness inducing a deviation from the
prediction at fluxes below
$2\times10^{-13}\,\mathrm{erg}\,\mathrm{cm}^{-2}\,\mathrm{s}^{-1}$. At
low fluxes, the blazar candidates dominate. Likely most or all blazars
with high fluxes have already been identified, unlike those at low
fluxes, which are therefore included in the candidate catalogs used to build the BLAZE catalog.
We can also compare the distributions of the subtypes of confirmed blazars and find that incompleteness effects
are seen at lower fluxes, with respect to the predictions by \citet{giommi:2015}.
At high fluxes the \textit{BLL}s agree quite well with their
prediction. For \textit{FSRQ}s the $\log N$-$\log S$ is found below
the theoretical curve but the slopes are consistent. For the
\textit{BCU}s, the $\log N$-$\log S$ distribution for most part lies
above that of \textit{FSRQ}s. All these sources either belong to the
\textit{BLL} or \textit{FSRQ} type but have not been identified yet,
explaining parts of the observed incompleteness \citep[the majority of
  \textit{BCU}s are thought be
  \textit{BLL}s;][]{kang:2019,pena-herazo:2020,chiaro:2021}.
\citet{giommi:2015}
also predict a population inversion between \textit{BLL}s and
\textit{FSRQ}s at a flux of $F_{\mathrm{X},\,0.3-3.5\,\mathrm{keV}}
\simeq
1\times10^{-13}\,\mathrm{erg}\,\mathrm{cm}^{-2}\,\mathrm{s}^{-1}$, resulting from the different slopes, which is related to the evolution and distribution of the source classes in the Universe.
Such an inversion is seen in the data as well, although at a slightly
higher flux, albeit this discrepancy could be due to the
incompleteness of our data and other causes can not be ruled out as well.

An earlier comparison by \citet{turriziani:2019} of data with the
\citet{giommi:2015} predictions shows a similar shape in $\log
N$-$\log S$, although due to incompleteness the $\log N$-$\log S$
distribution of confirmed blazars lies below the
prediction and reaches down to fainter fluxes as
  presented here. Candidates included by \citet{turriziani:2019}
resulted in upper limits for the distributions above
the $\log N$-$\log S$ expectation, therefore indicating that the
candidates included might be contaminated by other sources. This
observation is not reproduced here, due to a low level
  of contamination, however, since \citet{turriziani:2019} derived
  upper limits at low fluxes, both results could still be consistent.
  Given the lower flux limit of \citet{turriziani:2019} and
  the fact that their sample is a flux limited one, whereas the
  BlazEr1 catalog sample is not, a detailed comparison is not
  straightforward. Therefore, no disagreement with the previous
  results can be found.

\begin{table*}
    \centering
    \caption{ $\log N$-$\log S$ distributions fit results.}
    \renewcommand{\arraystretch}{1.1}
    \begin{tabular}{lllllll}
    \hline 
    \hline
      Parameter & All & Blazars & Blazar candidates & \textit{BLL} & \textit{FSRQ} & \textit{BCU}\\
     \hline
$\mathrm{C}\,[\mathrm{deg}^{-2}]$  & $54.6\pm2.0$  & $22.5\pm0.9$ & $234.4\pm27.2$ & $3.66\pm0.23$ & $57.5\pm6.8$ & $26.9\pm3.0$  \\
     $\alpha$ & $1.095\pm0.005$ & $1.002\pm0.005$ & $1.537\pm0.017$ & $0.825\pm0.008$ & $1.387\pm0.017$ & $1.232\pm0.015$ \\
     \hline
    \end{tabular}
    \label{tab:lognlogs_results}
    \tablefoot{Results of power law fits (Eq.~\ref{eq:logNlogS_fit}) to the $\log N$-$\log S$ of different samples, measured for fluxes above
    $3\times10^{-13}\,\mathrm{erg}\,\mathrm{cm}^{-2}\,\mathrm{s}^{-1}$,
      where incompleteness effects do not distort the distribution.}
\end{table*}
The slopes of $\log N$-$\log S$ distributions have implications for
the cosmological evolution of blazars
\citep[e.g.,][]{maccacaro:1987,langejahn:2020,marcotulli:2022}. A
slope of $\alpha=1.5$ implies no cosmological evolution, that is, the
sources are uniformly distributed in the Universe.
Less steep slopes
indicate a negative cosmological evolution, where sources are more
common in the local Universe and their density declines with increasing
redshift. On the other hand steeper slopes correspond to a
positive evolution and more or brighter sources at higher redshifts
than in the local Universe. We determined slopes for all the observed
distributions, taking flux uncertainties into account by using a
Monte-Carlo approach and assuming that fluxes follow a normal
distribution. We fitted a power law in the form of
\begin{equation}
    \label{eq:logNlogS_fit}
    \mathrm{N} = C \times \left(\frac{\ensuremath{F_{\mathrm{X}, 0.2-2.3\,\mathrm{keV}}}}{10^{-15}\,\mathrm{erg}\,\mathrm{cm}^{-2}\,\mathrm{s}^{-1}} \right)^{-\alpha} 
\end{equation}
to the 0.2--2.3\,keV $\log N$-$\log S$ distribution\footnote{The
conversion between flux bands does not impact the slope.}, limiting
ourselves to fluxes $F_{\mathrm{X},\,0.2-2.3\,\mathrm{keV}} \ge
3\times10^{-13}\,\mathrm{erg}\,\mathrm{cm}^{-2}\,\mathrm{s}^{-1}$ to
avoid incompleteness effects. The resulting parameters are listed in
Table~\ref{tab:lognlogs_results} and the best-fit functions are shown
in Fig.~\ref{fig:logn_logs_fin_fit}. 
The normalization
for the best fit to the \textit{BLL}s is too low compared to
the observed distribution, an effect due to the flattening caused by
incompleteness, beaming effects \citep{marcotulli:2022}, and the low
number of sources at high fluxes; however, the slope matches well. All
measured slopes are below or close to 1.5, therefore we find almost all subtypes and the overall \textit{eROSITA} observed blazars to be consistent with negative
cosmological evolution. For the blazar candidates the slope
is close to no evolution at all or a slightly positive one.
As the $\log N$-$\log S$ distribution possibly shows breaks, we also
considered a broken power law as a fit function for the entire sample
using the same fit approach. One possible break is found at
$F_{\mathrm{X, break},\,0.2-2.3\,\mathrm{keV}} =
(2.11\pm0.26)\times10^{-12}\,\mathrm{erg}\,\mathrm{cm}^{-2}\,\mathrm{s}^{-1}$,
although a second less significant solution exits with the break at a slightly smaller
flux value. The position of the break could coincide with the
flux region in which \textit{FSRQ}s start to be detected but is also
close to the previously described dip and could therefore be a survey
effect not related to blazars at all. Specific modeling of the $\log
N$-$\log S$ and an analysis of the cosmological evolution of blazars,
which would require a very detailed study of the luminosity function and modeling it with different evolutionary scenarios,
is beyond the scope of this paper.

Older studies were based on comparably
few sources, reducing the significance of some conclusions.
Previous work has
indicated a positive evolution for \textit{FSRQ}s \citep{wolter:2001,toda:2017,turriziani:2019,marcotulli:2022}. For \textit{BLL}s,
different trends have been observed, depending on the underlying
source sample. Slightly positive to no evolution has been found for
\textit{BLL}s of the LSP type \citep{stickel:1991}. HSPs have been
found to be consistent with negative evolution \citep{rector:2000}. In
hard X-ray selected samples, \textit{FSRQ}s are found to evolve
positively, whereas \textit{BLL}s show no or negative evolution
\citep{ajello:2009,toda:2017,marcotulli:2022}.
For $\gamma$-ray selected samples
\textit{BLL}s also follow a positive evolution, with the exception of
HSPs, which show a negative evolution \citep{ajello:2014}. However, the slope
reported by \citet{langejahn:2020} for a radio selected \textit{Swift}-BAT sample
is close to the one reported in our work, indicating a negative evolution as well. The hard X-ray $\log N$-$\log S$ by
\citet{langejahn:2020} shows a significant difference in slope compared to
their radio input sample. \citeauthor{langejahn:2020}
attributed this to different evolution of the X-ray and radio emission, with the
former having a peak at $z\sim 1.5$. Our results using
\textit{eROSITA} support this conclusion, as the soft X-rays seen by \textit{eROSITA} exhibit similar properties as the hard X-ray sources studied by \citet{langejahn:2020}, although the evolution found by them is different than in other studies \citep[e.g.,][]{toda:2017,marcotulli:2022}.

The number of blazars \textit{eROSITA} was supposed to detect has been predicted in
the literature assuming that soft and hard X-ray $\log N$-$\log S$-distributions are similar.
\citet{toda:2017} predicted \textit{eROSITA} to observe 13900 \textit{FSRQ}s and 1900
\textit{BLL}s on the entire sky within four years.
We did indeed detect slightly more \textit{FSRQ}s than \textit{BLL}s, since at low fluxes,
more \textit{FSRQ}s are observed by roughly an order of magnitude.
The prediction would require
the vast majority of unclassified sources to be \textit{FSRQ}s, which would be inconsistent with the expectation that \textit{BCU}s are more likely \textit{BLL}s. Considering that the actual \textit{eROSITA} survey is less sensitive than assumed, the numbers at least somewhat agree as a rough order-of magnitude estimate.
 
\citet{marcotulli:2022} predicted that in its first year \textit{eROSITA} would have
observed 230023 \textit{FSRQ}s with fluxes
$F_{\mathrm{X},\,0.2-2.0\,\mathrm{keV}} \ge
1\times10^{-14}\,\mathrm{erg}\,\mathrm{cm}^{-2}\,\mathrm{s}^{-1}$
across the entire sky and across a wide range of redshifts. The
BlazEr1 catalog does not contain even 1\% of this predicted amount of sources.
Considering the best fit to the $\log N$-$\log S$ distribution for the
\textit{FSRQ} sample, ${\sim}2.17\,\mathrm{deg}^{-2}$ are
expected to be observed, corresponding to 89654 \textit{FSRQ}s over the
entire sky, which only corresponds to 39\% of the predicted number.
The discrepancy is in part caused by \citet{marcotulli:2022} assuming a different, deeper, limiting flux level.
Additionally, the assumed slopes of the $\log N$-$\log S$ distribution are different and both estimates assume that the slope at higher fluxes is the same at low fluxes, which might not be the case.

\subsection{Past X-ray observations and properties of \textit{eROSITA} detected blazars}
\label{subsec:pastXray}
In this section, we compare our \textit{eROSITA} results with those from other
missions, for sources which are detected in both. Since earlier work estimated fluxes assuming
power law slopes with $\Gamma_\mathrm{X}=1.7$ or similar (OUSXB:
$\Gamma_\mathrm{X}=1.8$), when comparing results in this subsection, we consistently use \textit{eROSITA}-fluxes obtained assuming
$\Gamma_\mathrm{X}=1.7$, even in cases where we (or authors of other
works) were able to constrain the spectral shape.

\textit{ROSAT}, as the precursor of \textit{eROSITA}, sets the benchmark for all-sky
surveys at soft X-ray energies. Due to the different spatial
accuracies we matched the \textit{eROSITA} detected sources with a quite large
radius $\leq40\arcsec$, hence this comparison is more prone to
potential mismatches.
In total we obtained 1496 matches between \textit{eROSITA} and \textit{ROSAT} and in total 1865 \textit{ROSAT} sources coinciding with BLAZE catalog objects on the western Galactic hemisphere within $40''$.
Of the \textit{ROSAT} BLAZE catalog matches, 108 are included in the \emph{unverified} BlazEr1 catalog, while 261 sources detected by \textit{ROSAT} are not detected by \textit{eROSITA}.
Therefore, \textit{eROSITA} is able to recover 84\% of the blazars and blazar candidates detected by \textit{ROSAT}.
Since \textit{eROSITA} is sensitive over the entire sky down to a similar flux limit as \textit{ROSAT} \citep[$F_{\mathrm{X},\,0.1-2.4\,\mathrm{keV}} =
10^{-13}\,\mathrm{erg}\,\mathrm{cm}^{-2}\,\mathrm{s}^{-1}$,][]{boller:2016}, the sources not included in the BlazEr1 catalogs are probably missing due to variability. 
The fluxes measured by \textit{ROSAT} agree with the
\textit{eROSITA} fluxes within the \textit{eROSITA} uncertainties in only 9.5\% of the matches. Variability is present and the majority of sources are within the same
order of magnitude. There are 543 objects with \textit{ROSAT} fluxes higher by
more than an order of magnitude, whereas for the reverse, only five
cases are observed. This subsample of high-flux \textit{ROSAT} matches does not
grow or shrink in number with different angular separations between the source positions of the two missions. A similar
trend with respect to \textit{XMM-Newton} is reported by \citet[][their Fig.~25]{boller:2016}, probably caused by spurious \textit{ROSAT} detections
combined with systematic differences between the instruments, such as
energy coverage and sensitivity. The \textit{ROSAT} photon indices are
not well constrained, have large uncertainties and do not correlate with
the \textit{eROSITA} ones. This is likely due to the
different signal-to-noise levels and the restricted energy coverage of \textit{ROSAT}.

For sources both in the \textit{eROSITA} and the \textit{Swift}-XRT catalog
\citep{evans:2020}, we confidently detected 1249 of the 1700 matches within the eRASS1 footprint between \citet{evans:2020} and the BLAZE catalog , assuming a maximum separation of $8''$. 
451 sources are not part of the BlazEr1 catalog and \textit{eROSITA} picks up 73\% of the \textit{Swift}-XRT detected blazars and blazar candidates.
The fluxes are broadly consistent, although only
25\% of the matches agree with each other within the uncertainties.
However, many sources tend to have higher fluxes in \textit{Swift}-XRT than in
\textit{eROSITA}: for 57\% of the sources the flux in \textit{eROSITA} is at least 30\% less
than in \textit{Swift}-XRT. This higher flux is caused partly by variability, but
also due to the fact that the approach of \citet{evans:2020} uses all
available data for a given source position, such that these flux
values represent the combination of different flux states of a source,
with a bias toward brighter states. Many \textit{Swift}-XRT observations are triggered following flaring
states, explaining a bias toward brighter states. Sensitivity differences between
the instruments might also play a role. The photon indices mostly
agree with each other, with a slight trend toward somewhat softer photon
indices measured with \textit{eROSITA} ($\Delta\Gamma_{\mathrm{X}}\sim0.05$).

A similar picture presents itself when comparing our catalog to the
OUSXB \citep{giommi:2019}. Similar to \citet{evans:2020}, this catalog
uses all \textit{Swift}-XRT observations, however, the catalog offers data
for each individual observation. 
In terms of total numbers, OUSXB DR3 contains
1223 BLAZE catalog sources in the western Galactic hemisphere \citep[][2508 BLAZE catalog sources
are listed for the entire sky]{giommi:2019}, of which 1039 are confidently detected and an additional 85 objects are part of the \emph{unverified} catalog. 
This means that only 99 sources go completely undetected by \textit{eROSITA}.

\begin{figure}
\includegraphics[width=\linewidth]{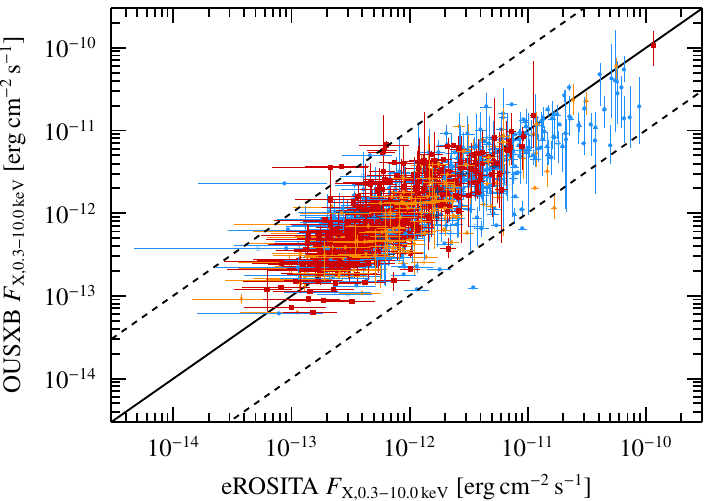}
\caption{\textit{Swift}-XRT flux reported by \citet{giommi:2019} as a function of
  \textit{eROSITA} flux, following the color scheme introduced in
  Fig.~\ref{fig:blazar_erass1_map_paper}. The mean values of all
  measurements for a source listed in \citet{giommi:2019} are shown.
  The uncertainties of the \textit{Swift}-XRT fluxes represent the minimum and
  maximum values provided by \citet{giommi:2019}. The solid line
  represents the unity line, while the dashed ones indicate a factor
  10 difference between the measurements.}
\label{fig:eROSITA_vs_OUSXB_flux}
\end{figure}

\begin{figure}
\includegraphics[width=\linewidth]{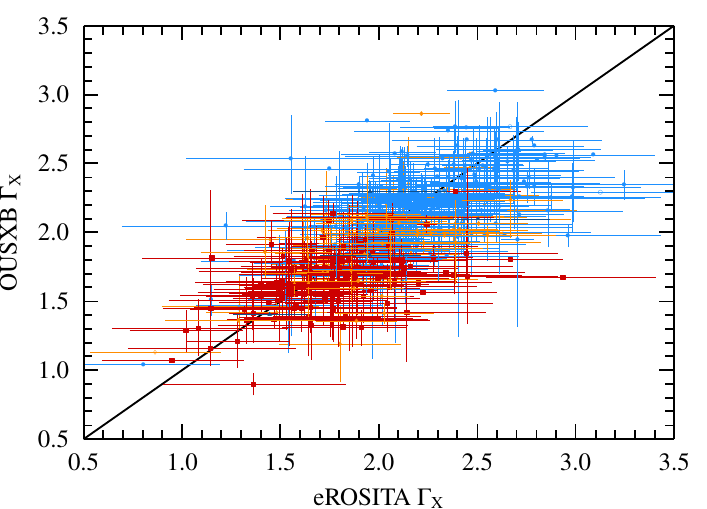}
\caption{Mean photon indices derived by \textit{Swift}-XRT using
  \citet{giommi:2019} and \textit{eROSITA}, following the color scheme introduced
  in Fig.~\ref{fig:blazar_erass1_map_paper}. The uncertainties of the
  \textit{Swift}-XRT photon indices represent the minimum and maximum values
  provided by \citet{giommi:2019}. The solid line represents unity.
  Differences are observed, indicating that blazars are variable in
  their spectral properties.}
\label{fig:eROSITA_vs_OUSXB_gamma}
\end{figure}

The BlazEr1 catalog lists 2106 confirmed blazars, therefore it is the
largest collection of X-ray detected blazars on the western Galactic
sky by a factor of 1.7. Taking the blazar candidates into
account, the BlazEr1 catalog encompasses the largest sample of X-ray
detected blazars and blazar-candidates to this day.
In Fig.~\ref{fig:eROSITA_vs_OUSXB_flux} we compare the mean flux value of
\citet{giommi:2019} with our flux. Generally, the fluxes are
consistent with each other within the \textit{eROSITA} uncertainties and the
maximum and minimum value provided by \citet{giommi:2019}.
In ${\sim}48\%$ of sources, we observe flux variations due to source variability.
However, due to a slightly different photon index being
used and due to the observational biases by \textit{Swift}-XRT in terms of
including all data and therefore also including flares, for more than
50\% of the sources the fluxes do not agree. 
The same has already been observed when comparing to the catalog by \citet{evans:2020}.
Generally, the photon
indices are consistent with each other with ${\sim}68\%$ of the values
agreeing within the uncertainties of \textit{eROSITA} and the maximum and minimum
values given in the OUSXB (Fig.~\ref{fig:eROSITA_vs_OUSXB_gamma}). Overall, there seems to be a
slight trend of deriving softer photon indices using \textit{eROSITA} compared to
\textit{Swift}-XRT (mean $\Delta\Gamma_{\mathrm{X}}\sim0.08$, with 17\% of the
sources having a diverging photon index of at least 1$\sigma$ deviation).
This may be explained by the harder-when-brighter trend in connection with the bias of \textit{Swift}-XRT observations toward flaring states.
Overall, the results derived from \textit{eROSITA} data are consistent with
previous studies.

For the hard X-rays we compare our soft X-ray fluxes with the hard
X-ray fluxes obtained by \textit{Swift}-BAT \citep{lien:2025} and \textit{NuSTAR}
\citep{middei:2022}. Very few sources are identified in the \textit{Swift}-BAT
data. The \textit{Swift}-BAT fluxes are all in a similar range and we test how
these hard X-ray fluxes correlate with the \textit{eROSITA} flux. 
\textit{Swift}-BAT picks up more \textit{FSRQ}s and \textit{BCU}s, and
they tend to have slightly higher fluxes in \textit{Swift}-BAT compared to
\textit{BLL}s, for which all correlations are insignificant. For the softer \textit{NuSTAR} band higher
fluxes in \textit{eROSITA} correspond to higher fluxes in \textit{NuSTAR}. In
the harder band, the trends are quite similar but appear with weaker significance. The photon indices of \textit{NuSTAR} and
\textit{Swift}-BAT are similar to those observed by \textit{eROSITA}, establishing a linear correlation
between indices in the soft and hard X-ray band.

Finally, we stress that many of the BlazEr1 sources were never observed with
other X-ray missions.
It contains 5611 and 5546 sources that have never been observed by \textit{XMM-Newton} and \textit{Chandra}, respectively.
Other missions show similar high numbers of unobserved sources (\textit{ASCA}: 5730,
\textit{NuSTAR}: 5772, \textit{Suzaku}: 5822, \textit{Swift}-XRT: 4069 and \textit{ROSAT}: 4331). We find
that 2966 BlazEr1 sources (51\% of the BlazEr1 catalog) have no
prior X-ray observations with any of the listed missions. Considering only
confirmed blazars, 31\% of these  were never observed with any pointed X-ray instrument.
The sample of previously unobserved blazars is dominated by
\textit{FSRQ}s and \textit{BCU}s. Since \textit{BLL}s are more easily
detected due to their higher X-ray fluxes, they might be targeted more
often, such that previous missions will have more exposure time 
at positions associated with \textit{BLL}s.
Due to the large amount of new X-ray data, \textit{eROSITA} enables us to study a more complete picture of the blazar population.

\subsection{Multiwavelength properties of the BlazEr1 sources}
\label{subsec:mwl}
In this
section, we combine \textit{eROSITA} data with multiwavelength data available
for BlazEr1 sources (see Sect.~\ref{sec:mwldata}).
We discuss the $\gamma$-ray, radio, infrared, and optical properties and 
how the properties in the different wavelength regimes are connected with the X-rays.

\subsubsection{The $\gamma$-ray emission of BlazEr1 sources}
\begin{figure}
\includegraphics[width=\linewidth]{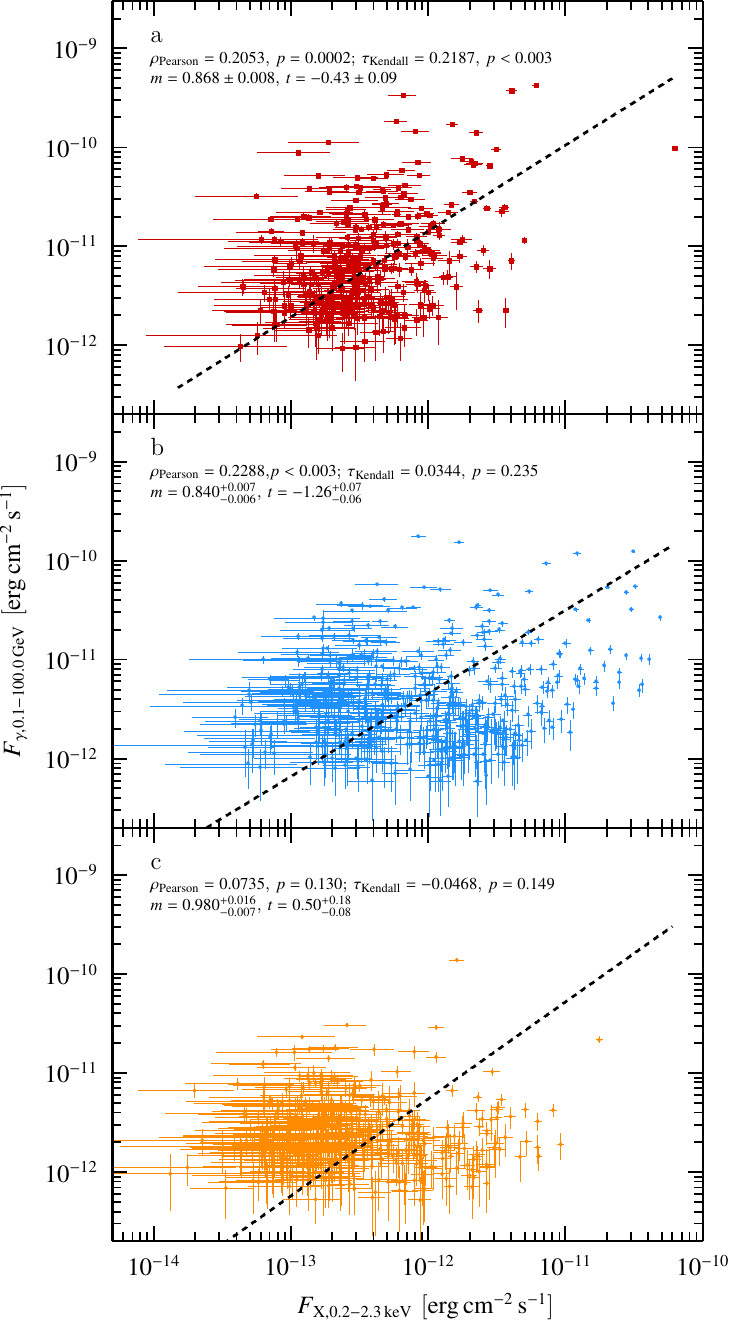}
\caption{\textit{Fermi}-LAT fluxes \citep[4FGL-DR4;][]{abdollahi:2022} compared with 
  \textit{eROSITA} fluxes. The three panels show the different subtypes, following
  the color scheme introduced in
  Fig.~\ref{fig:blazar_erass1_map_paper} (\textbf{a}: \textit{FSRQ}s, \textbf{b}:
  \textit{BLL}s, \textbf{c}: \textit{BCU}s). The \textit{Fermi}-LAT uncertainties
  are scaled to match the confidence limit of the \textit{eROSITA} data.}
\label{fig:eRO_LAT}
\end{figure}
In Figure~\ref{fig:eRO_LAT} we compare the X-ray and $\gamma$-ray fluxes of
the BlazEr1 sources. We fitted a linear function in log-space ($\log_{10}(F_{\mathrm{\gamma},0.2-100\,\mathrm{GeV}}) = m\,\times\,\log_{10}(F_{\mathrm{X},0.2-2.3\,\mathrm{keV}})\,+\,t$) to the
distributions, and calculated correlation coefficients. For
\textit{FSRQ}s, a dense cloud indicating a direct correlation of
X-ray and $\gamma$-ray fluxes can be seen, although the correlation is weak. 
The distributions of \textit{BLL}s and \textit{BCU}s
are more spread out and most likely not
correlated. Since almost all \textit{FSRQ}s belong to the class of
LSPs, a correlation between X-ray and $\gamma$-ray emission is somewhat
expected as the emission is expected to be produced by the same emission mechanism. For
\textit{BLL}s and \textit{BCU}s, which consist of a mix of LSPs, ISPs,
and HSPs, a correlation is less straightforward, but from
$F_{\mathrm{X},0.2-2.3\,\mathrm{keV}}\sim2\times10^{-12}\,\mathrm{erg}\,\mathrm{cm}^{-2}\,\mathrm{s}^{-1}$
on, sources with higher X-ray fluxes also seem to have higher $\gamma$-ray
fluxes.
Additionally, we observe that in general, sources with a $\gamma$-ray detection have higher X-ray fluxes than the non-$\gamma$-ray detected ones, which has also been previously reported by \citet{paliya:2017}.

\begin{figure}
\includegraphics[width=\linewidth]{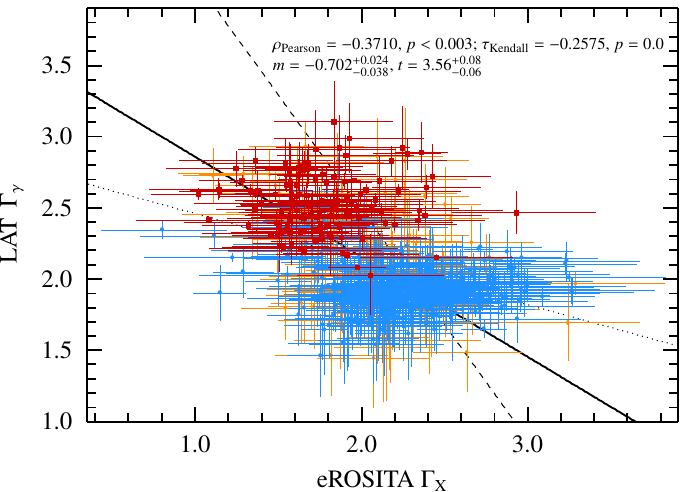}
\caption{\textit{Fermi}-LAT photon index \citep[4FGL-DR4;][]{abdollahi:2022} compared with the \textit{eROSITA}
        photon index, following the color scheme introduced in Fig.~\ref{fig:blazar_erass1_map_paper} (\textit{BLL}s: blue, \textit{FSRQ}s: red, \textit{BCU}s: orange). In solid, the linear best fit is shown, additionally the
        correlation between the two parameters found by
        \citet[][dashed]{yuan:2014} and \citet[][dotted]{fan:2012} are displayed. The uncertainties of $\Gamma_\gamma$ are extrapolated to the 90\% confidence interval.}
\label{fig:eROSITA_vs_LAT_gamma}
\end{figure}
In Fig.~\ref{fig:eROSITA_vs_LAT_gamma} we compare the \textit{Fermi}-LAT
photon indices to the \textit{eROSITA} photon indices. There is a clear separation
between \textit{BLL}s and \textit{FSRQ}s. On average, the X-ray index
of \textit{BLL}s is softer than the $\gamma$-ray index, while for
\textit{FSRQ}s the opposite is observed. This is partly due to the
different SED types these classes are mainly associated with.
However, roughly 4\% of the \textit{BLL}s are occupying the same
region as the \textit{FSRQ}s, which are mostly classified as
LSPs. The \textit{BCU}s are distributed among the different groups.
Previous studies \citep[e.g.,][]{comastri:1997,fan:2012,yuan:2014}
have found that the X-ray and $\gamma$-ray photon indices of the entire blazar population
are anti-correlated. This trend is seen for the overall sample as well
(see Fig.~\ref{fig:eROSITA_vs_LAT_gamma}), however, not as strongly
and there also seem to be differences for the subgroups. For the
entire sample we found a best fit with a linear
fit function ($\Gamma_{\gamma}=m\,\times\,\Gamma_{\mathrm{X}}\,+\,t$) with fit values between the ones
reported by
\citet{fan:2012} and \citet{yuan:2014}. We emphasize that the
correlation is relatively weak but significant. The subgroups show no correlation, with the exception of a very weak anticorrelation for \textit{BCU}s.

We note that the data compared here have not been taken
simultaneously, and we compare X-ray fluxes taken at a specific time
(during eRASS1) with $\gamma$-ray fluxes from the 4FGL-DR4 averaged over 14
years. Hence, variability can make the interpretation difficult.

\subsubsection{Radio emission of BlazEr1 sources}
\begin{figure}
\includegraphics[width=\linewidth]{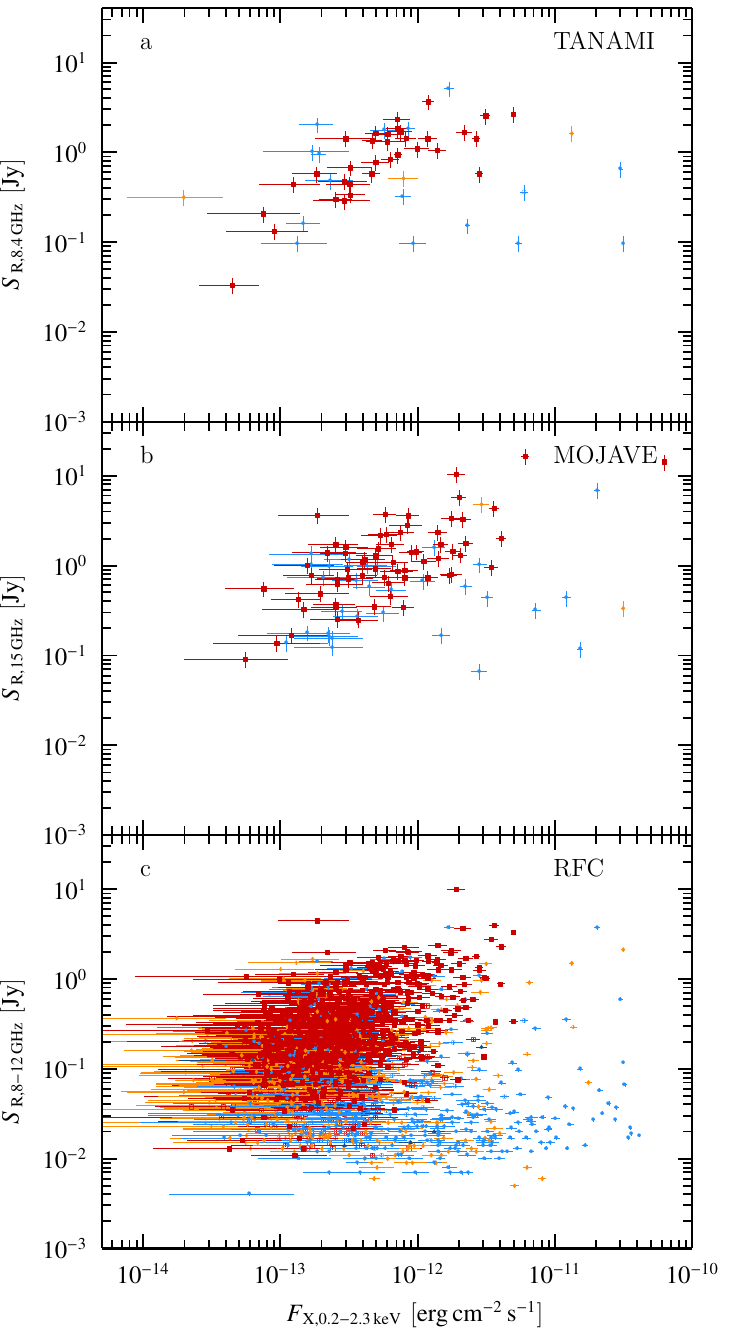}
\caption{Correlation of radio flux densities and \textit{eROSITA} fluxes,
  following the color scheme introduced in
  Fig.~\ref{fig:blazar_erass1_map_paper}, using \textbf{a} data from the
  \texttt{TANAMI} program, \textbf{b} data from \texttt{MOJAVE}, and \textbf{c}
  X-band flux densities from the \texttt{RFC} catalog.}
\label{fig:eRO_Radio}
\end{figure}
In Fig.~\ref{fig:eRO_Radio}, we show the flux densities obtained from
\texttt{TANAMI}, \texttt{MOJAVE}, and \texttt{RFC} as a function of
\textit{eROSITA} flux. Of the 52 \texttt{TANAMI} matches, the majority belongs to
the \textit{FSRQ} class, as this program targets $\gamma$-ray detected sources up to a limiting radio flux density, as well as highly variable jetted non-blazar AGNs. Flux densities in
\texttt{TANAMI} mostly increase as a function of \textit{eROSITA} flux and we find a strong correlation for \textit{FSRQ}s. Almost twice as many objects are
matched with \texttt{MOJAVE} (97); the overall trends are similar to
\texttt{TANAMI} and a correlation with the \textit{eROSITA} flux is found as well. The \texttt{RFC} contains radio
flux densities for 2620 BlazEr1 sources. Again a similar trend of the flux densities
being correlated as in \texttt{TANAMI} and \texttt{MOJAVE} is observed for \textit{FSRQ}s. Based on the \texttt{RFC} data, we
find that for \textit{BLL}s flux densities in the radio are
similar, regardless of the X-ray flux and no strong correlation is found.
This behavior is also observed for the \textit{BLL}s in \texttt{TANAMI} and \texttt{MOJAVE}, although the number of individual objects is very low compared to the \texttt{RFC} and the correlations found are not statistically significant.
Again, the different physical origin of the emission is likely driving the diverging behavior between the \textit{FSRQ}s and \textit{BLL}s.
Of the radio flux limited sample of HBL sources by \citet{giommi:2020}, five out of twelve sources on the western Galactic hemisphere are included in the BlazEr1 catalog.

\subsubsection{Infrared and optical properties of \textit{eROSITA}-detected blazars} 
\begin{figure}
\includegraphics[width=\linewidth]{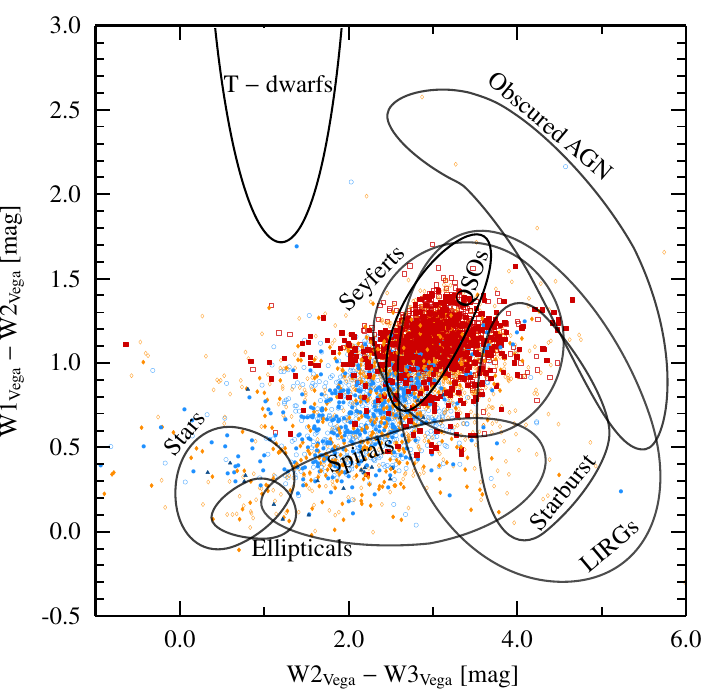}
\caption{WISE color-color diagram, following the color scheme
  introduced in Fig.~\ref{fig:blazar_erass1_map_paper}. Regions
  populated by different types of AGNs, as well as other objects, are shown. 
  Compare our results to \citet[][Fig.~9]{salvato:2018}, for a color-color diagrams of the 2RXS and XMMSL2 samples. }
\label{fig:wise_cols}
\end{figure}
Infrared colors are commonly used to characterize different types of AGNs
\citep[e.g.,][their Fig.~9]{salvato:2018}. In the infrared color-color diagram
(Fig.~\ref{fig:wise_cols}), blazars occupy a distinct region
due to the nonthermal emission of the jet, which enables good distinction from thermal emitters
\citep[e.g.,][]{massaro:2011,dabrusco:2012,massaro:2012,massaro:2012a,massaro:2016a,salvato:2018,demenezes:2019}.
Most of the sources in BlazEr1 are contained within the locus initially described by \citet{massaro:2011} to be associated with jet emission in the WISE color-color space displayed in Fig.~\ref{fig:wise_cols}. 
Using all WISE colors, the region containing $\gamma$-ray emitting blazars has been well characterized, establishing the WISE $\gamma$-ray strip \citep{dabrusco:2012,massaro:2012}.
The initial location by \citet{massaro:2011}, which has not been formally characterized, is considered a projection of the higher dimensional WISE $\gamma$-ray strip \citep{massaro:2016a}.
We therefore, find our blazars to be consistent with the initial location by \citet{massaro:2011} and the WISE $\gamma$-ray strip \citep{massaro:2016a}.
This region in the color-color diagram also covers the region usually occupied by QSOs and
Seyfert galaxies, and extends toward spiral and elliptical galaxies. 
Objects
located far from this region are mostly blazar candidates of some
kind, although sources associated with confirmed blazars are found in
other parts of the color-color diagram as well. A high number of
\textit{BCUCs} are found in the regions associated with starburst,
spiral, and elliptical galaxies, as well as stars. The infrared colors also
separate the \textit{BLL}s and \textit{FSRQ}s as they are found in
different regions. \textit{BLL}s tend to have bluer colors than the
\textit{FSRQ}s. \textit{BCU}s are found in similar loci as
\textit{BLL}s and \textit{FSRQ}s.

\begin{figure*}
\includegraphics[width=\linewidth]{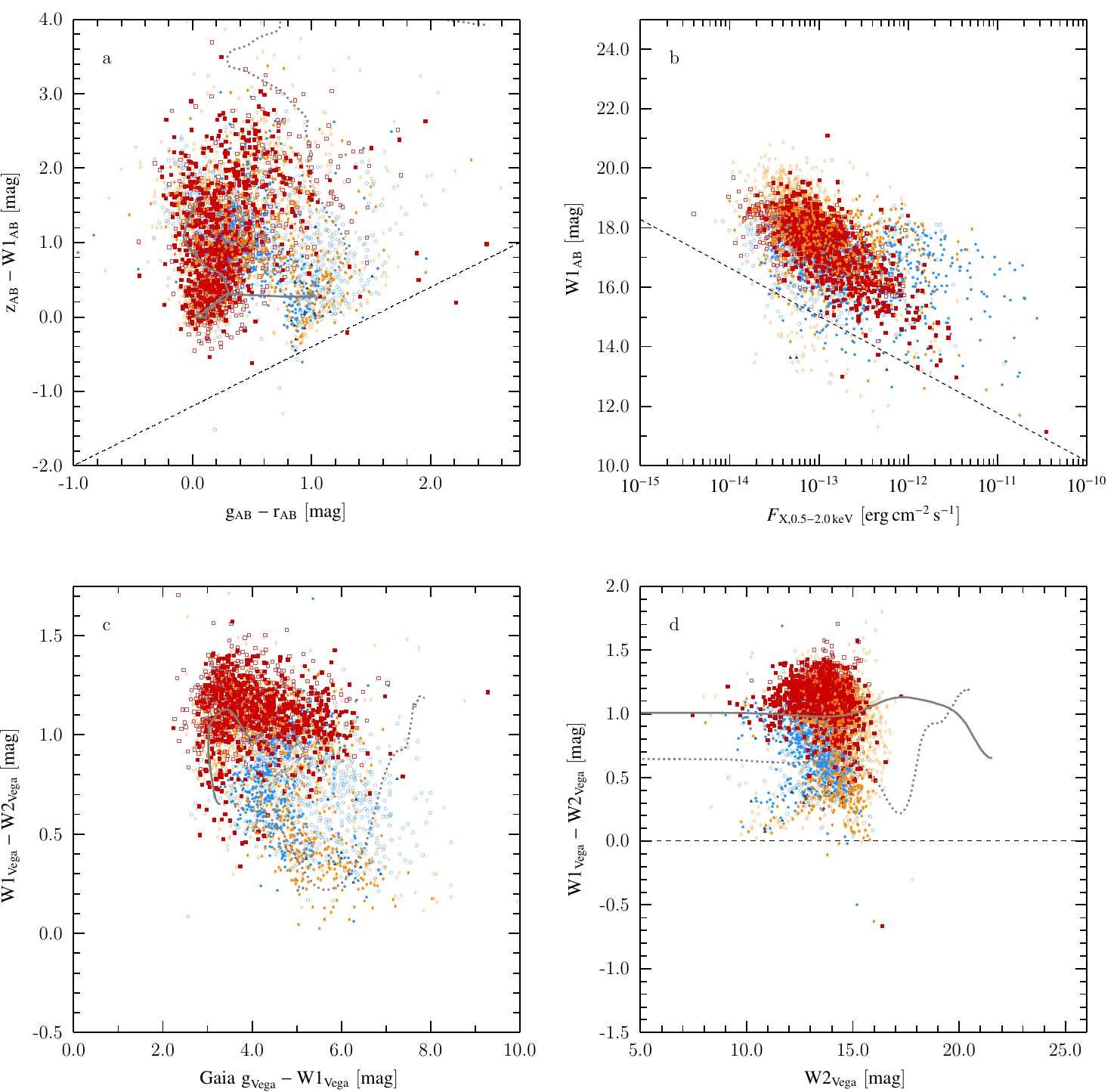}
\caption{Different photometric spaces combining optical, infrared and X-ray data, similar to
  \citet[][; Fig.~18]{salvato:2022} for point sources detected in eFEDS, using the
  color scheme introduced in
  Fig.~\ref{fig:blazar_erass1_map_paper}. Solid gray lines
  show positions usually occupied by quasars, whereas the gray dotted
  lines corresponded to Seyfert~2 galaxies. The dashed black lines in
  the top panels distinguish between Galactic and
  extragalactic sources.}
\label{fig:mara_combiplotfin}
\end{figure*}
We were able to combine the infrared data
with optical and X-ray information.
In Fig.~\ref{fig:mara_combiplotfin} we show four different photometric
spaces, which can be used to distinguish between Galactic and
extragalactic sources and see if objects exhibit similar properties as
other AGNs \citep{salvato:2022}. The (g$-$r)-(z$-$W1)
color-color diagram is shown in Fig.~\ref{fig:mara_combiplotfin}a.
Most BlazEr1 sources are above the dotted line and thus consistent with
being extragalactic. There are many candidates, as well as a
small number of confirmed sources below this line, hinting
at a low level of contamination (see Sect.~\ref{subsec:contamination}).

A sequence parallel to the separation line can be identified as well,
mainly consisting of candidates, \textit{BCU}s, and \textit{BZG}s,
which is typically the location of quiescent galaxies.
\citet{salvato:2022} show a similar trend with redshift, where sources
at low redshifts are more red. This trend is also observed in our
sample, since \textit{FSRQ}s are typically found at higher redshifts
and are also located at bluer colors. In Fig.~\ref{fig:mara_combiplotfin}b we
show the W1 magnitude as a function of X-ray flux. 
This information can be
used to check the extragalactic content, although the separation
line shown becomes a less robust indicator as the X-ray survey
increases in size \citep{salvato:2022}. While most sources are
consistent with an extragalactic origin, the objects below this line
are \textit{BCUC}s, again indicating that the catalog is subject to some degree of
contamination. The optical-infrared color-color diagram (Fig.~\ref{fig:mara_combiplotfin}c),
which can be used to distinguish stars from AGNs, does not exhibit a
sequence of most likely Galactic sources as seen in
\citet{salvato:2022}, maybe due to the incompleteness of
\textit{Gaia}. The tail at high values of g$-$W1
corresponding to inactive galaxies is less pronounced and mainly
occupied by candidates (\textit{BCUC}s and \textit{BLLC}s). The WISE color-magnitude diagram (Fig.~\ref{fig:mara_combiplotfin}d) can be used to identify stars (typically
W1$-$W2 $\sim 0.0$, marked with a dashed line) and
quasars. Most objects are quasar-like.

The BlazEr1 sources are located where we expect to observe AGNs and only very
few are located elsewhere and are inconsistent with an extragalactic
nature. Overall, sources are similar to QSOs (solid gray line) and extend toward
the loci of Seyferts (dotted gray line), where we also find a high number of
candidates.  The infrared and optical data show that the catalog might have some degree of contamination by other source types, although a detailed source-by-source analysis would be required to see if sources with data typical for non-blazars are in fact contaminants.

\subsubsection{Broadband spectral indices of the \textit{eROSITA} blazars}
\label{subsec:alphas}
\begin{figure}
\includegraphics[width=\linewidth]{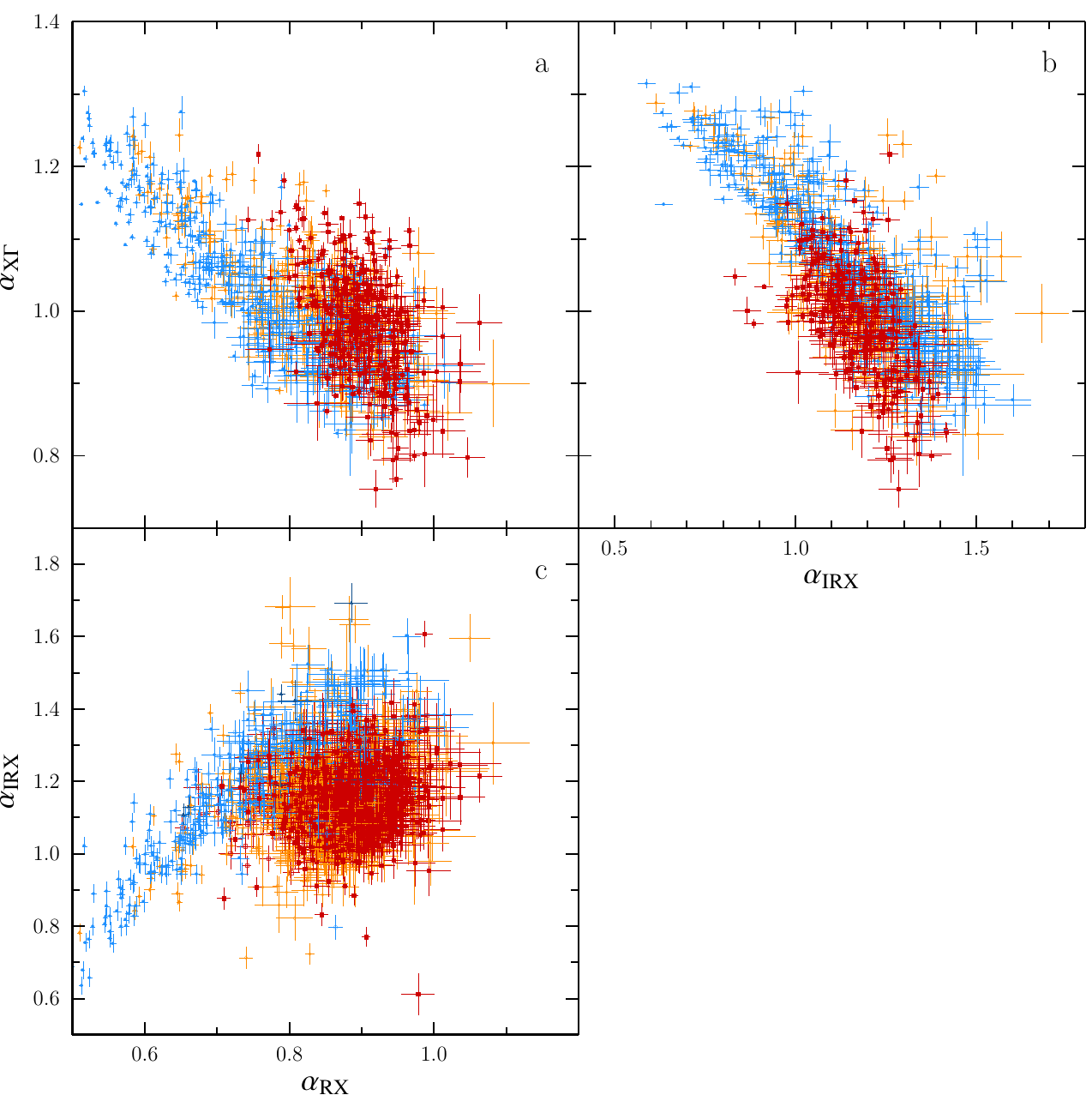}
\caption{Broadband spectral indices $\alpha_\mathrm{X\Gamma}$,
  $\alpha_\mathrm{IRX}$, and $\alpha_\mathrm{RX}$ against each other,
  following the color scheme introduced in
  Fig.~\ref{fig:blazar_erass1_map_paper}. Due to the similar values
  obtained for $\alpha_\mathrm{OX}$ and $\alpha_\mathrm{IRX}$, the
  former is not shown here.}
\label{fig:alpha_master_scatter_mod}
\end{figure}
As detailed in Sect.~\ref{sec:bb_spec_idx}, we computed broadband spectral indices,
$\alpha_{ij}$, for different bands from the radio to the $\gamma$ rays to relate the
X-rays to other bands. For
a given SED, the observed spectral index will depend on the peak
frequencies and the luminosities of the SED peaks. A value
$\alpha>1.0$ indicates a negative slope in ${\nu}F_{\nu}$ space, meaning
the SED decreases toward higher energies. 
We display the broadband spectral indices as calculated from various bands against each other in
Fig.~\ref{fig:alpha_master_scatter_mod}. For objects that are also in
the 4LAC, in Fig.~\ref{fig:4LAC_alpha}, we show the indices as a
function of Compton dominance, the ratio of the peak fluxes, and
low- and high-energy SED peak frequencies
\citep{ajello:2020,ajello:2022}. Additionally, we
calculated correlation coefficients for different parameter combinations
to assess the correlations between parameters for the entire sample and
the \textit{BLL} and \textit{FSRQ} sub samples. The results are listed
in Tables~\ref{tab:alpha_sed} and \ref{tab:alpha_cor} and mean values
obtained for the broadband spectral indices are shown in
Table~\ref{tab:alpha_means}.

\textit{FSRQ}s mostly show values of $\alpha_\mathrm{X\Gamma}\sim
1.0$, that is they show no significant increase or decrease from the
X-rays toward the $\gamma$ rays, which is expected, since X-rays are
more likely to cover the rise of the high-energy peak and the
$\gamma$ rays cover the fall of the same peak. The values of
$\alpha_\mathrm{X\Gamma}$ do not correlate significantly with the peak
positions but there is a significant weak anticorrelation with 
Compton dominance (see Fig.~\ref{fig:4LAC_alpha}i and
Table~\ref{tab:alpha_sed}). As indicated by the values of
$\alpha_\mathrm{OX}$, a decreasing slope from the optical toward the
X-rays is observed for a majority of \textit{FSRQ}s, and no
correlation with the SED parameters is found. Similar results are
found for $\alpha_\mathrm{IRX}$, however, the slopes are slightly
steeper and the dispersion is smaller (see
Table~\ref{tab:alpha_means}), perhaps due to \textit{FSRQ}s displaying
a big blue bump which adds flux in the infrared--UV range
\citep{krauss:2016}. A very weak correlation between the infrared X-ray
spectral index $\alpha_\mathrm{IRX}$ and the low-energy peak position is present.
Only extreme cases of \textit{FSRQ}s with low-energy peaks in the
radio range are found to decrease from the radio toward the X-rays,
however, the vast majority of sources show an increase. We find a very weak
anticorrelation between $\alpha_\mathrm{RX}$ and the low-energy peak
position.

For \textit{BLL}s both rising and falling indices from the X-rays to the
$\gamma$ rays are observed, but the mean value indicates a flat index.
Sources with decreasing slopes have their high-energy peak at energies below GeV $\gamma$ rays, where \textit{Fermi}-LAT is sensitive, and show low Compton
dominance. Larger values of $\alpha_\mathrm{X\Gamma}$ are linked to higher peak frequencies, as this parameter is moderately
and weakly correlated with the low- and high-energy peak
frequencies, respectively, and a lower Compton dominance, with a weak
anticorrelation being present. The values of $\alpha_\mathrm{OX}$
indicate that for \textit{BLL}s, on average a decreasing slope from
the optical toward the X-rays is observed. However, some sources
exhibit an increase that corresponds to extreme HSPs. In
general, a higher spread of values is seen than for \textit{FSRQ}s. In
most cases the low-energy peak occurs at energies lower than the
X-ray band and an increase in the value of $\alpha_\mathrm{OX}$
corresponds to a lower peak frequency for the low-energy peak, but this anticorrelation is very weak. An
anticorrelation with the peak frequency of the high-energy peak is observed, however, we find
no correlation with Compton dominance.

We observe a very similar picture for the infrared range. \textit{BLL}s are
found at more extreme decreasing and increasing slopes toward the X-ray band
than the \textit{FSRQ}s as indicated by the larger spread. We observe
only a weak anticorrelation with the position of the low-energy peak
and a moderate one with the high-energy peak. \textit{BLL}s show a
stronger increase from the radio band toward the X-rays than
\textit{FSRQ}s and a higher dispersion. We see a moderate
anticorrelation with the SED peak positions. Since in many
\textit{BLL}s the radio and X-ray emission are thought to originate
from the same process, a trend between peak positions and
$\alpha$-values is expected, as lower values in $\alpha$ correspond to
peak positions at higher energies. Candidates and confirmed blazars
show similar behavior without significant offsets. The \textit{BCU}s
are consistent with being blazars of either type.

For $\alpha_\mathrm{X\Gamma}$ we find a significantly strong or moderate
anticorrelation with all other indices, however, the correlation is
different for \textit{FSRQ}s and \textit{BLL}s, as for the
\textit{BLL}s the anticorrelations are stronger. There is a weaker
anticorrelation between $\alpha_\mathrm{X\Gamma}$ for
\textit{FSRQ}s and the other ranges, consistent with the X-rays
covering the dip of the low-energy peak or the rise of the high-energy
peak in the SED. For \textit{BLL}s the anticorrelation is stronger, and for lower $\alpha_\mathrm{X\Gamma}$ values, these
\textit{BLL}s have higher Compton dominance and lower peak frequencies
(LSP). The $\alpha_\mathrm{OX}$ and $\alpha_\mathrm{IRX}$ indices have
higher values compared to $\alpha_\mathrm{X\Gamma}$, which is the
opposite of $\alpha_\mathrm{RX}$. For the \textit{BLL}s a strong
correlation is found with the $\alpha_\mathrm{RX}$ and the infrared and
optical. For the \textit{FSRQ}s we find no or only very weak
correlations. 
This is expected, as
in many \textit{BLL}s these bands share the same emission process origin, showing strong
correlations, whereas
for \textit{FSRQ}s the X-rays originate from
the high-energy peak but the other bands are related to the low-energy
peak. The correlation between $\alpha_\mathrm{OX}$ and
$\alpha_\mathrm{IRX}$ is a strong or moderate one, with 
$\alpha_\mathrm{OX}$ having slightly higher values.
The weaker correlation for \textit{FSRQ}s is probably due to the big
blue bump \citep[][and references therein]{krauss:2016}.

The values derived for
$\alpha_\mathrm{OX}$ are similar to those found in the literature
\citep[see, e.g.,][]{giommi:1999,turriziani:2007}.
Broadband spectral
indices for the $\gamma$ rays and X-rays have also been provided for
example by \citet{comastri:1997} using \textit{EGRET}, however, their
values (\textit{FSRQ}s: $\alpha_\mathrm{X\Gamma}=0.58\pm0.12$,
\textit{BLL}s: $\alpha_\mathrm{X\Gamma}=0.83\pm0.18$) are lower than
ours, probably due to the lower energy band covered in the $\gamma$ rays.

No significant strong or moderate correlations can be found between the photon
indices and values of $\alpha$, however, the subclasses show different
trends. We found only weak
correlations for \textit{BLL}s with $\alpha_\mathrm{IRX}$ and
$\alpha_\mathrm{OX}$ and a weak anticorrelation with
$\alpha_\mathrm{X\Gamma}$.
Anticorrelations from some
samples with $\alpha_\mathrm{RX}$ are observed. 
Moderate correlations between the $\alpha$-value and redshift are observed.
The $\alpha_\mathrm{X\Gamma}$, decreases with redshift, while
$\alpha_\mathrm{RX}$ increases.

Since values of $\alpha$ correlate with the SED peaks and Compton
dominance,
selection cuts could be applied to
identify sources with different SED types or certain peak positions. 
In particular, $\alpha_\mathrm{IRX}$, $\alpha_\mathrm{OX}$, and
$\alpha_\mathrm{RX}$ could be useful to classify the sources into HSP
($\alpha_\mathrm{RX}<0.69$, $\alpha_\mathrm{IRX}<1.12$) and LSP
($\alpha_\mathrm{RX}>0.87$, $1.5>\alpha_\mathrm{IRX}>1.0$) categories.
For ISP it is not possible to find boundaries that separate them
from the other SED types well. 
For
the HSP constraints, 69\% of classifications agree with the SED
definition when compared to the 4LAC catalog; for LSPs
more than 96\% are
recovered. The BlazEr1 catalog contains 127 blazars within the HSP
constraints outlined, these  could potentially be TeV blazars \citep[see][]{metzger:2025}, and 812 sources within the proposed LSP criteria. Sources with values close to
$\alpha_\mathrm{IRX}\sim1.2$, $\alpha_\mathrm{OX}\sim1.2$, and
$\alpha_\mathrm{RX}\sim1.0$ could be MeV blazars, and are interesting targets
for future MeV missions \citep[e.g.,
  COSI;][]{tomsick:2019,tomsick:2024}.

\subsection{Why some sources go undetected}
\label{subsec:nondet}
\begin{table*}
\tiny
    \centering
\caption{Completeness of the BlazEr1 catalog.}
\label{tab:completness}
{\renewcommand{\arraystretch}{1.5}
    \begin{tabular}{l@{\hspace{0.2ex}}llllllllllllllll}
    \hline 
    \hline
            Catalog & \multicolumn{2}{l}{All} & \multicolumn{2}{l}{\textit{BLL}} & \multicolumn{2}{l}{\textit{BLLC}} & \multicolumn{2}{l}{\textit{BZG}} & \multicolumn{2}{l}{\textit{FSRQ}} & \multicolumn{2}{l}{\textit{FSRQC}} & \multicolumn{2}{l}{\textit{BCU}} & \multicolumn{2}{l}{\textit{BCUC}}\\
        &   & [\%] &  & [\%] & & [\%] & & [\%] & & [\%] & & [\%] & & [\%] & & [\%] \\
     \hline
        BLAZE  & $\left(\frac{5865}{41936}\right)$ & $14.0$  & $\left(\frac{597}{802}\right)$ & $74.4$  & $\left(\frac{954}{3188}\right)$ & $29.9$  & $\left(\frac{28}{59}\right) $ & $47.5$  & $\left(\frac{769}{929}\right) $ & $82.8$  & $\left(\frac{913}{1799}\right) $ & $51.3$  & $\left(\frac{712}{1241}\right) $ & $57.4$  & $\left(\frac{1892}{33918}\right) $ & $5.6$\\
        Gold sample    & $\left(\frac{2106}{3031}\right) $ & $69.5$  & $\left(\frac{597}{802}\right) $ & $74.4$  & -- & -- & $\left(\frac{28}{59}\right) $ & $47.5$  & $\left(\frac{769}{929}\right) $ & $82.8$  & -- & -- & $\left(\frac{712}{1241}\right) $ & $57.4$  & -- & -- \\
    4FGL      &  $\left(\frac{1293}{1873}\right) $ & $69.0$ & $\left(\frac{533}{672}\right) $ & $79.3$ & -- & -- & -- & -- & $\left(\frac{334}{376}\right) $ & $88.8$  & -- & -- & $\left(\frac{426}{825}\right) $ & $51.6$ & -- & -- \\
        BZCAT     & $\left(\frac{1209}{1551}\right) $ & $77.9$ & $\left(\frac{64}{130}\right) $ & $49.2$  & $\left(\frac{6}{12}\right) $ & $50.0$ & $\left(\frac{28}{59}\right) $ & $47.5$ & $\left(\frac{435}{553}\right) $ & $78.7$ & -- & --& $\left(\frac{28}{43}\right) $ & $65.1$ & -- & -- \\
        3HSP      & $\left(\frac{294}{400}\right) $ & $73.5$ & -- & --& -- & --& --& -- & -- & --& -- & --& $\left(\frac{258}{364}\right) $ & $70.9$ & -- & --\\
        HighZ     & $\left(\frac{6}{17}\right)$ & $ 35.3$ & -- & --& -- & --& -- & --& -- & --& -- & --& $\left(\frac{0}{9}\right)$ & $ 0.0$ & -- & -- \\
        Milliquas & $\left(\frac{958}{1304}\right) $ & $73.5$ & -- & --& $\left(\frac{72}{135}\right) $ & $53.3$ & -- & --& -- & --& -- & --& -- & --& -- & --\\
        KDEBLLACS & $\left(\frac{495}{2091}\right)$ & $ 23.7$ & -- & --& $\left(\frac{325}{1879}\right) $ & $17.3$ & -- & --& -- & --& -- & --& -- & --& -- & -- \\
        WIBRaLS2  & $\left(\frac{2406}{4133}\right) $ & $58.2$ & -- & -- & $\left(\frac{549}{1160}\right) $ & $47.3$ & -- & --& -- & --& $\left(\frac{913}{1799}\right) $ & $51.3$ & -- & --& $\left(\frac{138}{171}\right) $ & $80.7$ \\
        ABC       & $\left(\frac{313}{704}\right) $ & $44.5$ & -- & --& $\left(\frac{2}{2}\right) $ & $100.0$ & -- & --& -- & --& -- & --& -- & --& $\left(\frac{283}{651}\right) $ & $43.5$ \\
        BROS      & $\left(\frac{2659}{35016}\right) $ & $7.6$ & -- & --&
        -- & --& -- & --& -- & --& -- & --& -- & --& $\left(\frac{1471}{33096}\right) $ & $
        4.4$  \\
        \hline
    \end{tabular}}
    \tablefoot{The detection fractions of
        different catalogs are listed according to the class. The
        column All lists all sources regardless of class and each class has its own column.
        Note that the column All lists the fraction with respect to all sources
        within the respective input catalog used to compile the BLAZE catalog regardless if they are listed in the BLAZE catalog, whereas the class columns only consider sources
        which were taken from the respective catalogs and added to the BLAZE catalog. Only objects on the western Galactic hemisphere are considered.}
\end{table*} 
We went on to consider objects
that have not been detected by \textit{eROSITA}, or only at low significance.
In Table~\ref{tab:completness}, we list
the number of detected sources from each input catalog, 
also distinguishing among the different
classes, with respect to the
western Galactic hemisphere, and considering the quality cuts to the BLAZE catalog.  Overall, only ${\sim}14\%$ of all BLAZE catalog sources are
part of the BlazEr1 catalog, however, ${\sim}69\%$ of the gold sample (confirmed blazars in the BLAZE catalog) is detected.
Due to \textit{eROSITA}'s all-sky
survey strategy and higher sensitivity we miss less than one third of the confirm blazar population, a
significant improvement to \textit{Swift}-XRT \citep[43\%, ][]{giommi:2019}.

The BZCAT, 4FGL, and 3HSP catalogs exhibit the highest detection fraction
among all input catalogs,
as they contain the most extensive list of
confirmed blazars.
The fact that the 4FGL has such a high
detection rate implies that $\gamma$ rays are a good indicator for a subsequent X-ray detection, which is connected to the fact that the $\gamma$-ray detected sources have brighter X-ray fluxes.  The HighZ objects are less
likely to be detected, probably due to their faintness given their high redshifts.
The candidates
also have a lower detection rate, which is due to
contamination and these sources generally being fainter. 
The Milliquas catalog, which mainly adds candidates to the BLAZE catalog, has a high-detection rate as well, due to also containing many confirmed blazars.  Other candidate catalogs exhibit
detection rates between 8\% and 58\%.  The BROS catalog has the lowest
detection rate, despite having the highest number of input sources. 
A large contributing factor is the increased level of contamination compared to other catalogs.

Among the confirmed blazars, \textit{BLL}s and \textit{FSRQ}s have slightly
higher detection rates than \textit{BCU}s, of which only more than half are detected.  This could mean that \textit{BCU}s also
contain more contaminating objects, and that their X-ray fluxes are too low to be detected.  The different candidate classes have lower detection
rates.  \textit{eROSITA} is more likely to detect \textit{FSRQC}s than other candidate classes.

As a next step we used the \textit{eROSITA} sensitivity (see Sect.~\ref{subsec:sens}) to obtain the flux and luminosity upper limits for the undetected blazars and blazar candidates. 
The majority of nondetected sources have upper limits below the flux at which the all-sky survey is complete (Fig.~\ref{fig:erosita_sensitivity_minflux_lin}).
The upper limits agree with those of \citet{tubin-arenas:2024}.
Since \citet{tubin-arenas:2024} also considered the counts actually observed by \textit{eROSITA}, their upper
limits are sometimes higher than ours.
\begin{figure}
\includegraphics[width=\linewidth]{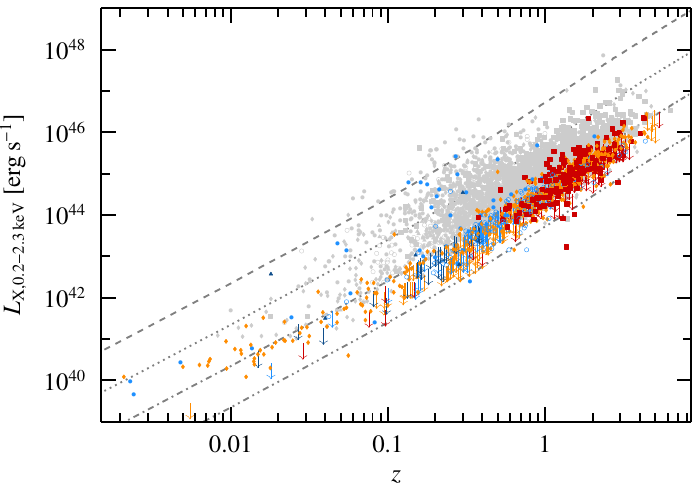}
\caption{X-ray luminosity as a function of redshift, following the color scheme introduced in Fig.~\ref{fig:blazar_erass1_map_paper}. BlazEr1 sources are
        shown in gray. The colored symbols are sources with low DET\_LIKE, quality issues or a high angular separation. The added lines are the same as in Fig.~\ref{fig:lum_z_log}. The upper limits have been taken from the BLAZE catalog.}
\label{fig:ul_lum_z_log}
\end{figure}
As a next step, we calculate luminosity upper limits for sources with
known redshifts (Fig.~\ref{fig:ul_lum_z_log}).
The exposure times, fluxes and luminosity limits are all listed in the BLAZE catalog.  At low redshifts,
\textit{BCU}s in particular are abundant.
We also observe \textit{FSRQ}s at higher redshifts than \textit{BLL}s as previously
seen in Sect.~\ref{subsec:lums}.
The dashed-dotted line in Fig.~\ref{fig:ul_lum_z_log}, which
corresponds to a flux of $F_{\mathrm{X},\,0.2-2.3\,\mathrm{keV}} =
1\times10^{-14}\,\mathrm{erg}\,\mathrm{cm}^{-2}\,\mathrm{s}^{-1}$, indicates that the
lower luminosity population is missed due to the
sensitivity limit of \textit{eROSITA}.

\begin{figure}
\includegraphics[width=\linewidth]{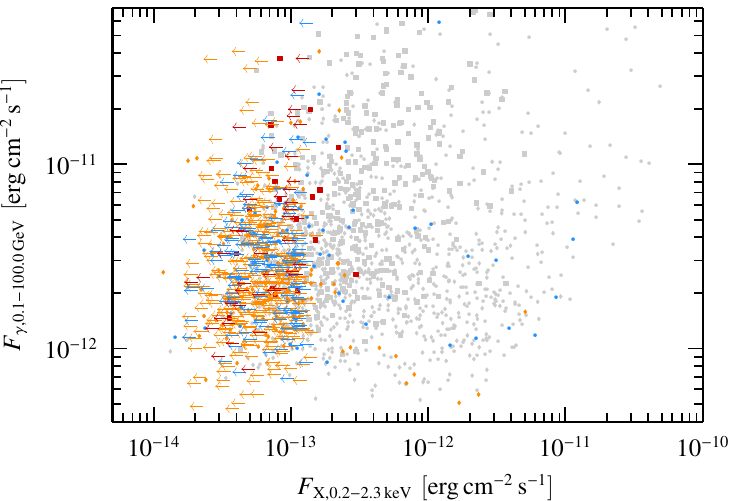}
\caption{\textit{Fermi}-LAT flux as a function of \textit{eROSITA} flux, following the color scheme introduced in Fig.~\ref{fig:blazar_erass1_map_paper}. The gray points are BlazEr1 sources, whereas sources without a significant detection are shown in color.
        Arrows indicate the upper limits.}
\label{fig:gamma_xray_ul}
\end{figure}
Similarly to Fig.~\ref{fig:eRO_LAT}, we compare the X-ray upper limits to $\gamma$-ray fluxes in
Fig.~\ref{fig:gamma_xray_ul}.  The majority of nondetections
correspond to \textit{BCU}s, found at lower $\gamma$-ray fluxes than detected sources.
Of the undetected blazars with
a counterpart in the 4LAC catalog,
the majority are LSPs (265\,objects), ISPs and HSPs are less
prominent (76 and 29, respectively).
LSPs are expected to
exhibit harder spectra, which hints at a potential bias toward
softer and more high peaked objects, which are easier to detect.
It is likely that all these sources were missed in eRASS1 due to being faint. These results indicate that subsequent all-sky survey releases by \textit{eROSITA}
will have higher detection rates than the BlazEr1 catalog, as the faintness of the
sources represents one of the main limitations for our catalog. 

\section{Summary and future work}
\label{sec:outlook}
We present the Blazars in
eRASS1 (BlazEr1) catalog of blazars
and blazar candidates detected by \textit{eROSITA} on the western Galactic hemisphere. This catalog
is available online through Vizier and at the \textit{eROSITA} DR1
webpage\footnote{\url{https://erosita.mpe.mpg.de/dr1/AllSkySurveyData_dr1/Catalogues_dr1/}}.
The catalog was derived by matching a master list of blazars and
blazar candidate sources from the literature (the BLAZE catalog,
released together with this paper) to the \textit{eROSITA} eRASS1 catalog
\citep{merloni:2024}. 
We used the classes given in the input catalogs the BLAZE
catalog has been constructed from, since the fraction of misclassifications, including the
changing-look objects, is expected to be only roughly 5\%. Considering the size of the BLAZE and BlazEr1 catalogs, this has a negligible impact
on our conclusions.
After applying quality cuts, a set of 5865
\textit{eROSITA} observed blazars and blazar candidates were detected. The \textit{eROSITA} sources removed during
processing steps have been released in a separate catalog. The
BlazEr1 catalog has been augmented with X-ray spectral and multiwavelength
information (see Appendix~\ref{sec:cataloginfo}). To date, this catalog is the most complete
compilation of confirmed X-ray detected blazars on the western
Galactic hemisphere and, to our knowledge, the largest catalog of X-ray
detected blazars and blazar candidates.
Overall, the
BlazEr1 catalog is expected to contain at most 633 contaminating
sources (Sect.~\ref{subsec:contamination}).

The main scientific results obtained from an analysis of the catalog
include the distribution of photon
indices for the spectroscopic sample (Sect.~\ref{subsec:photonindices}), which clearly show that the different blazar subtypes exhibit
distinct spectral properties, enabling tentative determinations of the
subtype based on the photon index. The relations between the spectral
properties and fluxes and luminosities are consistent with
expectations from the blazar sequence (Sect.~\ref{subsec:photonindices}). We find that \textit{eROSITA} is more
sensitive toward low-flux \textit{FSRQ}s than low-flux
\textit{BLL}s, probably due to the very soft spectra displayed by the
latter. Most of the brightest blazars detected are \textit{BLL}s, while the
fluxes of $\gamma$-ray detected sources are higher than those of the
non $\gamma$-ray detected ones.

The slopes of the
$\log N$-$\log S$ distributions constructed from the catalog imply a
negative cosmological evolution, both for the entire sample and for
different subsamples, meaning the objects are more common in the Local Universe; however, for the blazar candidates, the value could indicate a slight positive evolution (Sect.~\ref{subsec:lognlogs}). The observed distributions agree well with 
theoretical predictions \citep{giommi:2015}. We find evidence for a
population inversion between \textit{BLL}s and \textit{FSRQ}s at lower
fluxes.

The spectral properties, fluxes, and luminosities of the
\textit{BCU}s are similar to low flux \textit{BLL}s, consistent with
other publications \citep[Sect.~\ref{subsec:lums};][]{kang:2019,pena-herazo:2020,chiaro:2021}.
The multiwavelength data help to separate blazars into their
respective classes using
color-color diagrams 
and highlights the importance of the \textit{eROSITA}
data for building a comprehensive blazar catalog and studying their
properties, as well as for future SED modeling (Sect.~\ref{subsec:mwl}).

The BlazEr1 catalog can serve as a
benchmark for overall
blazar properties and single-object properties since it provides the
first X-ray observation for many blazars. For instance, the \textit{eROSITA} data will
be crucial to constrain the SEDs of a large number of objects, including
samples such as those from \texttt{TANAMI}, \texttt{MOJAVE}, or 4FGL. 
Earlier applications of the eRASS1 catalog included the identification of X-ray
counterparts for an ultra-high-energy neutrino event
\citep{km3netcollaboration:2025} and the identification of 
TeV blazar candidates for follow-up observations
\citep[][see also \citealt{marchesi:2025}]{metzger:2025}.

Additionally, the \textit{eROSITA} surveys allow for a
detailed variability study on timescales of hours-days for the brightest and on a half-year time line for all other sources. 
The BlazEr1 catalog
provides the framework for these studies.
Examples include work on X-ray blazar flares, as
demonstrated, for instance, for PKS\,0735$+$178
\citep{hammerich:2021,sahakyan:2022} or a $\gamma$-ray flare of
TXS\,0646$-$176 \citep{hammerich:2021a}.

A subsequent data release of 2.2 years of data (eRASS:5), will allows us to provide an updated, deeper version of the catalog.
The cumulative \textit{eROSITA} data will also deepen
the catalog, making the eRASS:5 version, which will
contain 2.2\,years of data, the deepest X-ray blazar catalog ever
constructed.

\section*{Data availability}
The BLAZE and BlazEr1 catalogs are available in electronic form at the CDS via anonymous ftp to \url{cdsarc.u-strasbg.fr} (\url{130.79.128.5}) or via \url{http://cdsweb.u-strasbg.fr/cgi-bin/qcat?J/A+A/} and the \textit{eROSITA} DR1 webpage \url{https://erosita.mpe.mpg.de/dr1/AllSkySurveyData_dr1/Catalogues_dr1/}.

\begin{acknowledgements}\label{thanx}
  S.H.\ is partly supported by the German Science Foundation (DFG
  grant numbers WI 1860/14-1 and 434448349). Mi.K. acknowledges
  support from DLR grant FKZ 50 OR 2307. Ma.K.\ and
  F.R.\ acknowledge funding by the Deutsche Forschungsgemeinschaft
  (DFG, German Research Foundation, grant 434448349).
  A.G.M.\ acknowledges support from Narodowe Centrum Nauki (NCN) grant
  2018/31/G/ST9/03224. 
  Ma.K.\, E.R.\, and J.W.\ gratefully acknowledge funding by the Deutsche Forschungsgemeinschaft (DFG, German Research Foundation), within the research unit 5195 “Relativistic Jets in Active Galaxies” under project number 443220636.
  This work is part of the M2FINDERS (ERC grant No. 101018682) and ASTROGEODESY (ERC grant No. 101076060) projects, funded by the European Research Council under the EU Horizon 2020 Research and Innovation Programme. 
  This work is based on data from eROSITA, the
  soft X-ray instrument aboard SRG, a joint Russian-German science
  mission supported by the Russian Space Agency (Roskosmos), in the
  interests of the Russian Academy of Sciences represented by its
  Space Research Institute (IKI), and the Deutsches Zentrum für Luft-
  und Raumfahrt (DLR). The SRG spacecraft was built by Lavochkin
  Association (NPOL) and its subcontractors, and was operated by NPOL
  with support from the Max Planck Institute for Extraterrestrial
  Physics (MPE). The development and construction of the \textit{eROSITA} X-ray
  instrument was led by MPE, with contributions from the Dr.\ Karl
  Remeis Observatory Bamberg and ECAP (FAU Erlangen-N\"urnberg), the
  University of Hamburg Observatory, the Leibniz Institute for
  Astrophysics Potsdam (AIP), and the Institute for Astronomy and
  Astrophysics of the University of Tübingen, with the support of DLR
  and the Max Planck Society. The Argelander Institute for Astronomy
  of the University of Bonn and the Ludwig Maximilians Universität
  Munich also participated in the science preparation for eROSITA. The
  \textit{eROSITA} data shown here were processed using the eSASS/NRTA software
  system developed by the German \textit{eROSITA} consortium. This research has
  made use of ISIS functions (ISISscripts) provided by ECAP/Remeis
  observatory and MIT
  (\url{https://www.sternwarte.uni-erlangen.de/isis/}). This research
  has made use of data, software and/or web tools obtained from the
  High Energy Astrophysics Science Archive Research Center (HEASARC),
  a service of the Astrophysics Science Division at NASA/GSFC and of
  the Smithsonian Astrophysical Observatory's High Energy Astrophysics
  Division, and of the Vizier and HEASARC database systems for
  querying objects and getting information from different catalogs. We
  thank Tess Jaffe (GSFC) for valuable input concerning our HEASARC
  queries. This work has also made use of data from the European Space
  Agency (ESA) mission \textit{Gaia}
  (\url{https://www.cosmos.esa.int/gaia}), processed by the
  \textit{Gaia} Data Processing and Analysis Consortium (DPAC,
  \url{https://www.cosmos.esa.int/web/gaia/dpac/consortium}). Funding
  for the DPAC has been provided by national institutions, in
  particular the institutions participating in the \textit{Gaia}
  Multilateral Agreement. The SIMBAD database, operated at CDS,
  Strasbourg, France, was also used to get additional information.
\end{acknowledgements}

\bibliographystyle{aa} \bibliography{BlazEr}
\begin{appendix}
\onecolumn
\section{Supplemental figures}
\begin{figure*}[h!]
\includegraphics[width=\linewidth]{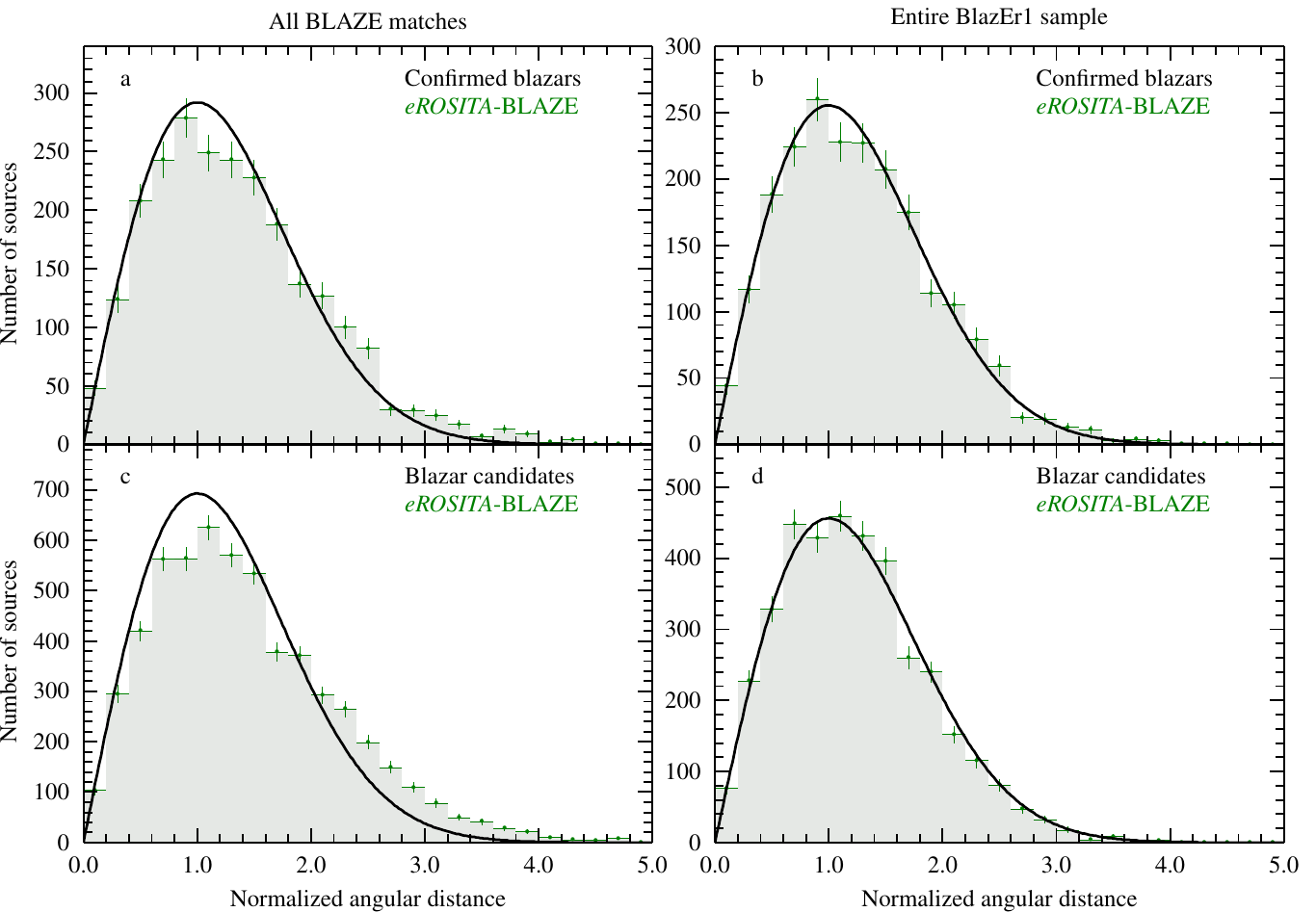}
\caption{Histograms of the angular separation between the positions provided in the BLAZE catalog normalized by the respective \textit{eROSITA} positional error. Top: \textit{eROSITA} observed blazars. Bottom: \textit{eROSITA} observed blazar candidates. Panels a and c:  All the initial matches. Panels b and d: BlazEr1 sample with applied quality and separation cuts. The theoretical Rayleigh distributions are shown as a black solid
line.}
\label{fig:mara_sep_combi_blaze}
\end{figure*}

\begin{figure*}[h!]
\includegraphics[width=\linewidth]{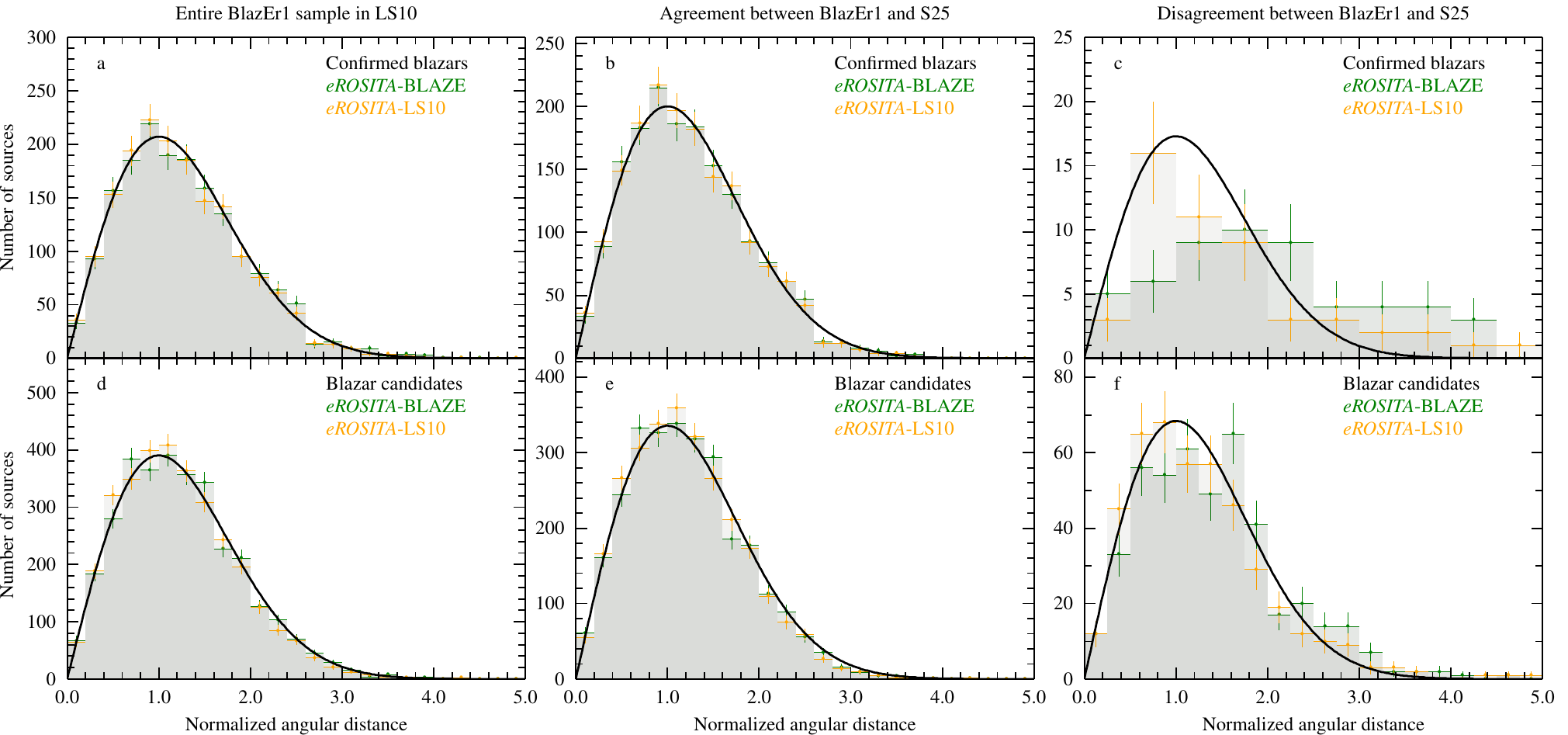}
\caption{Same as Fig.~\ref{fig:mara_sep_combi_blaze}, but with the
separation between \textit{eROSITA} and LS10 counterparts by S25 in yellow. All matches with the LS10 counterpart catalog are shown in panels a and d. Panels b--f only show sources where S25 provides the same or a different counterpart, respectively.}
\label{fig:mara_sep_combi_LS10}
\end{figure*}

\begin{figure*}[h!]
\includegraphics[width=\linewidth]{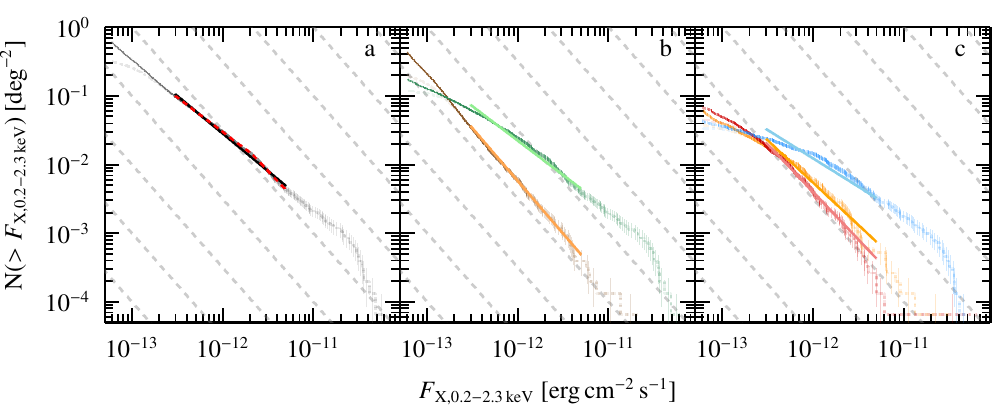}
\caption{$\log N$-$\log S$ distributions of different BlazEr1 subsamples, same as Fig.~\ref{fig:logn_logs_fin}, but for the \textit{eROSITA} main band and including the fit results listed in Table~\ref{tab:lognlogs_results}. The color scheme is the same as in Fig.~\ref{fig:logn_logs_fin}. The red line in the left panel shows the best fit broken power law for the entire sample ($\alpha_1=1.003\pm0.013$, $\alpha_2=1.39\pm0.07$, $F_{\mathrm{X, break},\,0.2-2.3\,\mathrm{keV}} =
(2.11\pm0.26)\times10^{-12}\,\mathrm{erg}\,\mathrm{cm}^{-2}\,\mathrm{s}^{-1}$, $\mathrm{C}=30.6\pm2.5\,\mathrm{deg}^{-2}$).}
\label{fig:logn_logs_fin_fit}
\end{figure*}

\begin{figure*}[h!]
\includegraphics[width=\linewidth]{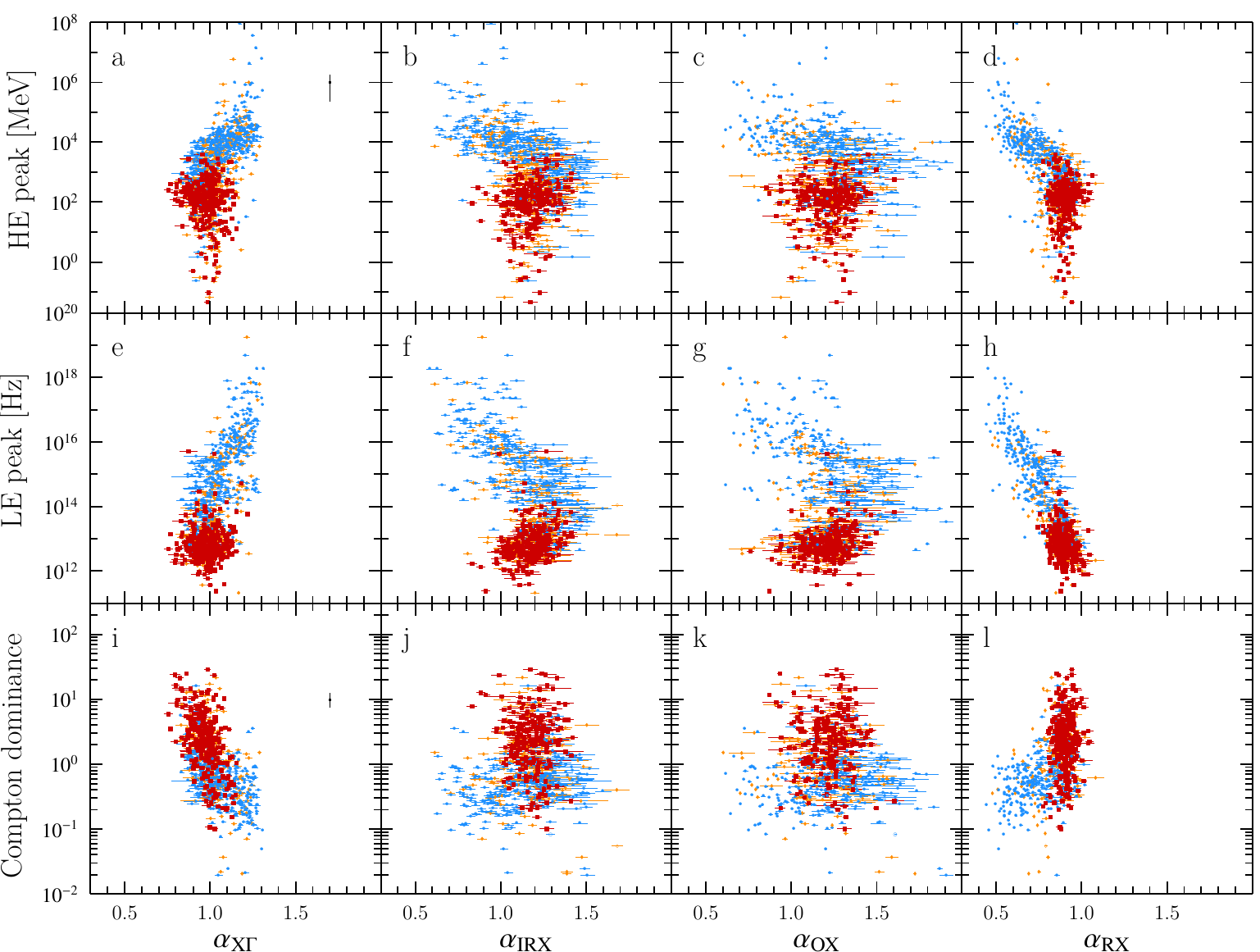}
\caption{SED parameters from the 4LAC catalog as a function of various broadband spectral indices, following the color scheme introduced in Fig.~\ref{fig:blazar_erass1_map_paper}. For the uncertainties of the parameters from the 4LAC we give the median uncertainty, when available, and in case of the Compton dominance by means of error propagation, as black data points in panels \textbf{a} and \textbf{i}.}
\label{fig:4LAC_alpha}
\end{figure*}

In this appendix, we extend the discussion of the main part of this paper with further diagnostic information. Figures~\ref{fig:mara_sep_combi_blaze} and~\ref{fig:mara_sep_combi_LS10} display the distributions of the normalized angular separation with respect to S25.
In Fig.~\ref{fig:logn_logs_fin_fit} the fits to the $\log N$-$\log S$ distributions are shown and Fig.~\ref{fig:4LAC_alpha} summarizes the relation between SED parameters and broadband spectral indices.
\FloatBarrier 

\section{Supplemental tables}
In this appendix, we present the tables summarizing the different correlations and the mean values of the broadband spectral indices.
Table~\ref{tab:correlations} lists all kinds of different parameter combinations and the respective correlation values, while Table~\ref{tab:alpha_sed} compares the broadband spectral indices with the SED parameters and Table~\ref{tab:alpha_cor} lists correlation coefficients between the different broadband spectral indices.
Mean values for the broadband spectral indices are given in Table~\ref{tab:alpha_means}.
\begin{table}[h!]
\tiny
    \centering
   \caption{Correlation coefficients for various parameter combinations discussed throughout the paper.}
    \label{tab:correlations}
    {\renewcommand{\arraystretch}{1.2}
    \begin{tabular}{lllllll}
    \hline
    \hline
     Parameter 1 & Parameter 2 & Sample & $\rho_\mathrm{Pearson}$ & $p$ & $\tau_\mathrm{Kendall}$ & $p$ \\
     \hline
     \textit{eROSITA} flux & \textit{eROSITA} $\Gamma_\mathrm{X}$ & \textit{BLL} & $-0.1217$ & 0.022 & $-0.1834$ & $<0.003$ \\
     \textit{eROSITA} $\Gamma_\mathrm{X}$ & $z$ & \textit{FSRQ} & $-0.4438$ & $<0.003$ & $-0.2789$ & $<0.003$ \\
     \textit{eROSITA} $\Gamma_\mathrm{X}$ & $z$ & \textit{FSRQC} & $-0.3248$ & $0.009$ & $-0.2360$ & $0.006$ \\
     \textit{Swift}-BAT flux & \textit{eROSITA} flux & \textit{FSRQ}& $0.9695$ & $0.0$ & $0.0719$ & $0.499$ \\
      \textit{Swift}-BAT flux & \textit{eROSITA} flux & \textit{BCU}& $0.9418$ & $<0.003$ & $0.3590$ & $0.100$ \\
      \textit{NuSTAR} soft flux & \textit{eROSITA} flux & all & $0.6950$ & $<0.003$ & $0.5382$ & $<0.003$ \\
      \textit{NuSTAR} hard flux & \textit{eROSITA} flux & all & $0.6060$ & $<0.003$ & $0.2707$ & $0.008$ \\
      \textit{NuSTAR} $\Gamma_\mathrm{X}$ & \textit{eROSITA} $\Gamma_\mathrm{X}$ & all & $0.6118$ & $<0.003$ & -- & -- \\
      \textit{Swift}-BAT $\Gamma_\mathrm{X}$ & \textit{eROSITA} $\Gamma_\mathrm{X}$ & all & $0.4794$ & $<0.003$ & -- & -- \\
      \textit{eROSITA} $\Gamma_\mathrm{X}$ & \textit{Fermi}-LAT $\Gamma_\gamma$ & \textit{BCU} & $-0.1288$ & $0.141$ & $-0.1999$ & $0.001$\\
      \texttt{TANAMI} flux density & \textit{eROSITA} flux & \textit{FSRQ} & $0.5503$ & $0.001$ & $0.5907$ & $<0.003$ \\
      \texttt{MOJAVE} flux density & \textit{eROSITA} & all & $0.5055$ & $<0.003$ & $0.2941$ & $<0.003$ \\
       \texttt{RFC} X-band flux density & \textit{eROSITA} & \textit{FSRQ} & $0.2513$ & $<0.003$ & $0.2958$ & $0.0$ \\
        \texttt{RFC} X-band flux density & \textit{eROSITA} & \textit{BLL} & $0.0455$ & $0.349$ & $-0.1395$ & $<0.003$ \\
      \textit{eROSITA} $\Gamma_\mathrm{X}$ & $\alpha_\mathrm{IRX}$ & \textit{BLL}& $0.2148$ & $<0.003$ & $0.2098$ & $<0.003$ \\
      \textit{eROSITA} $\Gamma_\mathrm{X}$ & $\alpha_\mathrm{OX}$ & \textit{BLL}& $0.2963$ & $<0.003$ & $0.2422$ & $<0.003$ \\
      \textit{eROSITA} $\Gamma_\mathrm{X}$ & $\alpha_\mathrm{X\Gamma}$ & \textit{BLL}& $-0.2305$ & $<0.003$ & $-0.2669$ & $<0.003$ \\
      \textit{eROSITA} $\Gamma_\mathrm{X}$ & $\alpha_\mathrm{RX}$ & all & $-0.4171$ & $<0.003$ & $-0.3437$ & $0.0$ \\
      $\alpha_\mathrm{X\Gamma}$ & $z$ & all & $-0.3973$ & $<0.003$ & $-0.3002$ & $0.0$ \\
      $\alpha_\mathrm{RX}$ & $z$ & all & $0.4405$ & $0.0$ & $0.3362$ & $0.0$ \\
     \hline
    \end{tabular}}
\end{table}
\begin{table}[h!]
\tiny
    \centering
   \caption{Correlation coefficients for parameters $\alpha$ and SED parameters from the 4LAC catalog.}
    \label{tab:alpha_sed}
    \begin{tabular}{llllllllllllll}
    \hline
    \hline
     & & HE peak & & &  & LE peak &  &  & & Compton dominance & & & \\
     \hline
    $\alpha$ & Class & $\rho_\mathrm{Pearson}$ & $p$ & $\tau_\mathrm{Kendall}$ & $p$ & $\rho_\mathrm{Pearson}$ & $p$ & $\tau_\mathrm{Kendall}$ & $p$ & $\rho_\mathrm{Pearson}$ & $p$ & $\tau_\mathrm{Kendall}$ & $p$ \\     
    \hline
    $\alpha_\mathrm{X\Gamma}$ & all & 0.0876 & 0.004 & 0.3860 & 0.0 & 0.0759 & 0.008 & 0.1893 & 0.0 & -- & -- & -0.2117 & 0.0 \\
     & \textit{BLL} & 0.0953 & 0.041 & 0.4768 & 0.0 & 0.0872 & 0.045 & 0.2957 & 0.0 & -- & -- & $-$0.1937 & <0.003 \\
     & \textit{FSRQ} & 0.0345 & 0.557 & $-$0.0770 & 0.050 & $-$0.0298 & 0.592 & $-$0.0128 & 0.731 & -- & -- & $-$0.2930 & <0.003 \\
    \hline
    $\alpha_\mathrm{IRX}$ & all & $-$0.0827 & 0.019 & $-$0.2409 & 0.0 & $-$0.1029 & 0.002 & $-$0.0829 & <0.003 & -- & -- & 0.0352 & 0.115 \\
     & \textit{BLL} & $-$0.0939 & 0.077 & $-$0.4666 & 0.0 & $-$0.1845 & <0.003 & $-$0.2943 & 0.0 & -- & -- & 0.0890 & 0.008 \\
     & \textit{FSRQ} & 0.1445 & 0.030 & 0.1366 & 0.002 & $-$0.0175 & 0.780 & 0.2180 & <0.003 & -- & -- & 0.0263 & 0.534\\
    \hline
    $\alpha_\mathrm{OX}$ & all & $-$0.0907 & 0.012 & $-$0.1285 & <0.003  & $-$0.0802 & 0.017 &  $-$0.0218 & 0.336 & -- & -- &  $-$0.0085 & 0.707 \\
     & \textit{BLL} & $-$0.1170 & 0.030 & $-$0.3393 & 0.0 & $-$0.1531 & 0.002 & $-$0.1924 & <0.003 & -- & -- & 0.0386 & 0.258\\
     & \textit{FSRQ} & 0.1033 & 0.128 & 0.0193 & 0.672 & 0.0170 & 0.789 & 0.1395 & 0.001 & -- & -- & 0.0185 & 0.667 \\
    \hline
    $\alpha_\mathrm{RX}$ & all & $-$0.0755 & 0.027 & $-$0.4319 & 0.0 & $-$0.2519 & <0.003 & $-$0.3725 & 0.0 & -- & -- & 0.1651 & <0.003 \\
     & \textit{BLL} & $-$0.0701 & 0.201 & $-$0.5639 & 0.0 & $-$0.2824 & <0.003 & $-$0.5004 & 0.0 & -- & -- & 0.1582 & <0.003\\
     & \textit{FSRQ} & $-$0.1242 & 0.036 & 0.0474 & 0.233 & $-$0.0930 & 0.097 & $-$0.2241 & <0.003 & -- & -- & 0.0925 & 0.014 \\
     \hline
    \end{tabular}
\end{table}
\begin{table}[h!]
\tiny
    \centering
   \caption{Correlation coefficients for parameters $\alpha$.}
    \label{tab:alpha_cor}
    \begin{tabular}{llllllllllllll}
    \hline
    \hline
     & & $\alpha_\mathrm{IRX}$  & & & & $\alpha_\mathrm{OX}$ & & & & $\alpha_\mathrm{RX}$ & & &  \\
     \hline
    $\alpha$ & Class & $\rho_\mathrm{Pearson}$ & $p$ & $\tau_\mathrm{Kendall}$ & $p$ & $\rho_\mathrm{Pearson}$ & $p$ & $\tau_\mathrm{Kendall}$ & $p$ & $\rho_\mathrm{Pearson}$ & $p$ & $\tau_\mathrm{Kendall}$ & $p$ \\     
    \hline
    $\alpha_\mathrm{X\Gamma}$ & all &  $-$0.7532 & <0.003 & $-$0.5500 & 0.0 &  $-$0.5296 & <0.003 & $-$0.3722 & 0.0 &  $-$0.7560 & <0.003 & $-$0.4981 & 0.0 \\
     & \textit{BLL} & $-$0.8633 & <0.003 & $-$0.6974 & 0.0 &  $-$0.7000 & <0.003 & $-$0.5526 & 0.0 &  $-$0.8582 & <0.003 & $-$0.6680 & 0.0 \\
     & \textit{FSRQ} & $-$0.5380 & <0.003 & $-$0.3879 & 0.0 & $-$0.2977 & <0.003 & $-$0.1975 & <0.003 & $-$0.4911 & <0.003 & $-$0.3080 & <0.003 \\
    \hline
    $\alpha_\mathrm{IRX}$ & all & -- & -- & -- & -- & 0.7786 & 0.0 & 0.5216 & 0.0 & 0.3500 & 0.0 & 0.1338 & 0.0\\
     & \textit{BLL} & -- & -- & -- & -- & 0.9292 & 0.0 & 0.7791 & 0.0 & 0.8374 & 0.0 & 0.6498 & 0.0 \\
     & \textit{FSRQ} & -- & -- & -- & -- & 0.5807 & 0.0 & 0.3663 & 0.0 & 0.1459 & 0.001 & 0.0943 & 0.001 \\
    \hline
    $\alpha_\mathrm{OX}$ & all & -- & -- & -- & -- & -- & -- & -- & --  &  0.1603 & <0.003 & 0.0465 & 0.003 \\
     & \textit{BLL} & -- & -- & -- & -- & -- & -- & -- & -- & 0.5997 & 0.0 & 0.4274 & 0.0 \\
     & \textit{FSRQ} & -- & -- & -- & -- & -- & -- & -- & -- & 0.0397 & 0.351 & 0.0518 & 0.068 \\
     \hline
    \end{tabular}
\end{table}
\begin{table}[h!]
    \centering
   \caption{Mean and standard deviation values for parameters of $\alpha$.}
    \label{tab:alpha_means}
    \begin{tabular}{lllllllll}
    \hline
    \hline
     Type & $\langle\alpha_\mathrm{X\Gamma}\rangle$ & $\sigma_{\alpha_\mathrm{X\Gamma}}$& $\langle\alpha_\mathrm{IRX}\rangle$ & $\sigma_{\alpha_\mathrm{IRX}}$ & $\langle\alpha_\mathrm{OX}\rangle$ & $\sigma_{\alpha_\mathrm{OX}}$ & $\langle\alpha_\mathrm{RX}\rangle$ & $\sigma_{\alpha_\mathrm{RX}}$ \\
    \hline
    All & 1.04 & 0.11 & 1.15 & 0.14  & 1.25 & 0.21  & 0.84 & 0.10  \\
    Confirmed & 1.04 & 0.11 & 1.13 & 0.16 & 1.22 & 0.21 & 0.84 & 0.12  \\
    Candidates & -- & -- & 1.17 & 0.13 & 1.27 & 0.21 & 0.85 & 0.07  \\
    \textit{BLL} & 1.07 & 0.12 & 1.12 & 0.21 & 1.21 & 0.24   & 0.75 & 0.13  \\
    \textit{BLLC} & -- & -- & 1.20 & 0.13 & 1.29 & 0.19 & 0.80 & 0.07  \\
    \textit{BZG} & -- & -- & 1.27 & 0.19  & 1.48 & 0.27  & 0.78 & 0.09  \\
    \textit{FSRQ} & 0.98 & 0.08 & 1.15 & 0.10  & 1.25 & 0.16  & 0.89 & 0.06  \\
    \textit{FSRQC} & -- & -- & 1.16 & 0.09  & 1.31 & 0.18  & 0.84 & 0.06  \\
    \textit{BCU} & 1.04 & 0.10 & 1.12 & 0.16  & 1.18 & 0.21  & 0.82 & 0.11  \\
    \textit{BCUC} & -- & -- & 1.15 & 0.15  & 1.25 & 0.24  & 0.86 & 0.06  \\
    \hline
    \end{tabular}
\end{table}
\FloatBarrier 
\twocolumn

\section{The BLAZE catalogs}
\label{sec:blazeinfo}
The BLAZE catalog is obtained by merging the catalogs listed in
Table~\ref{tab:blazar_catalogs} with the respective merging radii
(Sect.~\ref{sec:sample}). 
The BLAZE catalog contains redshifts from these input
catalogs. Based on \textit{eROSITA}'s sensitivity and exposure, we are able to
calculate flux and luminosity upper limits for all sources
in the western Galactic hemisphere.
The sources removed due to quality issues (see Sect.~\ref{subsec:quality}) are released as a separate file as well.
The catalogs
are released as FITS files with the following columns:
\begin{description}
\item[SRC\_NAME:] Common name of the source. 
\item[RAJ2000, DEJ2000:] Equatorial coordinates from input
  catalog [$\degr$, J2000.0].
\item[LII, BII:] Galactic coordinates from input catalog [$\degr$].
\item[BLAZAR\_CLASS:] Classification given by the input catalog.
\item[Z:] Redshift given in the input catalog.
\item[Z\_REF:] Reference for the redshift.
\item[ERO\_EXPO:] \textit{eROSITA} exposure time at source position, if available  [s].
\item[ERO\_UL\_FLUX:]  Upper limit 0.2--2.3\,keV flux, based on exposure time and sensitivity [$\mathrm{erg}\,\mathrm{cm}^{-2}\,\mathrm{s}^{-1}$].
\item[ERO\_UL\_LUM:] Upper limit for the 0.2--2.3\,keV luminosity band
  based on ERO\_UL\_FLUX
  [$\mathrm{erg}\,\mathrm{cm}^{-2}\,\mathrm{s}^{-1}$].
\item[REMOVE\_FLAG:] Only for the \emph{unverified} BLAZE catalog, reason for removal: 1: HECATE, 2: \citet{xie:2024}, 3: \citet{rakshit:2017}, 4: \citet{gordon:2023}, 5: known non-blazar.
\end{description}
\FloatBarrier

\section{The BlazEr1 catalogs}
\label{sec:cataloginfo}
In the following we describe all 661 FITS columns of the
BlazEr1 catalogs, that is, the \emph{unverified} and the standard BlazEr1 catalog. As
described in Sect.~\ref{subsubsec:xraycounterparts}, these catalogs  were
obtained by merging various input catalogs, which are matched with the
eRASS1 catalog \citep{merloni:2024}, and amending this list of sources
with additional information such as, for example, spectral fit results. In
the following the energy bands for X-ray and $\gamma$-ray fluxes and luminosities
are denoted with $n$, where $n$ is defined in
Table~\ref{tab:en_desig}.
\begin{table}
   \caption{X-ray and $\gamma$-ray energy band designators used in the
     BlazEr1 catalog.}
    \label{tab:en_desig}
    \begin{tabular}{ll}
    \hline
    \hline
    channel $n$ & energy band \\
    \hline
     0 & 0.2--2.3\,keV \\
     1 & 0.2--2.3\,keV \\
    P1 & 0.2--0.5\,keV \\ 
    P2 & 0.5--1.0\,keV \\
    P3 & 1.0--2.0\,keV \\
    P4 & 2.0--5.0\,keV \\
    P5 & 5.0--8.0\,keV \\
    P6 & 4.0--10.0\,keV \\
    P7 & 5.1--6.1\,keV \\ 
    P8 & 6.2--7.1\,keV \\
    P9 & 7.2--8.2\,keV \\
     S & 0.5--2.0\,keV  \\
     T & 0.2--10.0\,keV  \\
     U & 0.1--2.4\,keV \\
     V & 0.3--10.0\,keV \\
     X & 3.0--10.0\,keV \\
     Y & 10.0--30.0\,keV \\
     Z & 14.0--195.0\,keV \\
     A & 0.1--100.0\,GeV \\
     B & 1.0--100.0\,GeV \\
     \hline
    \end{tabular}
\end{table}
For the fluxes determined by fitting a power law with fixed photon
index to the spectra we denote the photon index $G$ used, as listed in
Table~\ref{tab:ph_desig}.
\begin{table}
    \caption{Photon index designators $G$ for flux values}
    \label{tab:ph_desig}
    \begin{tabular}{lllll}
    \hline
    \hline
       Photon index & 1.5 & 1.7 & 2.0 & 2.3\\
       \hline
       $G$ &  15 & 17 & 20 & 23 \\
       \hline
    \end{tabular}
\end{table}
The LS10 bands (G, R, I, Z, W1, W2, W3, and W4) are denoted by $X$ in the column names listed below. For
\texttt{RFC} values the bands (s, c, x, u, and k) are denoted by $Y$. Not all
parameters will have all of the bands available denoted with $X$.
Where parameters have uncertainties, they are consistently listed in a
column obtained by appending \_ERR to the column name. For symmetric
error bars the column is a standard FITS column, asymmetric confidence
intervals of the type $P^{+u}_{-l}$ are given as fixed length arrays
with two entries, $-l$ and $+u$, including the sign.
Sometimes symmetric and asymmetric uncertainties are given, since some methods return asymmetric uncertainties which then are approximately summarized as symmetric ones (\_ERR,
\_ASYM\_ERR). 
In the online Vizier version the asymmetric uncertainties are given as two columns which are prefixed by E\_ and e\_ for upper and lower uncertainties, respectively.
Overall, column names might slightly vary in the Vizier version.
Unless noted otherwise, confidence intervals are given
at the $1\sigma$ confidence level, except for spectral parameters
(e.g., photon indices, fluxes), where 90\% confidence intervals are
given, following the conventions of X-ray astronomy. Sometimes the
inverse variance (\_IVAR) is given instead of an uncertainty. Where
columns are taken from other catalogs, we sometimes copy in part the original
column descriptions in verbatim.
\begin{description} 
\item[SRC\_NAME:] Common name of the source.
\item[RAJ2000, DECJ2000:] Equatorial coordinates from input catalog [$\degr$].
\item[BLAZAR\_CLASS:] Blazar classification given by the input catalog.
\item[CONF\_BLAZAR:] 1 for confirmed blazars and 0 for candidates.
\item[FGL\_SRC\_NAME:] Name of source if listed in 4FGL \citep[4FGL-DR4;][]{abdollahi:2022}.
\item[BZCAT\_SRC\_NAME:] Name of source if listed in BZCAT \citep{massaro:2015}.
\item[HSP\_SRC\_NAME:] Name of source if listed in 3HSP \citep{chang:2019}.
\item[HIGHZ\_SRC\_NAME:] Name of source if listed in the HighZ sample \citep{sbarrato2026}
\item[MILLIQUAS\_SRC\_NAME:] Name of source if listed in Milliquas \citep{flesch:2023}.
\item[KDEBLLACS\_SRC\_NAME:] Name of source if listed in KDEBLLACS \citep{dabrusco:2019}.
\item[WIBRALS2\_SRC\_NAME:] Name of source if listed in WIBRaLS2 \citep{dabrusco:2019}.
\item[ABC\_SRC\_NAME:] Name of source if listed in ABC \citep{paggi:2020}.
\item[BROS\_SRC\_NAME:] Name of source if listed in BROS \citep{itoh:2020}.
\item[ANGSEP:] Angular separation between catalog and \textit{eROSITA} position [$\degr$].
\item[DETUID:] \textit{eROSITA} unique detection ID.
\item[IAUNAME:] Official IAU name of the \textit{eROSITA} source.
\item[SKYTILE:] \textit{eROSITA} sky tile ID.
\item[ID\_SRC:] \textit{eROSITA} source ID in each sky tile.
\item[UID:] Integer unique detection ID. See \citet{merloni:2024} for details.
\item[UID\_HARD:] Hard catalog UID of the source with a strong association, or -UID if the association is weak. 0 means no counterpart found in the Hard catalog.
\item[ID\_CLUSTER:] Group ID of sources simultaneously modeled during source detection.
\item[RA(\_ERR), DEC(\_ERR):] Equatorial coordinates and associated uncertainties of \textit{eROSITA} source (ICRS) [$\degr$].
\item[RA\_RAW(\_ERR), DEC\_RAW(\_ERR):] Equatorial coordinates of \textit{eROSITA} source (ICRS), without applying the \textit{eROSITA} aspect correction [$\degr$].
\item[POS\_ERR:] $1\sigma$ position uncertainty [$''$].
\item[RADEC\_ERR:] Combined positional error, raw output from PSF fitting [$''$].
\item[LII, BII:] Galactic coordinates of \textit{eROSITA} source with applied position correction (not identical to the coordinates listed by \citealt{merloni:2024}) [$\degr$].
\item[ELON, ELAT:] Ecliptic coordinates of \textit{eROSITA} source [$\degr$].
\item[MJD:] Modified Julian Date of the observation of the source nearest to the optical axis of \textit{eROSITA}.
\item[MJD\_MIN, MJD\_MAX:] Modified Julian Date of the sources' first
  and last \textit{eROSITA} observation.
\item[EXT(\_ERR , \_ASYM\_ERR):] Source extent parameter and
  associated uncertainty [$''$].
\item[EXT\_LIKE:] Extent likelihood.
\item[DET\_LIKE\_$n$:] Detection likelihood.
\item[ML\_CTS\_$n$(\_ERR , \_ASYM\_ERR):] ML source net counts and associated uncertainty [$\mathrm{cts}$].
\item[ML\_RATE\_$n$(\_ERR , \_ASYM\_ERR):] ML source count rate and associated uncertainty [$\mathrm{cts}\,\mathrm{s}^{-1}$].
\item[ML\_FLUX\_$n$(\_ERR , \_ASYM\_ERR):] ML source flux and associated uncertainty [$\mathrm{erg}\,\mathrm{cm}^{-2}\,\mathrm{s}^{-1}$].
\item[ML\_BKG\_$n$:] ML background counts at the source position [$(')^{-2}$].
\item[ML\_EXP\_$n$:] ML vignetted exposure time at the source position [$\mathrm{s}$].
\item[ML\_EEF\_$n$:] ML enclosed energy fraction.
\item[APE\_CTS\_$n$:] Total counts extracted within the aperture [$\mathrm{cts}$].
\item[APE\_BKG\_$n$:] Background counts extracted within the aperture, excluding nearby sources using the source map [$\mathrm{cts}$].
\item[APE\_EXP\_$n$:] Exposure map value at the given position [$\mathrm{s}$].
\item[APE\_RADIUS\_$n$:] Extraction radius [pixel, 1\,pixel corresponds to $4''$].
\item[APE\_POIS\_$n$:] Poisson probability that the extracted counts (APE\_CTS) are background fluctuation.
\item[FLAG\_SP\_SNR:] If 1, source may lie within an overdense region near a supernova remnant.
\item[FLAG\_SP\_BPS:] If 1, source may lie within an overdense region near a bright point source.
\item[FLAG\_SP\_SCL:] If 1, source may lie within an overdense region near a stellar cluster.
\item[FLAG\_SP\_LGA:] If 1, source may lie within an overdense region near a local large galaxy.
\item[FLAG\_SP\_GC\_CONS:] If 1, source may lie within an overdense
  region near a galaxy cluster.
\item[FLAG\_NO\_RADEC\_ERR:] If 1, source contained no RA\_DEC\_ERR in
  the preprocessed version of the catalog.
\item[FLAG\_NO\_EXT\_ERR:] If 1, source contained no EXT\_ERR in the
  preprocessed version of the catalog.
\item[FLAG\_NO\_CTS\_ERR:] If 1, source contained no ML\_CTS\_0\_ERR
  in the preprocessed version of the catalog.
\item[FLAG\_OPT:] If 1, source matched within $15''$ with a bright
  optical star, likely contaminated by optical loading.
\item[CTS\_$n$\_(\_ERR):] Counts from spectrum and associated
  uncertainty  \citep{gehrels:1986} [$\mathrm{cts}$].
\item[SPEC\_EXPOTIME:] Exposure time as given in the spectrum
  [$\mathrm{s}$].
\item[EXPOTIME:] Exposure time from the \textit{eROSITA} exposure map for the
  0.2--2.3\,keV band [$\mathrm{s}$].
\item[MIN\_FLUX\_95:] Minimum detectable flux at source position to
  guarantee 95\% detection rate with a DET\_LIKE\_0 of 10 or greater
  in the 0.2--2.3\,keV band
  [$\mathrm{erg}\,\mathrm{cm}^{-2}\,\mathrm{s}^{-1}$].
\item[AVERAGE\_NH:] Average $N_\mathrm{H}$ from the
  \citet{hi4picollaboration:2016} [$\mathrm{cm}^{-2}$].
\item[WE\_AVERAGE\_NH:] Weighted average $N_\mathrm{H}$ from the \citet{hi4picollaboration:2016} [$\mathrm{cm}^{-2}$].
\item[HR12(\_ERR):] Hardness ratio between 0.2--0.7\,keV and
  0.7--1.2\,keV and associated uncertainty.
\item[HR23(\_ERR):] Hardness ratio between 0.7--1.2\,keV and
  1.2--2.3\,keV and associated uncertainty.
\item[HR13(\_ERR):] Hardness ratio between 0.2--0.7\,keV and
  1.2--2.3\,keV and associated uncertainty.
\item[PWL$G$\_NORM(\_ERR):] Power law norm using a fixed $N_\mathrm{H}$
  and an index of $G$ at 1\,keV and associated uncertainty [$\mathrm{photons}\,\mathrm{keV}^{-1}\,\mathrm{cm}^{-2}\,\mathrm{s}^{-1}$].
\item[PWL$G$\_FLUX\_(UNABS , ABS)\_$n$(\_ERR):] Fitted unabsorbed and
  observed flux using a fixed $N_\mathrm{H}$ and fixed power law index
  $G$ and associated uncertainty
  [$\mathrm{erg}\,\mathrm{cm}^{-2}\,\mathrm{s}^{-1}$].
\item[PWL$G$\_C\_VARIANCE:] Variance of Cash statistics of fixed
  $N_\mathrm{H}$ and fixed power law index fit according to
  \citet{kaastra:2017}.
\item[PWL$G$\_CASH:] Cash statistics of fixed $N_\mathrm{H}$ and fixed
  power law index fit.
\item[PWL$G$\_CSTAT\_THEORY:] Theoretical Cash statistics while using
  a fixed $N_\mathrm{H}$ and fixed power law index fit according to \citet{kaastra:2017}.
\item[PWL$G$\_REDCASH:] Reduced Cash statistics from fit with fixed
  $N_\mathrm{H}$ and fixed power law index.
\item[PWL$G$\_CASHBINS:] Number of bins used during the fixed
  $N_\mathrm{H}$ and fixed power law index fit.
\item[PWL$G$\_CASHPAR:] Number of parameters used during the fixed
  $N_\mathrm{H}$ and fixed power law index fit.
\item[PWL$G$\_NUM\_BINS:] Number of bins used to calculate
  \citet{kaastra:2017} values of fixed $N_\mathrm{H}$ and fixed power
  law index.
\item[Flux\_JY(\_ERR):] Flux density at 1\,keV derived from the norm
  of the fixed $\Gamma_\mathrm{X}=2.0$ fit [$\mathrm{Jy}$].
\item[(FREE\_ , FIXED\_)NH(\_ERR):] Freely fitted or fixed
  $N_\mathrm{H}$ and associated uncertainty [$\mathrm{cm}^{-2}$].
\item[(FREE\_ , FIXED\_)GAMMA(\_ERR):] Power law photon index using a
  free or fixed $N_\mathrm{H}$ and associated uncertainty.
\item[(FREE\_ , FIXED\_)NORM(\_ERR):] Power law norm using a free or fixed
  $N_\mathrm{H}$ at $1\,\mathrm{keV}$ and associated uncertainty
  [$\mathrm{photons}\,\mathrm{keV}^{-1}\,\mathrm{cm}^{-2}\,\mathrm{s}^{-1}$].
\item[(FREE\_ , FIXED\_)FLUX\_(UNABS , ABS)\_$n$(\_ERR):] Fitted
  unabsorbed and observed flux for a free or fixed $N_\mathrm{H}$
  power law fit and associated uncertainty
  [$\mathrm{erg}\,\mathrm{cm}^{-2}\,\mathrm{s}^{-1}$].
\item[(FREE\_ , FIXED\_)C\_VARIANCE:] Variance of Cash statistics of
  free or fixed $N_\mathrm{H}$ power law fit according to
  \citet{kaastra:2017}.
\item[(FREE\_ , FIXED\_)CASH:] Cash statistics of free or fixed
  $N_\mathrm{H}$ power law fit.
\item[(FREE\_ , FIXED\_)CSTATS\_THEORY:] Theoretical Cash statistics
  while using a free or fixed $N_\mathrm{H}$ power law fit according
  to \citet{kaastra:2017}.
\item[(FREE\_ , FIXED\_)REDCASH:] Reduced Cash statistics from fit
  with free or fixed $N_\mathrm{H}$.
\item[(FREE\_ , FIXED\_)CASHBINSs:] Number of bins used during the
  free or fixed $N_\mathrm{H}$ power law fit.
\item[(FREE\_ , FIXED\_)CASHPAR:] Number of parameters used during the
  free or fixed $N_\mathrm{H}$ power law fit.
\item[(FREE\_ , FIXED\_)NUM\_BINS:] Number of bins used to calculate
  \citet{kaastra:2017} values of free or fixed $N_\mathrm{H}$ power
  law fit.
\item[RXS\_RA , RXS\_DEC:] Equatorial position of 2RXS source
  \citep{boller:2016} [$\degr$].
\item[RXS\_NH\_GAL:] Galactic $N_\mathrm{H}$ assumed by
  \citet{boller:2016} [$\mathrm{cm}^{-2}$].
\item[RXS\_NH:] $N_\mathrm{H}$ derived from fit by \citet{boller:2016}
  [$\mathrm{cm}^{-2}$].
\item[RXS\_GAMMA(\_ERR):] $\Gamma_\mathrm{X}$ from fit and associated
  uncertainty by \citet{boller:2016}.
\item[RXS\_FLUX\_$n$:] Flux from power law fit by \citet{boller:2016}
  [$\mathrm{erg}\,\mathrm{cm}^{-2}\,\mathrm{s}^{-1}$].
\item[XRT\_SRC\_NAME:] Name given in \citet{evans:2020}.
\item[XRT\_RA, XRT\_DEC:] Equatorial coordinates from
  \citet{evans:2020} [$\degr$].
\item[XRT\_XRT\_ANGSEP:] Angular separation between blazar catalog
  source and closest source in \citet{evans:2020} [\arcsec].
\item[XRT\_FLAG:] If 1 the blazar and the source listed by
  \citet{evans:2020} are located within $8''$ of each other.
\item[XRT\_NH(\_ERR):] Galactic $\mathrm{N}_\mathrm{H}$ and associated
  uncertainty from \citet{evans:2020} [$\mathrm{cm}^{-2}$].
\item[XRT\_GAMMA(\_ERR):] Power law photon index and associated
  uncertainty from \citet{evans:2020}.
\item[XRT\_(ABS , UNABS)\_FLUX\_$n$(\_ERR):] Unabsorbed and observed
  flux and associated uncertainty from \citet{evans:2020}
  [$\mathrm{erg}\,\mathrm{cm}^{-2}\,\mathrm{s}^{-1}$].
\item[OUSXB\_RA, OUSXB\_DEC:] Equatorial coordinates of \textit{Swift}-XRT sources
  from \citet{giommi:2019} [$\degr$].
\item[OUSXB\_(MIN, MAX, MEAN, MEDIAN)\_FLUX\_$n$:] Minimum, maximum,
  median and mean sources flux from \citep{giommi:2019}
  [$\mathrm{erg}\,\mathrm{cm}^{-2}\,\mathrm{s}^{-1}$].
\item[OUSXB\_N\_FLUX\_$n$:] Number of observations of source in
  \citet{giommi:2019} with a flux.
\item[OUSXB\_(MIN, MAX, MEAN, MEDIAN)\_GAMMA:] Minimum, maximum,
  median and mean power law photon index \citep{giommi:2019}.
\item[OUSXB\_N\_GAMMA:] Number of observations of source in
  \citet{giommi:2019} with a power law photon index.
\item[NUSTAR\_RA, NUSTAR\_DEC:] Equatorial coordinates of \textit{NuSTAR}
  sources from \citet{middei:2022} [$\degr$].
\item[NUSTAR\_GAMMA(\_ERR):] Photon index of \textit{NuSTAR} sources from
  \citet{middei:2022}.
\item[NUSTAR\_FLUX\_$n$(\_ERR):] Flux of \textit{NuSTAR} sources from
  \citet{middei:2022} [$\mathrm{erg}\,\mathrm{cm}^{-2}\,\mathrm{s}^{-1}$].
\item[BAT\_SRC\_NAME:] Name of source given by \citet{lien:2025}.
\item[BAT\_RA, BAT\_DEC:] Equatorial coordinates of \textit{Swift}-BAT sources from
  \citet{lien:2025} [$\degr$].
\item[BAT\_TYPE:] Source type provided by \citet{lien:2025}.
\item[BAT\_( , MIN\_ , MAX\_)FLUX\_$n$:] Flux, minimum and maximum
  flux of \textit{Swift}-BAT sources from \citet{lien:2025}
  [$\mathrm{erg}\,\mathrm{cm}^{-2}\,\mathrm{s}^{-1}$].
\item[BAT\_( , MIN\_ , MAX\_)GAMMA:] Photon index of \textit{Swift}-BAT sources
  from \citet{lien:2025}.
\item[ASCAMASTER\_EXPOTIME:] \textit{ASCA} exposure time according to
  Heasarc's ascamaster table [$\mathrm{s}$].
\item[CHANMASTER\_EXPOTIME] \textit{Chandra} exposure time according to the
  chanmaster table [$\mathrm{s}$].
\item[NUMASTER\_EXPOTIME:] \textit{NuSTAR} exposure time according to the numaster
  table [$\mathrm{s}$].
\item[SUZAMASTER\_EXPOTIME:] \textit{Suzaku} exposure time according to the
  suzamaster table [$\mathrm{s}$].
\item[XMMMASTER\_EXPOTIME:] \textit{XMM-Newton} exposure time according to the xmmmaster
  table [$\mathrm{s}$].
\item[SWIFTMASTR\_EXPOTIME:] \textit{Swift}-XRT exposure time according to
  the swiftmastr table in [$\mathrm{s}$].
\item[ROSMASTER\_EXPOTIME:] \textit{ROSAT} exposure time according to the
  rosmaster table in [$\mathrm{s}$].
\item[FGL\_COMMON\_SRC\_NAME:] Name of identified or likely associated source from the 4FGL catalog \citep[4FGL-DR4;][]{abdollahi:2022}.
\item[FGL\_FLUX\_$n$(\_ERR):] Integral photon flux and associated
  uncertainty for the B band \citep[4FGL-DR4;][]{abdollahi:2022}
  [$\mathrm{photons}\,\mathrm{cm}^{-2}\,\mathrm{s}^{-1}$].
\item[FGL\_FLUX\_$n$(\_ERR):] Energy flux and associated uncertainty
  for the A band \citep[4FGL-DR4;][]{abdollahi:2022}
  [$\mathrm{erg}\,\mathrm{cm}^{-2}\,\mathrm{s}^{-1}$].
\item[FGL\_SPEC\_TYPE:] Spectral type in the global model (Power Law,
  LogParabola, PLSuperExpCutoff) \citep[4FGL-DR4;][]{abdollahi:2022}.
\item[FGL\_PL\_FLUXDEN(\_ERR):] Differential flux at pivot energy of
  the power law fit and associated uncertainty \citep[4FGL-DR4;][]{abdollahi:2022}
  [$\mathrm{cm}^{-2}\,\mathrm{MeV}^{-1}\,\mathrm{s}^{-1}$].
\item[FGL\_PL\_INDEX(\_ERR):] Photon index of power law fit and
  associated uncertainty \citep[4FGL-DR4;][]{abdollahi:2022}.
\item[FGL\_LP\_FLUXDEN(\_ERR):] Differential flux at pivot energy of
  the LogPar fit and associated uncertainty \citep[4FGL-DR4;][]{abdollahi:2022}
  [$\mathrm{cm}^{-2}\,\mathrm{MeV}^{-1}\,\mathrm{s}^{-1}$].
\item[FGL\_LP\_INDEX(\_ERR):] Photon index at pivot energy for LogPar
  fit and associated uncertainty \citep[4FGL-DR4;][]{abdollahi:2022}.
\item[FGL\_LP\_BETA(\_ERR):] Curvature parameter of LogPar fit and
  associated uncertainty \citep[4FGL-DR4;][]{abdollahi:2022}.
\item[FGL\_LP\_EPEAK(\_ERR):] Peak energy of LogPar fit and associated
  uncertainty \citep[4FGL-DR4;][]{abdollahi:2022} [$\mathrm{MeV}$].
\item[FGL\_LP\_SIGCURV:] Significance of fit improvement between Power
  Law and LogParabola \citep[4FGL-DR4;][]{abdollahi:2022} [$\sigma$].
\item[FGL\_PLEC\_FLUXDEN(\_ERR):] Differential flux at pivot energy of
  the PLSuperExpCutoff fit and associated uncertainty
  \citep[4FGL-DR4;][]{abdollahi:2022} [$\mathrm{cm}^{-2}\,\mathrm{MeV}^{-1}\,\mathrm{s}^{-1}$].
\item[FGL\_PLEC\_INDEX(\_ERR):] Photon index at pivot energy for
  PLSuperExpCutoff fit and associated uncertainty \citep[4FGL-DR4;][]{abdollahi:2022}.
\item[FGL\_PLEC\_EXPFACTOR(\_ERR):] Spectral curvature at pivot energy
  for PLSuperExpCutoff fit and associated uncertainty
  \citep[4FGL-DR4;][]{abdollahi:2022}.
\item[FGL\_PLEC\_EXPINDEX(\_ERR):] Exponential index of
  PLSuperExpCutoff and associated uncertainty \citep[4FGL-DR4;][]{abdollahi:2022}.
\item[FGL\_PLEC\_EPEAK(\_ERR):] Peak energy of PLSuperExpCutoff fit
  and associated uncertainty \citep[4FGL-DR4;][]{abdollahi:2022} [$\mathrm{MeV}$].
\item[FGL\_PLEC\_SIGCURV:] Significance of fit improvement between
  Power Law and PLSuperExpCutoff \citep[4FGL-DR4;][]{abdollahi:2022} [$\sigma$].
\item[LAC\_SED\_CLASS:] SED classes listed in the 4LAC catalog by
  \citet{ajello:2022}.
\item[LAC\_HE\_EPEAK(\_ERR):] High-energy peak position listed in the
  4LAC catalog by \citet{ajello:2022} and associated uncertainty [MeV].
\item[LAC\_HE\_NUFNU(\_ERR):] High-energy peak flux listed in the 4LAC
  catalog by \citet{ajello:2022}
  [$\mathrm{MeV}\,\mathrm{cm}^{-2}\mathrm{s}^{-1}$].
\item[LAC\_LE\_FREQ:] Low-energy peak position listed in the 4LAC
  catalog by \citet{ajello:2022} [$\mathrm{Hz}$].
\item[LAC\_LE\_NUFNU:] Low-energy peak flux listed in the 4LAC
  catalog by \citet{ajello:2022}
  [$\mathrm{erg}\,\mathrm{cm}^{-2}\,\mathrm{s}^{-1}$].
\item[RFC\_SRC\_NAME:] Name of source in \texttt{RFC} catalog
  \citep{petrov:2025}.
\item[RFC\_$Y$BAND\_RES:] Total $Y$-band flux density integrated over
  entire map \citep{petrov:2025} [$\mathrm{Jy}$].
\item[RFC\_$Y$UPPER\_RES:] Blank for valid $Y$-band resolved total
  flux density, $-$ for invalid and $<$ for upper limits
  \citep{petrov:2025}.
\item[RFC\_$Y$BAND\_UNRES:] Unresolved $Y$-band flux density at long
  VLBA baselines \citep{petrov:2025} [$\mathrm{Jy}$].
\item[RFC\_$Y$UPPER\_UNRES:] Blank for valid $Y$-band unresolved flux density at VLBA baselines, $-$ for invalid and $<$ for upper limits \citep{petrov:2025}.
\item[TANAMI\_SRC\_NAME:] Name within \texttt{TANAMI} program \citep{ojha:2010,muller:2018}. 
\item[TANAMI\_CLASS:] Class given by \texttt{TANAMI}
  \citep{ojha:2010,muller:2018}.
\item[TANAMI\_FLUX:] Flux density at 8.4\,GHz from \texttt{TANAMI}
  \citep{ojha:2010,muller:2018} [$\mathrm{Jy}$].
\item[MOJAVE\_SRC\_NAME:] Name of source in the \texttt{MOJAVE}
  program \citep{lister:2021,homan:2021}.
\item[MOJAVE\_CLASS:] Class as listed by \texttt{MOJAVE}
  \citep{lister:2021,homan:2021} (Q or B).
\item[MOJAVE\_FLUX:] Flux density at 15\,GHz from \texttt{MOJAVE}
  \citep{lister:2021,homan:2021} [$\mathrm{mJy}$].
\item[N\_MATCH:] Number of potential counterparts (S25).
\item[LS10\_sep:] Angular separation between BLAZE catalog and LS10 source
  [\arcsec].
\item[LS10\_use:] 1 for agreement on the counterpart with S25,
  0 otherwise.
\item[LS10\_RELEASE:] LS10 release of counterpart (S25).
\item[LS10\_BRICKID:] LS10 brick ID of counterpart (S25).
\item[LS10\_OBJID:] LS10 object ID of counterpart (S25).
\item[LS10\_RA(\_IVAR) , LS10\_DEC(\_IVAR):] Position of LS10
  counterpart (S25) [$\degr$ , $(\degr)^{-2}$].
\item[LS10\_Xray\_proba:] Probability of LS10 counterpart being an
  X-ray emitter (S25).
  \item[LS10\_FULLID:] LS10 catalog ID (S25).
\item[ERO\_LS10\_ID:] Combined \textit{eROSITA} and LS10 catalog ID (S25).
\item[Separation\_LS10\_ERO:] Separation between LS10 counterpart and
  \textit{eROSITA} position (S25) [$''$].
\item[NCAT:] Number of catalogs matched (S25).
\item[DIST\_BAYESFACTOR:] Logarithm of ratio from prior and posterior
  distance matching (S25).
\item[DIST\_POST:] Distance between posterior probability of a match
  compared to no matches (S25).
  \item[BIAS\_LS10\_XRAY\_PROBA:] Probability of counterpart being an X-ray emitter (S25).
\item[P\_SINGLE:] Similar to DIST\_POST including weighting by prior
  (S25).
\item[P\_ANY:] Probability of any found match being correct (S25).
\item[P\_I:] Match probability (S25).
\item[MATCH\_FLAG:] Set to 1 for most probable match, 2 for other
  matches (S25).
\item[TYPE:] LS10 morphology model given by S25.
\item[LS10\_DCHISQ:] Difference in $\chi^2$ between successively more-complex model fits separated by \_ (S25).
\item[EBV:] $E(\mathrm{B}-\mathrm{V})$ at source position (S25)
  [$\mathrm{mag}$] .
\item[FLUX(\_IVAR)\_$X$:] LS10 band fluxes and associated inverse
  variance (S25) [$\mathrm{nanomaggy}$ ,
    $\mathrm{nanomaggy}^{-2}$].
\item[MW\_TRANSMISSION\_$X$:] LS10 Galactic transmission at position
  of source (S25).
\item[ANYMASK\_$X$:] Bitmask information if in any image of a band the
  central pixel satisfies the LS10 catalog criteria (S25).
\item[ALLMASK\_$X$:] Bitmask information if in all image of a band the
  central pixel satisfies the LS10 catalog criteria (S25).
\item[PSFSIZE\_$X$:] Size of FWHM of the LS10 PSF (S25)
  [$''$].
\item[PSFDEPTH\_$X$:] LS10 $5\sigma$ detection limit for a certain
  band (S25) [$\mathrm{nanomaggy}^{-2}$].
\item[GALDEPTH\_$X$:] Same as PSFSIZE\_$X$ assuming a galaxy (S25) [$\mathrm{nanomaggy}^{-2}$].
\item[WISEMASK\_$X$:] Bitmask information (S25).
\item[W(1 , 2 , 3 , 4)\_MAG\_AB:] WISE magnitude in AB system
  calculated from FLUX\_$X$ using MW\_TRANSMISSION\_$X$
  [$\mathrm{mag}$].
\item[W(1 , 2 , 3 , 4)\_MAG:] WISE magnitude in the Vega system calculated
  from FLUX\_$X$ using MW\_TRANSMISSION\_$X$ [$\mathrm{mag}$].
\item[SHAPE\_R(\_IVAR):] LS10 half-light radius for given TYPE (S25) [$''$ , $''^{-2}$].
\item[SHAPE\_(E1 , E2)(\_IVAR):] LS10 ellipticity component of galaxy
  (S25).
  \item[SERSIC(\_IVAR):] LS10 power law index of S\'ersic profile and
  associated inverse variance (S25).  
\item[REF\_CAT:] LS10 reference catalog for stars (S25).	 
\item[REF\_ID:]	 LS10 reference catalog id for star (S25).
\item[REF\_EPOCH:] LS10 reference epoch of REF\_CAT (S25) [$\mathrm{a}$].
\item[GAIA\_PHOT\_G\_MEAN\_MAG:] LS10 Gaia G band magnitude (S25) [$\mathrm{mag}$].
\item[GAIA\_PHOT\_G\_MEAN\_FLUX\_OVER\_ERROR:] LS10 Gaia G band
  signal-to-noise ratio (S25).
\item[GAIA\_PHOT\_BP\_MEAN\_MAG:] LS10 Gaia BP band magnitude (S25)
 [$\mathrm{mag}$].
\item[GAIA\_PHOT\_BP\_MEAN\_FLUX\_OVER\_ERROR:] LS10 Gaia BP band
  signal-to-noise ratio (S25).
\item[GAIA\_PHOT\_RP\_MEAN\_MAG:] LS10 Gaia RP band magnitude (S25)
  [$\mathrm{mag}$].
\item[GAIA\_PHOT\_RP\_MEAN\_FLUX\_OVER\_ERROR:] LS10 Gaia RP band
  signal-to-noise ratio (S25).
  \item[GAIA\_ASTROMETRIC\_EXCESS\_NOISE:] LS10 Gaia astrometric excess noise (S25).
  \item[GAIA\_PHOT\_BP\_RP\_EXCESS\_FACTOR:] LS10 Gaia BP/RP excess factor (S25).
  \item[GAIA\_ASTROMETRIC\_SIGMA5D\_MAX:] LS10 Gaia longest semi-major axis of the 5-d error ellipsoid (S25).
  \item[GAIA\_DUPLICATED\_SOURCE:] LS10 Gaia duplicate source flag (S25).
\item[INALLLS10:] Nominal LS10 survey depth reached in all bands (S25).
\item[INANYLS10:] Nominal LS10 survey depth reached in one band (S25).
\item[DERED\_MAG\_$X$:] Dereddend magnitude (S25) [$\mathrm{mag}$].
\item[SOFTFLUX\_S:] \textit{eROSITA} flux (S25)
  [$\mathrm{erg}\,\mathrm{cm}^{-2}\,\mathrm{s}^{-1}$].
\item[SALVATO18\_W1\_X\_LINEDISTANCE:] Distance from separation line
  between Galactic and extra-galactic sources in the W1-X-ray plane
  (S25).
  \item[SALVATO22\_ZW1\_GR\_LINEDISTANCE:] Distance from separation line
  between Galactic and extra-galactic sources in the ZW1-GR plane
  (S25).
\item[(G , Z , W1)\_MINUS\_(R , W1 , W2):] Color combinations (S25)
  [$\mathrm{mag}$].
\item[HAS\_SALVATO18\_AGN\_COLORS:] Information if source has colors
  consistent with being an AGN based on \citet{salvato:2018}, from S25.
\item[GAIA\_MOVING\_5SIGMA:] Information if source shows movement
  at $5\sigma$ significance in Gaia (S25).
\item[HAS\_SALVATO22\_AGN\_COLORS:] Information if source has colors
  consistent with being an AGN based on \citet{salvato:2022}, from
  S25.
\item[SALVATO\_MAIN\_ID\_SIMBAD:] SIMBAD ID (S25).
\item[SALVATO\_(RA , DEC)\_SIMBAD:] SIMBAD position (S25)
  [$\degr$].
  \item[EXGAL\_PROB\_STAREX:] Probability of source being extra-galactic
  by STAREX with Gaia data (S25).
\item[EXGAL\_PROB\_STAREX:] Probability of source being extra-galactic
  by STAREX without Gaia data (S25).
\item[CLASS\_GAL\_EXGAL:] Galactic/extra-galactic class assigned by S25.
\item[SALVATO\_HECATE:] 1 for sources within HECATE source (S25) [$\degr$].
\item[SALVATO\_HEC\_OBJNAME:] Name of HECATE source (S25).
\item[SALVATO\_HEC\_RA , SALVATO\_HEC\_DEC:] Position of HECATE source
  (S25) [$\degr$].
\item[SALVATO\_HEC\_R1:] Semi major axis of HECATE source (S25) [$\arcmin$].
\item[SALVATO\_HEC\_R2:] Semi minor axis of HECATE source (S25) [$\arcmin$].
\item[SALVATO\_HEC\_PA:] Position angle of HECATE source (S25) [$\degr$].
\item[PURITY(6 , 7 , 8):] Purity obtained for $\mbox{DET\_LIKE\_0}>(6
  , 7 , 8)$ at source position (S25).
\item[COMPLETENESS(6 , 7 , 8):] Completeness obtained for
  $\mbox{DET\_LIKE\_0}>(6 , 7 , 8)$ at source position (S25).
\item[COMPUR(6 , 7 , 8):] Intersection between completeness and purity
  obtained for $\mbox{DET\_LIKE\_0}>(6 , 7 , 8)$ at source position
  (S25).
\item[THRESHOLD(6 , 7 , 8):] P\_ANY value at intersection obtained for
  \mbox{DET\_LIKE\_0>(6 , 7 , 8)} at source position (S25).
\item[Z\_PHOTO(\_S3L , \_S3U , \_S1L , \_S1U):] $3\sigma$ and
  $1\sigma$ upper and lower value of photometric redshift (S25).
\item[Z\_SPEC:] Spectroscopic redshift from \citet{kluge:2024}
  provided by S25.
\item[SPEC\_REF:] Source with Z\_SPEC from
  \citet{kluge:2024} provided by S25.
\item[Z\_ABC:] Redshift listed in the ABC catalog \citep{paggi:2020}.
\item[Z\_HSP:] Redshift listed in the 3HSP catalog \citep{chang:2019}.
\item[Z\_F\_HSP:] Redshift flag listed in the 3HSP catalog \citep{chang:2019}.
\item[Z\_MILLIQUAS:] Redshift listed in the Milliquas catalog
  \citep{flesch:2023}.
\item[Z\_HIGHZ:] Redshift listed in the HighZ sample.
\item[Z\_HIGHZ\_REF:] Paper reference for high-redshift sources.
\item[Z\_MOJAVE:] Redshift listed in \texttt{MOJAVE}.
\item[Z\_4LAC:] Redshift listed in the 4LAC catalog \citep{ajello:2022}.
\item[Z\_BZCAT:] Redshift listed from the BZCAT catalog
  \citep{massaro:2015}.
\item[Z\_U\_BZCAT:] Redshift flag from the BZCAT catalog
  \citep{massaro:2015}.
\item[Z\_CGRABS:] Redshift listed in the CGRaBS catalog
  \citep{healey:2008}.
\item[Z\_F\_CGRABS:] Redshift flag listed in CGRaBS catalog
  \citep{healey:2008}.
\item[Z\_VERONCAT:] Redshift listed in the VERONCAT catalog
  \citep{veron-cetty:2010}.
\item[Z\_VERONCAT\_QSO:] Redshift listed in the VERONCAT QSO catalog
  \citep{veron-cetty:2010}.
\item[Z\_SIMBAD:] Redshift listed on SIMBAD.
\item[Z\_MASTER:] Reliable redshift adopted for population studies
  (for selection see Sect.~\ref{subsec:optdata}).
\item[Z\_MASTER\_REF:] Source of Z\_MASTER.
\item[Z\_MASTER\_PHOTO:] Reliable redshift including the photometric
  ones from S25.
\item[Z\_MASTER\_PHOTO\_REF:] Source where Z\_MASTER\_PHOTO
  was taken from.
\item[XRAY\_LUM\_$n$(\_ERR):] X-ray luminosity calculated from the
  fixed $\Gamma_\mathrm{X}=2.0$ fit and Z\_MASTER
  [$\mathrm{erg}\,\mathrm{s}^{-1}$].
\item[XRAY\_LUM\_PHOTO\_$n$(\_ERR):] X-ray luminosity calculated from
  the fixed $\Gamma_\mathrm{X}=2.0$ fit and Z\_MASTER\_PHOTO
  [$\mathrm{erg}\,\mathrm{s}^{-1}$].
\item[alpha\_XG(\_ERR):] $\alpha_{\mathrm{X}\Gamma}$ using
  PWL20\_FLUX\_UNABS\_0 and FGL\_FLUX\_A and associated uncertainty.
\item[alpha\_IRX(\_ERR):] $\alpha_{\mathrm{IRX}}$ using
  PWL20\_FLUX\_UNABS\_0 and FLUX\_W1 combined with
  MW\_TRANSMISSION\_W1 and associated uncertainty, including a
  systematic uncertainty.
\item[alpha\_OX(\_ERR):] $\alpha_{\mathrm{OX}}$ using
  PWL20\_FLUX\_UNABS\_0 and FLUX\_R combined with MW\_TRANSMISSION\_R
  and associated uncertainty, including a systematic uncertainty.
\item[alpha\_RX(\_ERR):] $\alpha_{\mathrm{RX}}$ using
  PWL20\_FLUX\_UNABS\_0 and RFC\_XBAND\_RES and associated
  uncertainty.
\item[alpha\_TANAMI(\_ERR):] $\alpha_{\mathrm{TANAMI}}$ using
  PWL20\_FLUX\_UNABS\_0 and TANAMI\_FLUX and associated uncertainty.
\item[alpha\_MOJAVE(\_ERR):] $\alpha_{\mathrm{MOJAVE}}$ using
  PWL20\_FLUX\_UNABS\_0 and MOJAVE\_FLUX and associated uncertainty.
\item[SIMBAD\_SRC\_NAME:] Name of closest source match on SIMBAD.
\item[SIMBAD\_TYPE:] Type of source, if a blazar or AGN classification
  or a more precise type is listed as alternative, this one is
  selected.
\item[SIMBAD\_ALT\_TYPES:] All alternate types as listed in SIMBAD.
\item[SIMBAD\_TYPE2:] Type of source as listed by SIMBAD.
\item[SIMBAD\_ANGSEP:] Angular separation between BLAZE catalog source and
  SIMBAD source [$\degr$]
\item[CAMBRIDGE\_SRC\_NAME:] Name in Third Cambridge catalog of radio sources as listed in SIMBAD.
\item[PARKS\_SRC\_NAME:] Name in Parkes catalog of radio sources as
  listed in SIMBAD.
\item[TEXAS\_SRC\_NAME:] Name in Texas survey of radio sources as listed
  in SIMBAD.
\item[EINSTEIN\_SRC\_NAME:] Name in Einstein 2E catalog of IPC X-Ray sources as listed in SIMBAD.
\item[COMMON\_SRC\_NAME:] Most commonly used source name from the
  above as listed in SIMBAD.
\item[BZCAT\_RMAG:] R-band magnitude from the BZCAT catalog
  \citep{massaro:2015} [$\mathrm{mag}$].
\item[BZCAT\_FR:] Radio flux density at 1.4\,GHz or if unknown at
  0.843\,GHz from the BZCAT catalog \citep{massaro:2015}
  [$\mathrm{mJy}$].
\item[BZCAT\_F143:] Flux density at 143\,GHz from the Planck Compact
  Source Catalog Public Release 1 from the BZCAT catalog
  \citep{massaro:2015} [$\mathrm{mJy}$].
\item[BZCAT\_FLUX\_U:] X-ray flux in the 0.1--2.4\,keV band from the
  BZCAT catalog \citep{massaro:2015}
  [$\mathrm{fW}\mathrm{m}^{-2}$].
\item[BZCAT\_FLUX\_B:] $\gamma$-ray flux in the 1--100\,GeV band from the
  BZCAT catalog \citep{massaro:2015}
  [$\mathrm{photons}\,\mathrm{cm}^{-2}\,\mathrm{s}^{-1}$].
\item[BZCAT\_ALPHA\_RO:] The radio-to-optical spectral index of the
  blazar from the BZCAT catalog \citep{massaro:2015}.
\item[XIE\_MORPH\_BZCAT:] Radio morphology for BZCAT sources as listed
  in \citet{xie:2024}.
\item[XIE\_MORPH\_KDEBLLACS:] Radio morphology for KDEBLLACS sources
  as listed in \citet{xie:2024}.
\item[XIE\_MORPH\_WIBRALS:] Radio morphology for WIBRaLS sources as
  listed in \citet{xie:2024}.
\end{description}
\end{appendix}
\end{document}